\documentclass[11pt,a4paper,fleqn]{article}
\usepackage[utf8]{inputenc}
\usepackage{authblk}
\usepackage{geometry}
\usepackage{xcolor}
\usepackage{natbib}
\usepackage{graphicx}
\usepackage{pdflscape}
\usepackage{appendix}
\usepackage{setspace}
\usepackage{colortbl}

\usepackage{threeparttable}
\usepackage{booktabs}

\usepackage{amsmath}
\usepackage{amsfonts}
\usepackage{bm}

\usepackage{xpatch}
\xapptocmd\normalsize{%
 \abovedisplayskip= 6pt plus 3pt minus 9pt
 \abovedisplayshortskip=0pt plus 3pt
 \belowdisplayskip=6pt plus 3pt minus 9pt
 \belowdisplayshortskip=0pt plus 3pt
}{}{}

\usepackage{enumitem}
\newlist{steps}{enumerate}{1}
\setlist[steps, 1]{label = Step \arabic*:}

\usepackage{dcolumn}
\newcolumntype{d}[1]{D{.}{.}{#1}}



\usepackage{caption}
\usepackage{subcaption}
\definecolor{nblue}{HTML}{000660}
\usepackage[colorlinks=true,urlcolor=nblue,linkcolor=nblue,citecolor=nblue]{hyperref}

\title{\LARGE \textbf{Fast and Flexible Bayesian Inference in Time-varying Parameter Regression Models}}
\author[1]{\MakeUppercase{Niko Hauzenberger}}
\author[1]{\MakeUppercase{Florian Huber}\thanks{Corresponding author: Florian Huber. Department of Economics, University of Salzburg. \textit{Address}: M\"{o}nchsberg 2a, 5020 Salzburg, Austria. \textit{Email}: \href{mailto:florian.huber@sbg.ac.at}{florian.huber@sbg.ac.at}. We would like to thank the participants of the $6^{th}$ NBP Workshop on Forecasting (Warsaw, 2019), the $11^{th}$ European Seminar on Bayesian Econometrics (Madrid, 2021) and internal seminars at the University of Salzburg, the FAU Erlangen-Nuremberg and the ECB, four anonymous referees as well as Anna Stelzer, Michael Pfarrhofer and Paul Hofmarcher  for helpful comments and suggestions. The first two authors gratefully acknowledge financial support by the Austrian Science Fund (FWF): ZK 35 and by funds of the Oesterreichische Nationalbank
(Austrian Central Bank, Anniversary Fund, project number: 18127). This paper, subject to minor editorial changes, will appear in the Journal of Business \& Economic Statistics.}}
\author[2]{\MakeUppercase{\\Gary Koop}}
\author[3,4]{\MakeUppercase{Luca Onorante}}
\affil[1]{\textit{University of Salzburg}}
\affil[2]{\textit{University of Strathclyde}}
\affil[3]{\textit{Joint Research Centre, European Commission}}
\affil[4]{\textit{European Central Bank}}
\date{\today}

\begin{document}

\maketitle\thispagestyle{empty}\normalsize\vspace*{-2em}\small
\begin{center}
\begin{minipage}{0.85\textwidth}
\noindent\small In this paper, we write the time-varying parameter (TVP) regression model involving $K$ explanatory variables and $T$ observations as a constant coefficient regression model with $KT$ explanatory variables.
In contrast with much of the existing literature which assumes coefficients to evolve according to a random walk,  a hierarchical mixture model on the TVPs is introduced. The resulting model closely mimics a random coefficients specification which groups the TVPs into several regimes. These flexible mixtures allow for TVPs that feature a small, moderate or large number of structural breaks. We develop computationally efficient Bayesian econometric methods based on the singular value decomposition of the $KT$ regressors.  In artificial data, we find our methods to be accurate and much faster than standard approaches in terms of computation time. In an empirical exercise involving inflation forecasting using a large number of predictors, we find our models to forecast better than alternative approaches and document different patterns of parameter change than are found with approaches which assume random walk evolution of parameters.
\\\\ 
\textbf{JEL}: C11, C30, E3, E44 \\
\textbf{Keywords}: Time-varying parameter regression, singular value decomposition, clustering, hierarchical priors\\
\end{minipage}
\end{center}
\doublespacing\normalsize\renewcommand{\thepage}{\arabic{page}}

\newpage

\section{Introduction}\label{sec:intro}

Time-varying parameter (TVP) regressions and Vector Autoregressions (VARs) have shown their usefulness in a range of applications in macroeconomics \citep[e.g.,][]{cogley2005drifts, primiceri2005,giannone2013}. Particularly when the number of explanatory variables is large, Bayesian methods are typically used since prior information can be essential in overcoming over-parameterization concerns. These priors are often hierarchical and ensure parsimony by automatically shrinking coefficients. Examples include \cite{bkk}, \cite{kg2014}, \cite{bitto_fs} and \cite{hko2019}. Approaches such as these have two characteristics that we highlight so as to motivate the contributions of our paper. First, they use Markov Chain Monte Carlo (MCMC) methods which can be computationally demanding. They are unable to scale up to the truly large data sets that macroeconomists now work with. Second, the regression coefficients in these TVP models are assumed to follow random walk or autoregressive (AR) processes. In this paper, we develop a new approach which is computationally efficient and scaleable. Furthermore, it allows for more flexible patterns of time variation in the regression coefficients.    

We achieve the computational gains by writing the TVP regression as a static regression with a particular, high dimensional, set of regressors. Using the singular value decomposition (SVD) of this set of regressors along with conditionally conjugate priors yields a computationally fast algorithm which scales well in high dimensions. One key feature of this approach is that no approximations are involved.  This contrasts with other computationally-fast approaches to TVP regression which achieve computational gains by using approximate methods such as variational Bayes \citep{koopkorobilisvb}, message passing \citep{korobilis2019high} or expectation maximization \citep{rockova}. 

Our computational approach avoids large-scale matrix operations altogether and exploits the fact that most of the matrices involved are (block) diagonal. In large dimensional contexts, this allows fast MCMC-based inference and thus enables the researcher to compute highly non-linear functions of the time-varying regression coefficients while taking parameter uncertainty into account. Compared to estimation approaches based on forward-filtering backward-sampling \citep[FFBS, see][]{carterkohn, fs1994} algorithms, the computational burden is light. In particular, we show that it rises {(almost) linearly} in the number of covariates. For quarterly macroeconomic datasets that feature a few hundred observations, this allows us to estimate and forecast, exploiting all available information without using dimension reduction techniques such as principal components.

Computational tractability is one concern in high dimensional TVP regressions. The curse of dimensionality associated with estimating large dimensional TVP regressions is another. To solve over-parameterization issues and achieve a high degree of flexibility in the type of coefficient change, we use a sparse finite mixture representation \citep[see][]{malsiner2016} for the time-varying coefficients.  This introduces shrinkage on the amount of time variation by pooling different time periods into a (potentially) small number of clusters.  We also use shrinkage priors which allow for the detection of how many clusters are necessary. Shrinkage towards the cluster means is then introduced by specifying appropriate conjugate priors on the regression coefficients. At a general level, this model is closely related  to random coefficient models commonly used in microeconometrics \citep[see, e.g.,][]{allenby1998heterogeneity, lenk2000bayesian}. We propose {three} different choices for this prior. The first of these is based on Zellner's g-prior \citep{Zellnergprior}. The second is based on the Minnesota prior \citep{doan1984forecasting, litterman1986} and the final one is a ridge-type prior \citep[see, e.g.,][]{griffin2013some}.  As opposed to a standard TVP regression which assumes that the states evolve smoothly over time, our model allows for abrupt changes (which might only happen occasionally) in the coefficients. This   resembles the behavior of regime switching models \citep[see, e.g.,][]{Hamilton1989, FS}. Compared to those, our approach has two additional advantages: it  remains agnostic on the precise law of motion of the coefficients, and it endogenously finds the number of regimes.\footnote{Other approaches which remain agnostic on the transition distribution of the coefficients are, e.g., \cite{kalli2019bayesian} and \cite{kapetanios2019large}.}

We investigate the performance of our methods using two applications. Based on synthetic data, we first illustrate computational gains if $K$ and $T$ become large. We then proceed to show that our approach effectively  recovers key properties of the data generating process. In a real-data application, we model US inflation dynamics. Our framework provides  new insights on how the relationship between unemployment and inflation evolves over time. Moreover, in an extensive forecasting exercise we show that our proposed set of models performs well relative to a wide range of competing models. Specifically, we find that our model yields precise point and density forecasts for one-step-ahead and four-step-ahead predictions. Improvements in forecast accuracy are especially pronounced during recessionary episodes.

The remainder of the paper is structured as follows. Section \ref{sec:model} introduces the static representation of the TVP regression model while Section \ref{sec: SVD} shows how the SVD can be used to speed up computation. Section \ref{sec:prior} provides an extensive discussion of our prior setup.  The model is then applied to synthetic data in Section \ref{artdata} and real data in Section \ref{forecasting}. Finally, the last section summarizes and concludes the paper and the Online Appendix provides additional details on computation and further empirical findings.

\section{A Static Representation of the TVP  Model}\label{sec:model}
Let $\{y_t\}_{t=1}^T$ denote a scalar response variable\footnote{This setup can be easily extended to VAR models. In particular, recent papers \citep[see, e.g.,][]{carriero2019large, kpp2019,tsionas2019bayesian, cadonna2020triple, hko2019, huberkastner2020, carriero2021corrigendum} work with a structural VAR specification which allows for the equations to be estimated separately. Accordingly, the size of the system does not penalize the estimation time.  This extension is part of our current research agenda.}
that is described by a TVP regression given by
\begin{equation}
y_t = \bm x_t' \bm \beta_t + \sigma \eta_t, \quad \eta_t \sim \mathcal{N}(0,1), \label{eq: tvp_reg}
\end{equation}
where $\bm x_{t}$ is a $K$-dimensional vector of regressors, $\bm \beta_t$ is a set of $K$  time-varying regression coefficients and $\sigma^2$ is the error variance. For now, we assume homoskedastic errors, but will relax this assumption later in the paper. 

The TVP regression can be written as a static regression model as follows:
\begin{equation}
\underbrace{\begin{pmatrix} 
  y_1  \\
  y_2  \\
  \vdots \\
  y_T
\end{pmatrix}}_{\bm y} = 
\underbrace{\begin{pmatrix} 
  \bm x_1'  & \bm 0_{K \times 1}'   & \dots  & \bm 0_{K \times 1}' \\
  {\bm  \phi}'_2    & {\bm  x}'_2 & \dots  & \bm 0_{K \times 1}' \\ 
  \vdots    & \vdots   & \ddots & \vdots \\
  {\bm  \phi}'_T    & {\bm  \phi}'_T   & \dots  & \bm x_T'
\end{pmatrix}}_{\bm Z} 
\underbrace{\begin{pmatrix} 
  \bm \beta_1 \\
  \bm \beta_2 \\  
  \vdots      \\
  \bm \beta_T
\end{pmatrix}}_{\bm \beta} +  
\sigma 
\underbrace{\begin{pmatrix} 
  \eta_1 \\
  \eta_2 \\  
  \vdots      \\
  \eta_T
\end{pmatrix}}_{\bm \eta}. \label{eq: statreg}
\end{equation}
\autoref{eq: statreg}  implies that the dynamic regression model in \autoref{eq: tvp_reg} can be cast in the form of a standard linear regression model with $KT$ predictors stored in a $T \times KT$-dimensional design matrix $\bm Z$. Notice that the rank of $\bm Z$ is equal to $T$ and inverting $\bm Z' \bm Z$ is not possible. We stress that, at this stage, we are agnostic on the evolution of $\bm \beta_{t}$ over time. {A common assumption in the literature is that the latent states evolve according to a random walk. Such behavior can be achieved by setting $\bm \phi_t = \bm x_t$ for all $t$, implying a lower triangular matrix $\bm Z$. If $\bm \phi_t = \bm 0_{K \times 1}$ for all $t $, we obtain a block-diagonal matrix $\bm Z$ which, in combination with a Gaussian prior on $\bm \beta$ would imply a white noise state equation.}

The researcher may want to investigate whether any explanatory variable has a time-varying, constant or a zero coefficient. In such a case, it proves convenient to work with a different parameterization of the model which decomposes $\bm \beta$ into a time-invariant ($\bm \gamma$) and a time-varying part ($\tilde{\bm \beta}$):
\begin{equation}
	\bm y = \bm X \bm \gamma + \bm Z \tilde{\bm \beta} +  \sigma \bm \eta,
\label{static}
\end{equation}
with $\bm X=(\bm x_1, \dots, \bm x_T)'$ denoting a $T\times K$ matrix of stacked covariates and $\bm \beta_t = \bm \gamma + \tilde{\bm \beta}_t$, with $ \tilde{\bm  \beta}_t$ being the relevant  elements of $\tilde{\bm \beta}$.\footnote{{In the case of lower triangular $\bm Z$, the $\tilde{\bm \beta}_t$'s can be interpreted as the shocks to the latent states with the actual value of the TVPs in time $t$ given by $\sum_{s = 1}^t \tilde{\bm \beta}_s$.}} 

Thus, we have written the TVP regression as a static regression, but with a huge number of explanatory variables. That is, $\tilde{\bm \beta}$ is a $KT$-dimensional vector with $K, T$ both being potentially large numbers. 

This representation is related to a non-centered parameterization \citep{fs_wagner} of a state space model. The main intuition behind \autoref{static} is that parameters tend to fluctuate around a time-invariant regression component $\bm \gamma$, with deviations being driven by $\tilde{\bm \beta}_t$. This parameterization, in combination with the static representation of the state space model, allows us to push the model towards a time-invariant specification during certain points in time, if necessary. This behavior closely resembles characteristics of  mixture innovation models \citep[e.g.,][]{GiordaniKohn}, and allows the model to decide the points in time when it is necessary to allow for parameter change.

In the theoretical discussion which follows, we will focus on the time-varying part of the regression model:
\begin{equation*}
		\hat{\bm y} = \bm y - \bm X \bm \gamma = \bm Z \tilde{\bm  \beta} + \sigma \bm \eta,
\end{equation*}
since sampling from the conditional posterior of $\bm \gamma$ (under a Gaussian shrinkage prior and conditional on $\bm Z \tilde{\bm  \beta}$) is straightforward. In principle, any shrinkage prior can be introduced on $\bm \gamma$. In our empirical work, we use a hierarchical Normal-Gamma prior of the form:
\begin{equation*}
\gamma_j | \tau_j \sim \mathcal{N}(0, \tau_j), \quad \tau_j|\psi \sim \mathcal{G}(\vartheta, \vartheta \psi/2),\quad \psi \sim \mathcal{G}(a_0, a_1),
\end{equation*}
where $\gamma_j$ is the $j^{th}$ element of $\bm \gamma$ for $j=1,\dots,K$. We set $\vartheta=0.1$ and $a_0=a_1=0.01$ and use MCMC methods to learn about the posterior for these parameters. The relevant posterior conditionals are given in \cite{griffin2010inference} and Section \ref{app:post} of the Online Appendix.

\section{Fast Bayesian Inference using SVDs}\label{sec: SVD}
\subsection{The Homoskedastic Case}
In static regressions with huge numbers of explanatory variables, there are several methods for ensuring parsimony that involve compressing the data. Traditionally principal components or factor methods have been used \citep[see][]{stockwatson2011}. Random compression methods have also been used with TVP models \citep[see][]{kpp2019}. 


The SVD of our matrix of explanatory variables, $\bm Z$, is 
\begin{equation*}
\underbrace{\bm Z}_{T \times KT} = \underbrace{\bm U}_{T \times T~} \underbrace{\bm \Lambda}_{T \times T} \underbrace{\bm V'}_{~T \times KT}
\end{equation*}
whereby $\bm U$ and $\bm V$ are orthogonal matrices and $\bm \Lambda$ denotes a diagonal matrix with the singular values, denoted by $\bm \lambda$, of $\bm Z$ as diagonal elements. 
 
The usefulness and theoretical soundness of the SVD to compress regressions is demonstrated in \cite{trippe2019}. They use it as an approximate method in the sense that, in a case with $K$ regressors, they only use the part of the SVD corresponding to the largest $M$ singular values, where $M<K$. In such a case, their methods become approximate.

In our case, we can exploit the fact that $\text{rank}(\bm Z) = T$  ($T \ll KT$) and utilize the SVD of $\bm Z$ as in \cite{trippe2019}. But we do not truncate the SVD using only the $M$ largest singular values, instead we use all $T$ of them.  But since the rank of $\bm Z$ is $T ( \ll KT)$, our approach translates into an exact low rank structure implying no loss of information through the SVD.

Thus, using the SVD we can exactly recover the big matrix $\bm Z$. The reason for using the SVD instead of $\bm Z$ is that we can exploit several convenient properties of the SVD that speed up computation. To be specific, if we use a  Gaussian prior, this leads to a computationally particularly convenient expression of the posterior distribution of $\tilde{\bm \beta}$ which avoids complicated matrix manipulations such as inversion and the Cholesky decomposition of high-dimensional matrices. Hence, computation is fast. 

We assume a conjugate prior of the form: 
\begin{equation*}
\tilde{\bm \beta}|\sigma^2 \sim \mathcal{N}(\bm b_0, \sigma^2 \bm D_0),
\end{equation*}
 with $\bm D_0 = \bm I_T \otimes \bm \Psi$ being a $KT$-dimensional diagonal prior variance-covariance matrix, where $\bm I_T$ denotes a $T$-dimensional identity matrix and $\bm \Psi$ a $K \times K$-dimensional diagonal matrix that contains covariate-specific shrinkage parameters on its main diagonal. Our prior will be hierarchical so that $\bm \Psi$ will depend on other prior hyperparameters $\bm \theta$ to be defined later. 

Using textbook results for the Gaussian linear regression model with a conjugate prior (conditional on the time-invariant coefficients $\bm \gamma$), the posterior is
\begin{equation}
\tilde{\bm \beta}|Data, \bm \gamma, \sigma^2, \bm \theta \sim \mathcal{N}(\bm \mu_{\tilde{\beta}}, \sigma^2 \bm V_{\tilde{\beta}}).\label{eq: betatilde}
\end{equation}

In conventional regression contexts, the computational bottleneck is typically the $KT \times KT$ matrix $\bm V_{\tilde{\beta}}$. However, with our SVD regression, \cite{trippe2019}, show this to take the form:
\begin{align}
\bm V_{\tilde{\beta}} &= \left( \bm D^{-1}_{0} + \bm V~\text{diag}~(\bm \lambda \odot \bm \lambda) \bm V'\right)^{-1} \nonumber \\
&= \bm D_0 - \bm D_0 \bm V~\left(\text{diag}~(\bm \lambda \odot \bm \lambda)^{-1} +  \bm V' \bm D_0 \bm V \right)^{-1}\bm V'\bm D_0, \label{eq:SIG_btvp} \\
\bm \mu_{\tilde{\beta}} &= \bm V_{\tilde{\beta}} (\bm Z' \hat{\bm y}+\bm D_0^{-1} \bm b_0),\label{eq:mu_btvp}
\end{align}
with $\odot$ denoting the dot product. Crucially, the matrix $\bm \Xi = \left(\text{diag}~(\bm \lambda \odot \bm \lambda)^{-1} +  \bm V' \bm D_0 \bm V \right)^{-1}$ is a diagonal matrix {if $\bm Z$ is block-diagonal} and thus trivial to compute. For a lower triangular matrix $\bm Z$ and a general prior covariance matrix, this result does not hold. However, if we set $\bm D_0 = \theta \times \bm I_{KT}$ {(i.e., assume a ridge-type prior)} the matrix $\bm \Xi$ again reduces to a diagonal matrix.\footnote{Notice that if the condition number (i.e., the ratio of the largest and the smallest element in $\bm \lambda$) is very large, numerical issues can arise. This is the case if $\bm x_t \approx \bm 0$. In our simulations and real-data exercises, we never encountered computational issues. If these arise, a simple solution would be to use a truncated SVD and discard eigenvalues smaller than a threshold very close to zero.} The main computational hurdle boils down to computing $\bm V \bm \Xi \bm V'$, but, for a block-diagonal $\bm Z$ it is a {sparse matrix and efficient algorithms can be used. In case we use a lower triangular $\bm Z$ coupled with a ridge-prior,  computation can be sped up enormously by noting that $\bm \mu_{\tilde{\bm \beta}} = \left[\bm V \text{diag}\left(\frac{\bm \lambda}{\theta^{-1} \bm \iota_{T} + \bm \lambda^2}\right)\right] \bm U' \tilde{\bm y} + \bm D^{-1}_0\bm b_0$}. The resulting computation time, conditional on a fixed $T$, rises approximately linearly in $K$ because most of the matrices involved are (block) diagonal and sparse. The key feature of our algorithm is that we entirely avoid inverting a full matrix. The only inversion involved is the one of $\bm \Xi$ which can be carried out in O($T$) steps. 

{To efficiently simulate  $\tilde{\bm \beta} \sim \mathcal{N}(\bm \mu_{\tilde{\beta}},\sigma^2   \bm V_{\tilde{\beta}})$ using \autoref{eq:SIG_btvp}, we exploit Algorithm 3 proposed in \cite{cong2017fast}. In the first step, this algorithm samples $\bm a \sim \mathcal{N}(\bm 0_{TK}, \bm D_0)$ and $\bm b \sim \mathcal{N}(\bm 0_T, \text{diag}~(\bm \lambda \odot \bm \lambda)^{-1})$. In the second step, a valid draw of $\tilde{\bm \beta}$ is obtained by computing $\tilde{\bm \beta} = \bm \mu_{\tilde{\beta}} + \sigma (\bm a - \bm D_0 \bm V \bm \Xi ( \bm V' \bm D_0 \bm a + \bm b)$. Step 2 is trivial since $\bm \Xi$ is diagonal for a block-diagonal $\bm Z$ and also for a lower triangular $\bm Z$ with a ridge-prior. Hence, sampling  of $\bm \tilde{\bm \beta}$ is fast and scalable  to large dimensions.}



In this sub-section, we have described computationally efficient methods for doing Bayesian estimation in the homoskedastic Gaussian linear regression model when the number of explanatory variables is large. They can be used in any Big Data regression model, but here we are using them in the context of our TVP regression model written in static form as in \autoref{static}. These methods involve transforming the matrix of explanatory variables using the SVD. If the matrices of prior hyperparameters, $\bm b_0$ and 
$\bm D_0$, were known and if homoskedasticity were a reasonable assumption, then textbook, conjugate prior, results for Bayesian inference in the Gaussian linear regression model are all that is required. Analytical results are available for this case and there would be no need for MCMC methods. This is the case covered by \cite{trippe2019}. However, in macroeconomic data sets, homoskedasticity is often not a reasonable assumption. And it is unlikely that the researcher would be able to make sensible choices for $\bm b_0$ and $\bm D_0$ in this high-dimensional context. Accordingly, we will develop methods for adding stochastic volatility and propose a hierarchical prior for the regression coefficients. 

\subsection{Adding Stochastic Volatility}
Stochastic volatility typically is an important feature of successful macroeconomic forecasting models \citep[e.g.,][]{clark2011}. 
We incorporate this by replacing $\sigma^2$ in \autoref{static} with $\bm \Sigma=\text{diag}(\sigma_1^2, \dots, \sigma_T^2) \otimes \bm I_K$. This implies that the prior on $\tilde{\bm \beta}$ is 
\begin{equation*}
\tilde{\bm \beta}|\bm \Sigma \sim \mathcal{N}(\bm b_0, \bm \Sigma \bm D_0).
\end{equation*} 
Note that the prior in a specific period is given by
\begin{equation*}
\tilde{\bm \beta}_t|\sigma^2_t \sim \mathcal{N}(\bm b_{0 t}, \sigma^2_t  \bm \Psi),
\end{equation*}
with $\bm b_{0 t}$ being the relevant block associated with the $t^{th}$ period. Thus, it can be seen that the degree of shrinkage changes with $\sigma_t^2$, implying less shrinkage in more volatile times. From a computational perspective, assuming that $\sigma_t^2$ scales the prior variances enables us to factor $\bm \Sigma$ out of the posterior covariance matrix  and thus obtain computational gains because $\bm D_0$ does not need to be updated for every iteration of the MCMC algorithm. From an econometric perspective, the feature that shrinkage decreases if error volatilities are large implies that, in situations characterized by substantial uncertainty, our approach naturally allows for large shifts in the TVPs and thus permits swift adjustments to changing economic conditions. Our forecasting results suggest that this behavior improves predictive accuracy in turbulent times such as the global financial crisis.

We assume that $h_t = \log(\sigma^2_t)$ follows an AR(1) process:
\begin{equation*}
h_t = \mu_h + \rho_h (h_{t-1}-\mu_h) + \sigma_h v_t, \quad v_t \sim \mathcal{N}(0, 1), \quad h_0 \sim \mathcal{N}\left(\mu, \frac{\sigma_h^2}{1-\rho_h^2}\right).
\end{equation*}
In our empirical work, we follow \cite{kastner2014ancillarity} and specify a Gaussian prior on the unconditional mean $\mu_h \sim \mathcal{N}(0, 10)$, a Beta prior on the (transformed) persistence parameter $\frac{\rho_h+1}{2}\sim \mathcal{B}(25, 5)$ and a non-conjugate Gamma prior on the process innovation variance $\sigma_h^2 \sim \mathcal{G}(1/2, 1/2)$.  {Bayesian estimation of the volatilities proceeds using MCMC methods based on the algorithm of \cite{kastner2014ancillarity}. A small alteration to this algorithm needs to be made due to the dependency of the prior of $\tilde{\bm \beta_t}$ on $\sigma_t$ (see the Online Appendix for details).}
\subsection{Posterior Computation}
Conditional on the specific choice of the prior on the regression coefficients (discussed in the next section) we carry out posterior inference using a relatively straightforward MCMC algorithm. Most steps of this algorithm are standard and we provide exact forms of the conditional posterior distributions, the precise algorithm and additional information on MCMC mixing in the Online Appendix. Here it suffices to note that we repeat our MCMC algorithm $30,000$ times and discard the first $10,000$ draws as burn-in.

\section{A Hierarchical Prior for the Regression Coefficients}\label{sec:prior}
\subsection{General Considerations}
With hierarchical priors, where $\bm b_0$ and/or $\bm D_0$ depend on unknown parameters, MCMC methods based on the full conditional posterior distributions are typically used. In our case, we would need to recompute the enormous matrix $\bm V_{\tilde{\beta}}$ and its Cholesky factor at every MCMC draw. This contrasts with the non-hierarchical case with fixed $\bm b_0$ and $\bm D_0$ where $\bm V_{\tilde{\beta}}$ is calculated once. Due to this consideration, we wish to avoid using MCMC methods based on the full posterior conditionals. 

Many priors, including the {three} introduced here, have $\bm D_0$ depending on a small number of prior hyperparameters. These can be simulated using a Metropolis Hastings (MH) algorithm. With such an algorithm, updating of $\bm V_{\tilde{\beta}}$ only takes place for accepted draws (in our forecasting exercise roughly $30\%$ of draws are accepted). Since priors which feature closed form full conditional posteriors for the hyperparameters imply that $\bm V_{\tilde{\beta}}$ needs to be recomputed for each iteration in our posterior simulator, this reduces computation time appreciably.


\subsection{The Prior Covariance Matrix}
In this paper, we consider {three} different hierarchical priors for $\tilde{\bm \beta}$.  Since our empirical application centers on forecasting inflation, the predictors $\bm x_t$ will be structured as follows $\bm x_t = (y_{t-1}, \dots, y_{t-p_y}, \bm d'_{t-1}, \dots, \bm d'_{t-p_d}, 1)',
$ with $\bm d_t$ denoting a set of $N$ exogenous regressors and $p_y$ and $p_d$ being the maximum number of lags for the response and the exogenous variables, respectively. In what follows, we will assume that $p = p_y = p_d$. In principle, using different lags is easily possible.

The first prior is inspired by the Minnesota prior \citep[see][]{litterman1986}. It captures the idea that own lags are typically more important than other lags and, thus, require separate shrinkage. It also captures the idea that more distant lags are likely to be less important than more recent ones. Our variant of the Minnesota prior translates these ideas to control the amount of time-variation, implying that coefficients on own lags might feature more time-variation while parameters associated with other lags feature less time-variation. The same notion carries over to coefficients related to more distant lags which should feature less time-variation a priori.

This prior involves two hyperparameters to be estimated: $\bm \theta = (\zeta_1, \zeta_2)'$. These prior hyperparameters are used to parameterize $\bm \Psi$ to match the Minnesota prior variances:
\begin{equation*}
[\bm \Psi]_{ii} = \begin{cases} \frac{\zeta_1^2}{l^2} \text{\hspace{.3cm} on the coefficients associated  with $y_{t-l}~(l=1,\dots, p)$  }\\
\frac{\zeta_2^2}{l^2} \frac{\hat{\sigma}^2_y}{\hat{\sigma}^2_j} \text{ on the coefficients related to  $d_{jt-l}$ } \\
\zeta_2^2 \text{ on the intercept term}.
\end{cases}
\end{equation*}
Here, we let $[\bm \Psi]_{ii}$ denote the $(i, i)^{th}$ element of $\bm \Psi$, $d_{jt}$ refers to the $j^{th}$ element of $\bm d_t$, $\hat{\sigma}^2_y$, $\hat{\sigma}^2_j$ denotes the OLS variance obtained by estimating an AR($p$) model in $y_t$ and $d_{jt}$, respectively. 
The hyperpriors on $\zeta_1$ and $\zeta_2$  follow a Uniform distribution: 
\begin{equation*}
\zeta_j \sim \mathcal{U}(\mathfrak{s}_{0, j},\mathfrak{s}_{1, j}) \quad \text{for} \quad j=1,2.
\end{equation*} 
 


The second prior we use is a variant of the g-prior involving a single prior hyperparameter: $\theta = \xi$. This specification amounts to setting $\bm \Psi = \xi \times \bm \Omega,$ 
where ${\bm \Omega}$ is a diagonal matrix with the $(i, i)^{th}$ element being defined as $[{\bm \Omega}]_{ii} = \hat{\sigma}^2_y/\hat{\sigma}^2_j$. For reasons outlined in \cite{doan1984forecasting}, we depart from using the diagonal elements of $(\bm X' \bm X)^{-1}$ to scale our prior and rely on the OLS variances of an AR($p$) model as in the case of the Minnesota-type prior. 
{The third prior is a ridge-type prior which simply sets $\bm \Psi = \xi \times \bm I_{K}$.  This specification is used in the case of a lower triangular $\bm Z$ for reasons outlined in Section \ref{sec: SVD}. While being simple, this prior has been shown to work well in a wide range of applications \citep{griffin2013some}.}

Similar to the Minnesota prior we again use a Uniform prior on $\xi$ in both cases:
\begin{equation*}
\xi \sim \mathcal{U}(\mathfrak{s}_0, \mathfrak{s}_1). 
\end{equation*}

Since we aim to infer $\xi, \zeta_1$ and  $\zeta_2$ from the data we set $\mathfrak{s}_0 = \mathfrak{s}_{0,1} = \mathfrak{s}_{0,2} = 10^{-10}$ close to zero and $\{\mathfrak{s}_1, \mathfrak{s}_{1,1}, \mathfrak{s}_{1,2}\}$ is specified as follows:
\begin{equation}
\mathfrak{s}_1 = \mathfrak{s}_{1,j} = \kappa \frac{T}{K^2} \quad \text{for} \quad j = 1, 2.
\label{kappa}
\end{equation}
Here, $\kappa$ is a constant being less or equal than unity to avoid excessive overfitting in light of large $K$ and $T$.  Since large values of $\kappa$ translate into excessive time variation in $\tilde{\bm \beta}_t$, we need to select $\kappa$ carefully. The hyperparameters of this prior are inspired by the risk inflation criterion put forward in \cite{foster1994risk} which would correspond to setting $\xi = 1/K^2$. Since this prior was developed for a standard linear regression model it would introduce too little shrinkage in our framework (or, if we set $\xi = 1/(TK)^2$ too much shrinkage, ruling out any time-variation). Our approach lets the data speak but essentially implies that the bound of the prior is increasing in $T$ and decreasing in the number of covariates. Intuitively speaking, our prior implies that if the length of the time series increases, the prior probability of observing substantial structural breaks also increases slightly. 

In the empirical application, we infer $\kappa$ over a grid of values and select the $\kappa$ that yields the best forecasting performance in terms of log predictive scores. Further discussion of and empirical evidence relating to $\kappa$ (and $G$) is given in Section \ref{sec: empAppendix} of the Online Appendix.

The methods developed in this paper will hold for any choice of prior covariance matrix, $\bm D_0$, although assuming it to be diagonal greatly speeds up computation. In this sub-section, we have proposed {three} forms for it which we shall (with some abuse of terminology) refer to as the Minnesota, g-prior and ridge-prior forms, respectively, in the following material.

\subsection{The Prior Mean}

As for the prior mean, $\bm b_0$, it can take a range of possible forms. The simplest option is to set it to zero. After all, from \autoref{static} it can be seen that $\tilde{\bm \beta}_t$ measures the deviation from the constant coefficient case which, on average, is zero. This is what we do with the Minnesota prior {and if we set $\bm Z$ to be lower triangular.}\footnote{Using the Minnesota prior in combination with the clustering specification introduced in this sub-section is less sensible. That is, its form, involving different treatments of coefficients on lagged dependent variables and exogenous variables and smaller prior variances for longer lag length already, in a sense, clusters the coefficients into groups. {A similar argument holds for a lower triangular matrix $\bm Z$ since that would translate into a random walk with a (potentially) time-varying drift term.}}
 However, it is possible that we can gain estimation accuracy through pooling information across coefficients by adding extra layers to the prior hierarchy. In this paper, we do so using a sparse finite location mixture of Gaussians and adapt the methods of \cite{malsiner2016} to the TVP regression context. Sparse finite mixtures, relative to Dirichlet process mixtures, have the advantage of being finite dimensional while allowing the number of clusters to be random a priori. The number of groups can then be inferred during MCMC sampling by counting the number of non-empty regimes.\footnote{For a detailed discussion on the relationship between sparse finite mixtures and Dirichlet process mixtures, see \cite{fruhwirth2019here}.} 
 
 In the discussion below, we refer to these two treatments of the prior mean as non-clustered and clustered, respectively. With the g-prior, we consider both clustered and non-clustered approaches. 

We emphasize that both of these specifications for the prior mean are very flexible and let the data decide on the form that the change in parameters takes. This contrasts with standard TVP regression models, where it is common to assume that the states evolve according to random walks. This implies that the prior mean of $\bm \beta_{t}$ is $\bm \beta_{t-1}$.

With the clustered approach, we assume that each $\tilde{\bm \beta}_t$ has a prior of the following form:
\begin{equation*}
f_\mathcal{N}(\tilde{\bm \beta}_{t}|\bm \mu_1, \dots, \bm \mu_G, \bm w, \sigma_t^2, \bm \Psi) = \sum_{g=1}^G w_g f_{\mathcal{N}}(\tilde{\bm \beta}_{t}|\bm \mu_g,  \sigma_t^2 \bm \Psi),
\end{equation*}
where $f_\mathcal{N}$ denotes the density of a Gaussian distribution and $\bm w$ are component weights with $\sum_{g=1}^G w_g = 1$ and $w_g \ge 0$ for all $g$.
$\bm \mu_g~(g=1,\dots,G)$ denotes  $G$ component-specific means with $G$ being a potentially large integer that is much smaller than $T$ (i.e., $G \ll T)$.

An equivalent representation, based on auxiliary variables $\delta_t$, is
\begin{equation}
\tilde{\bm \beta_t}|\delta_t =g \sim \mathcal{N}(\bm \mu_g , \sigma_t^2 \bm \Psi), \label{eq: priorBETAT}
\end{equation}
with $\text{Pr}(\delta_t = g) = w_g$ being the probability that $\tilde{\bm \beta_t}$ is assigned to group $g$. Equation \ref{eq: priorBETAT} can be interpreted as a state evolution equation which resembles a hierarchical factor model since each $\tilde{\bm \beta_t}$ clusters around the different component means $\bm \mu_g$. As opposed to assuming a random walk state evolution, which yields smoothly varying TVPs, this model provides more flexibility by pulling $\tilde{\bm \beta_t}$ towards $G \le T$ prior means. Under the prior in \autoref{eq: priorBETAT}, our model can be interpreted as a random coefficients model \citep[for a Bayesian treatment, see, e.g.,][]{fruhwirth2004bayesian}. 

Before proceeding to the exact prior setup, it is worth noting that the mixture model is not identified with respect to relabeling the latent indicators. In the forecasting application, we consider functions of the states which are not affected by label switching. Thus, we apply the random permutation sampler of \cite{FS} to randomly relabel the states in order to make sure that our algorithm visits the different modes of the posterior. In what follows, we define $\bm m_t = \bm \mu_g$ if $\delta_t = g$. Using this notation, the prior mean is given by $\bm b_0 = (\bm m_{1}', \dots, \bm m'_{T})'$.

For the weights $\bm w = (w_1, \dots, w_G)'$, we use a symmetric Dirichlet prior:
\begin{equation*}
\bm w|\pi \sim \text{Dir}(\pi, \dots, \pi).
\end{equation*}
Here, $\pi$ denotes the intensity parameter that determines how the model behaves in treating superfluous components. If $\pi \le K/2$, irrelevant components are emptied out while if $\pi>K/2$, the model tends to duplicate component densities to handle overfitting issues. This implies that careful selection of $\pi$ is crucial since it influences the number of breaks in $\tilde{\bm \beta}_t$. The literature suggests different strategies based on using traditional model selection criteria or reversible jump MCMC algorithms to infer $G$ from the data. Our approach closely follows \cite{malsiner2016} and uses a shrinkage prior on $\pi$. The prior we adopt follows a Gamma distribution:
\begin{equation*}
\pi \sim \mathcal{G}(a, a G),
\end{equation*}
with $a=10$ being a hyperparameter that determines the tightness of the prior \citep{malsiner2016}. The prior on $\bm w$ and $\pi$ can be rewritten as:
\begin{equation*}
    \bm w \sim \text{Dir}(a/G, \dots, a/G), \quad a \sim \mathcal{G}(10, 10).
\end{equation*}
\cite{fruhwirth2020generalized} and \cite{greve2020spying} analyze this prior choice and show that it performs well.\footnote{The \texttt{R} package \texttt{fipp}, which is available on CRAN, allows for investigating how influential the prior on $a$ is and whether alternative specifications substantially change the posterior of the number of non-empty groups.} 

To assess which elements in $\bm \mu_g$ determine the group membership, we use yet another shrinkage prior on the component means:
\begin{equation*}
\bm \mu_g| \bm \Pi, \tilde{\bm \beta} \sim \mathcal{N}(\bm \mu_0, \bm \Pi),
\end{equation*}
whereby $\bm \Pi = \bm \Upsilon \bm R \bm \Upsilon$ with $\bm \Upsilon = \text{diag}(\sqrt{\upsilon_1}, \dots, \sqrt{\upsilon_K})$ and $\bm R = \text{diag}(R_1^2, \dots, R_K^2)$. We let $R_j$ denote the range of $\tilde{\bm \beta}_{j}= (\tilde{\beta}_{j1}, \dots,\tilde{\beta}_{jT})'$. The  prior on $\upsilon_j~(j=1,\dots,K)$ follows a Gamma distribution:
\begin{equation*}
\upsilon_j \sim \mathcal{G}(c_0, c_1),
\end{equation*}
translating into the Normal-Gamma prior of \cite{griffin2010inference}. In the empirical application we set $c_0 = c_1 = 0.6$, with $c_0 < 1$ being crucial for pushing the idiosyncratic group means $\bm \mu_g$ strongly towards the common mean $\bm \mu_0$ \citep{malsiner2016}. For $\bm \mu_0$, we use an improper Gaussian prior with mean set equal zero and infinite variance.

This location mixture model is extremely flexible in the types of parameter change that are possible. It allows us to capture situations where the breaks in parameters are large or small and frequent or infrequent. It can effectively mimic the behavior of break point/Markov switching models, standard time-varying parameter models, mixture innovation models and many more. The common variance factor implicitly affects the tightness of the prior and ensures (conditional) conjugacy.

Compared to a standard time-varying parameter model which assumes a random walk state evolution, our prior on $\bm \beta_t$ is invariant with respect to time, up to a scaling factor $\sigma_t$. If $\sigma_t$ is constant,  $(\bm \beta_1, \dots, \bm \beta_T)$ has the same prior distribution as $(\bm \beta_{\rho(1)}, \dots, \bm \beta_{\rho(T)})$ for any permutation $\rho$. In our general case, temporal dependence is not an assumption, but  arises through  appropriately choosing $\bm x_t$ and by allowing for prior dependence on $\sigma_t$. In the extreme case where $\bm x_t$ does not include lagged values of $y_t$ (we include several lags of $y_t$  in our empirical work) and $\sigma_t$ is constant, the dynamic nature of the model is lost since the model is invariant to reordering the time series with respect to $t$ and no dependency is imposed.

\section{Illustration Using Artificial Data}\label{artdata}
In this section we illustrate our modeling approach that utilizes the g-prior and clustering by means of synthetic data simulated from a simple  data generating process (DGP). 

\begin{figure}[t!]
\centering
\caption{Runtime comparison: SVD, FFBS and AWOL}\label{fig: SVD_comp}
\begin{minipage}[c]{1\linewidth}
\centering (a) for different $K$ and $T=200$
\end{minipage}
\begin{minipage}[c]{1\linewidth}
\centering
\includegraphics[scale=0.63]{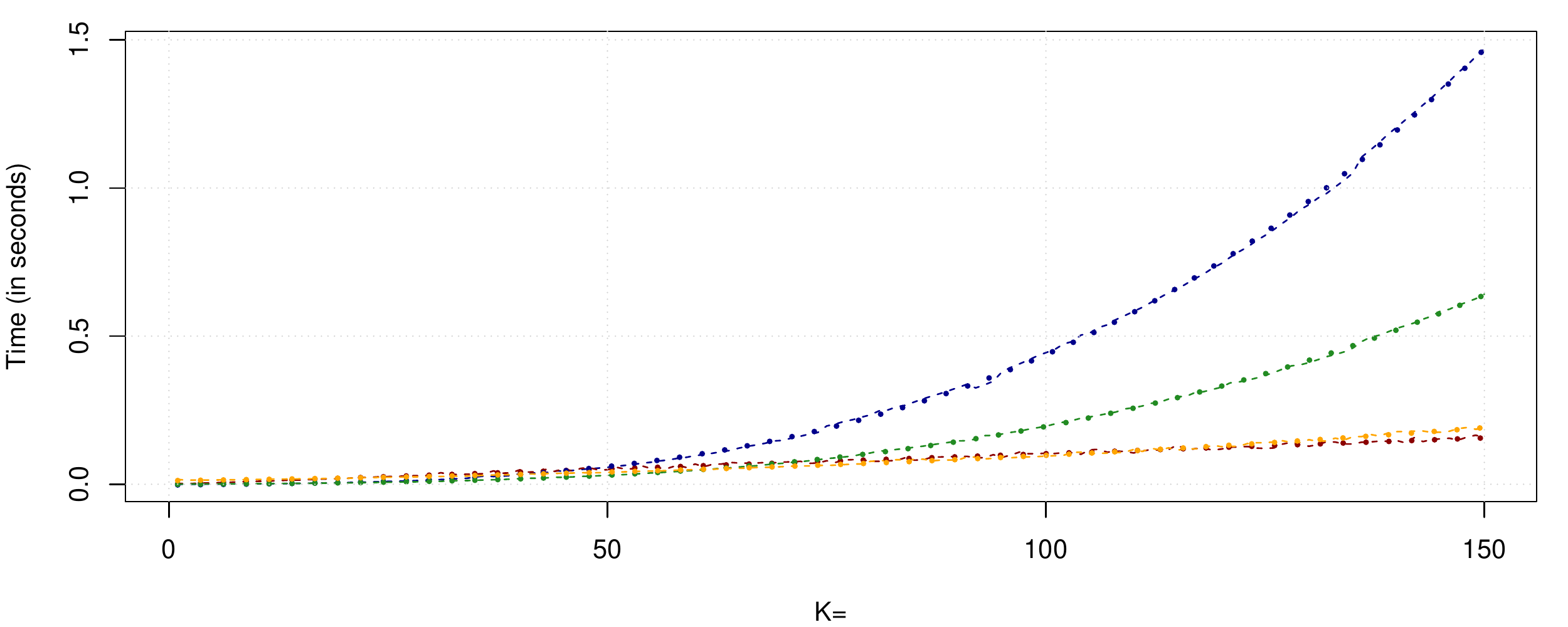}
\end{minipage}
\begin{minipage}[c]{1\linewidth}
\centering (b) for different $T$ and $K=100$
\end{minipage}
\begin{minipage}[c]{1\linewidth}
\includegraphics[scale=0.63]{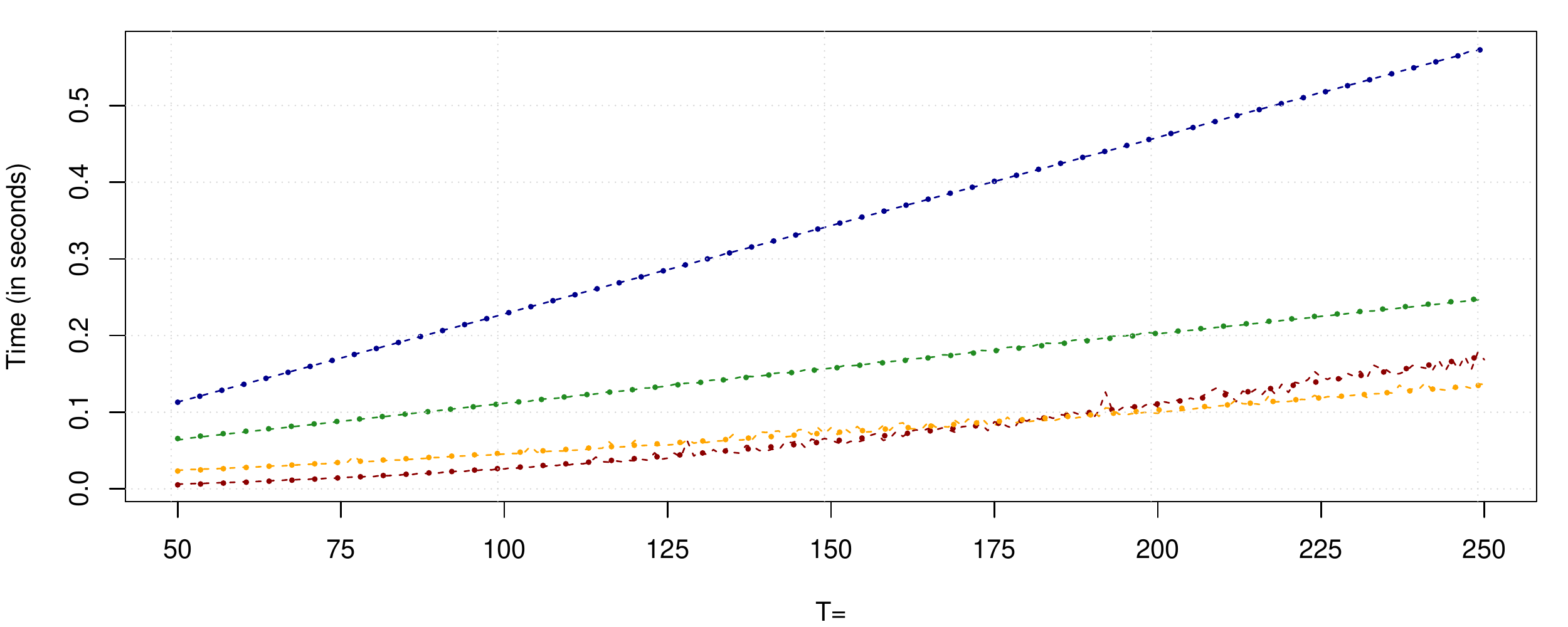}
\end{minipage}
\begin{minipage}[c]{\linewidth}
\footnotesize{\textbf{Notes}: The figure shows the actual  and theoretical time necessary to obtain a draw of $\tilde{\bm \beta}$ using our proposed SVD algorithm for $\bm Z$ being block-diagonal and lower triangular, an AWOL sampler \citep[implemented in \texttt{R} through the \texttt{shrinkTVP} package of][]{bitto2019shrinkage} and the FFBS algorithm. The dashed red lines refer to the SVD approach with a lower triangular $\bm Z$ and a ridge-prior, the orange dashed line refers to the SVD algorithm with block-diagonal $\bm Z$, the dashed green lines refer to the AWOL sampler, while the dashed blue lines indicate the FFBS. The dots refer to theoretical run times. Here, we fit a non-linear trend on the empirical estimation times.}
\end{minipage}
\end{figure}

We begin by illustrating the computational advantages arising from using the SVD, relative to a standard Bayesian approach to TVP regression which involves random walk evolution of the coefficients and the use of FFBS as well as a model estimated using the precision sampler \textit{all without a loop} \citep[AWOL, see][]{chan2009efficient, MCCAUSLAND2011199, kastner2014ancillarity}. \autoref{fig: SVD_comp}(a) shows a comparison of the time necessary to generate a draw from $p(\tilde{\bm \beta}|Data, \bm \gamma, \sigma^2)$ using our algorithm based on the SVD, the FFBS algorithm and the AWOL sampler as a function of $K \in \{1, 2, \dots, 150\}$  and for  $T=200$.\footnote{The AWOL sampler is implemented in  \texttt{R} through the \texttt{shrinkTVP} package  \citep{bitto2019shrinkage}.} 

 To illustrate how computation times change with $T$, \autoref{fig: SVD_comp}(b) shows computation times as a function of $T \in \{50, \dots, 250\}$ for $K=100$. The dashed lines refer to the actual time (based on a cluster with $400$ IntelE5-2650v3 2.3 GHz cores) necessary to simulate from the full conditional of the latent states while the dots indicate theoretical run times through a (non-)linear trend. 

In panel (a), we fit a (non-)linear trend on the empirical estimation times of the different approaches. This implies that while the computational burden is cubic in the number of covariates $K$ for the FFBS approach, our technique based on using the SVD  suggests that runtimes increase (almost) linearly in $K$. Notice that the figure  clearly shows that traditional algorithms based on FFBS quickly become infeasible in high dimensions. {Up to $K \approx 50$, our algorithm (for both choices of $\bm Z$) is slightly slower while the computational advantage increases remarkably with $K$, being more than four times as fast for $K=100$ and over nine times as fast for $K=150$. When we compare the SVD  to the  AWOL algorithm we also observe sizeable improvements in estimation times. For $K=150$, our proposed approach is almost four times faster. This performance is even more impressive given that our SVD approach is implemented in \texttt{R}, a high level interpreted language, while both  FFBS and AWOL are efficiently implemented in \texttt{Rcpp} \citep{eddelbuettel2011rcpp}.}

Panel (b) of the figure shows that, for fixed $K$, computation times increase  linearly for most approaches  if $T$ is varied. {The main exception is the case of a lower triangular $\bm Z$, with computation times growing non-linearly in $T$. This is because this approach relies on several non-sparse matrix-vector products. Since $T$ is typically moderate in macroeconomic data this does not constitute a main bottleneck of the algorithm for general matrices $\bm Z$.} It is, moreover, noteworthy that the slope of the line referring to FFBS is  steeper than the ones associated with the  SVD (for block-diagonal $\bm Z$) and AWOL approaches. This reflects the fact that one needs to perform a filtering (that scales linearly in $T$) and smoothing step (that is also linear in $T$).  This brief discussion shows that the SVD algorithm scales well and renders estimation of huge dimensional models feasible.

We now assume that $y_t$ is generated by the following DGP:
\begin{align*}
y_t = \tilde{\beta}_t + \varepsilon_t,\quad  \varepsilon_t \sim \mathcal{N}(0, 0.1^2),
\end{align*}
for $t=1,\dots, 160$, $\gamma = 0$, and $\tilde{\beta}_t \sim \mathcal{N}(m_t, 0.1^2)$. $\tilde{\beta}_t$ depends on $m_t$ which evolves according to the following law of motion:
\begin{equation*}
    m_t = 3 \times {I}(t \le 60) + 1 \times {I}(60 < t \le 85) - 3 \times {I}(86 < t \le 120) - 1 \times  {I}(t > 120),
\end{equation*}
with $I(\bullet)$ being the indicator function that equals $1$ if its argument is true. 

Analyzing this stylized DGP allows us to illustrate how our approach can be used to infer the number of latent clusters that determine the dynamics of  $\tilde{\beta}_t$.  In what follows, we simulate a single path of $y_t$ and use this for estimating our model. We estimate the model using the g-prior with clustering and set $G=12$.   In this application, we show quantities that depend on the labeling of the latent indicators. This calls for appropriate identifying restrictions and we introduce the restriction that $\mu_1 < \dots < \mu_G$. This is not necessary if interest centers purely on predictive distributions and, thus, we do not impose this restriction in the forecasting section of this paper. 

Before discussing how well our model recovers the true state vector $\tilde{\beta}_t$, we show how our modeling approach can be used to infer the number of groups $G$. Following \cite{malsiner2016}, the number of groups is estimated during MCMC sampling as follows:
\begin{equation*}
G_0^{(j)} = G - \sum_{g=1}^G I\left(T_{g}^{(j)} = 0 \right)
\end{equation*}
with $T_{g}^{(j)}$ denoting the number of observations in cluster $g$ for the $j^{th}$ MCMC draw. This yields a posterior distribution for $G_0$. Its posterior mode can be used as a point estimate of $G$.

In \autoref{tab: G_post},  we report the posterior probability of a given number of regimes by simply computing the fraction of draws with $G_0=g$ for $g=1,\dots, 12$. The table suggests that the probability that $G_0=4$ is around $66$ percent. This indicates that our algorithm successfully selects the correct number of groups, since the mode of the posterior distribution equals four. It is also worth noting that the posterior mean of $\pi$ is very small at $0.09$, suggesting that our mixture model handles irrelevant components by emptying them instead of replicating them (which would be the case if $\pi$ becomes large). Notice, however, that $G_0=5$ also receives some posterior support. We have a probability of about $26$ percent associated with a too large number of regimes. In the present model, this slight overfitting behavior might be caused by additional noise driven by the shocks to the states $\tilde{\beta}_t$, with our mixture model trying to fit the noise.
\begin{table}[t]
\centering
\caption{Posterior probabilities for a given number of groups $G(=12)$}\label{tab: G_post}
\begin{tabular}{rrrrrrrrrrrrr}
  \toprule
 $G_0=$& 1 & 2 & 3 & 4 & 5 & 6 & 7 & 8 & 9 & 10 & 11 & 12 \\ 
  \midrule
 & 0.00 & 0.00 & 0.00 & 0.66 & 0.26 & 0.07 & 0.01 & 0.00 & 0.00 & 0.00 & 0.00 & 0.00 \\ \bottomrule
\end{tabular}
\end{table}

\begin{figure}[ht] 
\centering 
\caption{Posterior distribution of $\tilde{\beta}_t$ and $m_t$}\label{fig: simBeta}
\includegraphics[scale=.38]{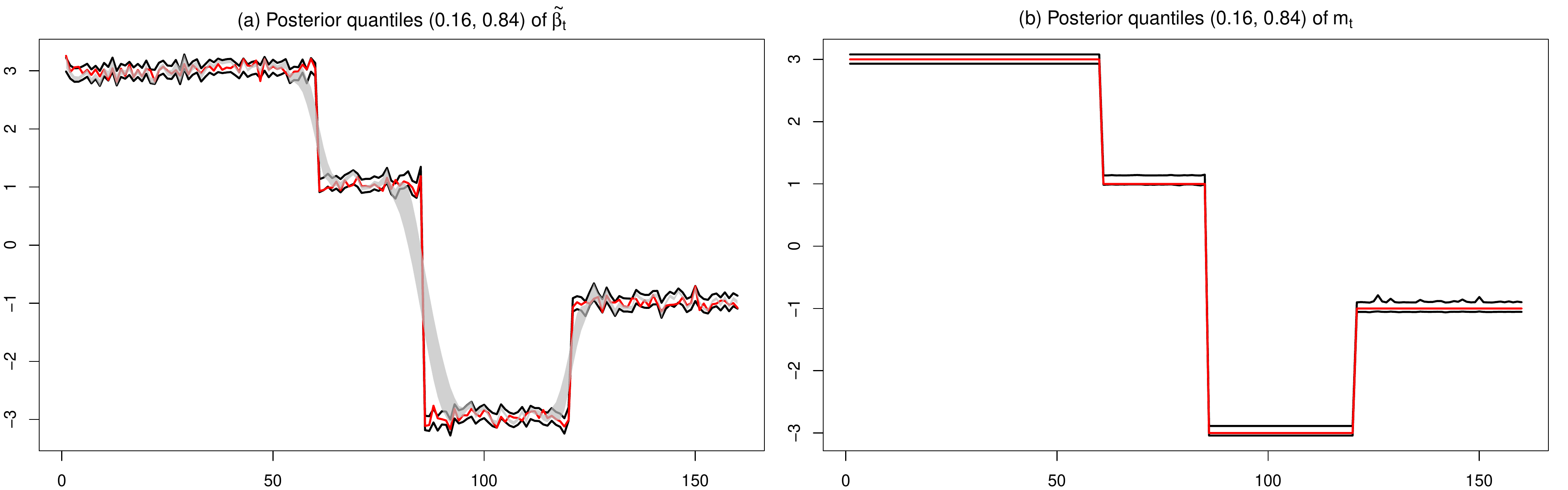}
\begin{minipage}{\linewidth}
\footnotesize \textbf{Notes}: Panel (a) shows $16^{th}$/$84^{th}$ posterior percentiles of $\tilde{\beta}_t$ for our proposed model (solid black lines) and a standard TVP regression with random walk state equation (gray shaded area). The red line denotes the actual outcome. Panel (b) shows the $16^{th}$/$84^{th}$ percentiles of the posterior distribution of $m_t$ (in solid black) and the true value of $m_t$ (in solid red).
\end{minipage}
\end{figure}

Next, we assess whether our model is able to recover $\tilde{\beta}_t$ and $m_t$.  \autoref{fig: simBeta} shows the pointwise $16^{th}$ and $84^{th}$ percentiles of the posterior distribution (in solid black) of $\tilde{\beta}_t$ (see panel (a)) and $m_t$  (see panel (b)) over time.  The gray shaded areas represent the $16^{th}$ and $84^{th}$ percentiles of the posterior of $\tilde{\beta}_t$  obtained from estimating a standard TVP regression model with random walk state equations and stochastic volatility. Apart from the assumption of random walk evolution of the states, all other specification choices are made so as to be as close as possible to our SVD approach. In particular, this model features the hierarchical Normal-Gamma prior \citep[see][]{griffin2010inference} on both the time-invariant part of the model and the signed square root of the state innovation variances \citep{bitto_fs}. It is estimated using a standard FFBS algorithm. We refer to this model as TVP-RW-FFBS.  

In \autoref{fig: simBeta} the red lines denote the true value of $\tilde{\beta}_t$ and $m_t$, respectively.  Panel (a) clearly shows that our model successfully detects major breaks in the underlying states, with the true value of $\tilde{\beta}_t$ almost always being located within the credible intervals. Our modeling approach not only captures low frequency movements but also successfully replicates higher frequency changes. By contrast, the posterior distribution of the TVP-RW-FFBS   specification is not capable of capturing abrupt breaks in the latent states. Instead of capturing large and infrequent changes, the TVP-RW-FFBS approach yields a smooth evolution of $\tilde{\beta}_t$ over time, suggesting that our proposed approach performs comparatively better in learning about sudden breaks in the regression coefficients. 

Considering panel (b) of \autoref{fig: simBeta} reveals a similar picture. Our approach yields credible sets that include the actual outcome of $m_t$ for all $t$. This discussion shows that our model also handles cases with infrequent breaks in the regression coefficients rather well. As compared to standard TVP regressions that imply a smooth evolution of the states, using a mixture model to determine the state evolution enables us to capture large and abrupt breaks.

\begin{figure}[ht]
\centering 
\caption{Posterior distribution of $\tilde{\beta}_t$ and $m_t$}\label{fig: simBetaRW}
\includegraphics[scale=.38]{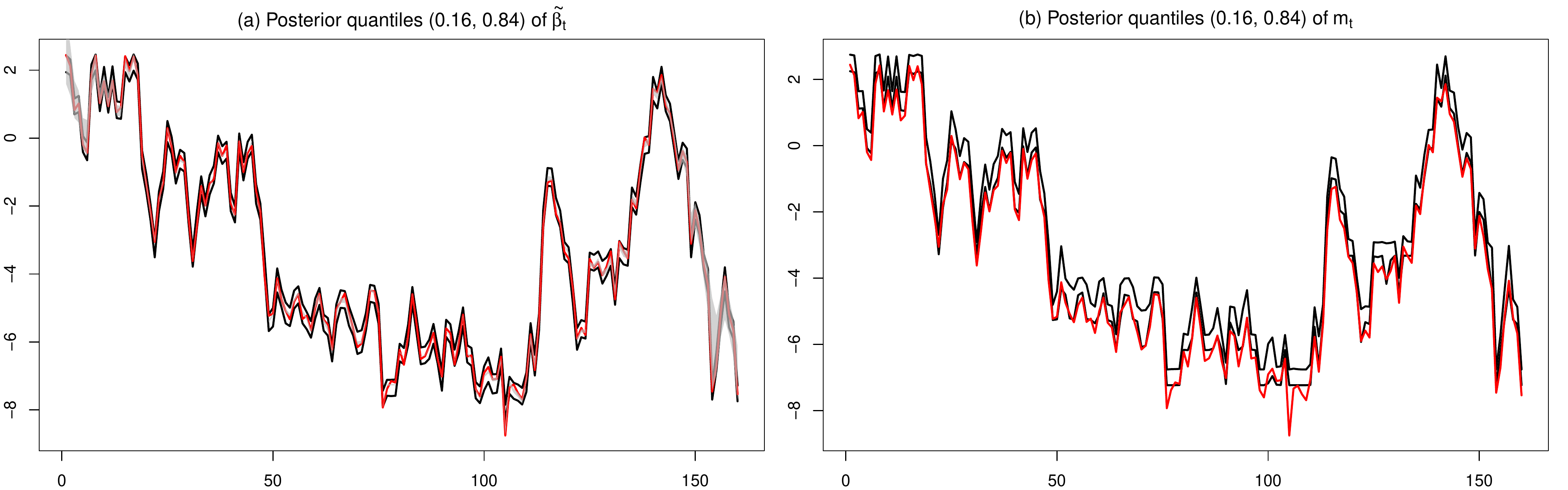}
\begin{minipage}{\linewidth}
\footnotesize \textbf{Notes}: Panel (a) shows $16^{th}$/$84^{th}$ posterior percentiles of $\tilde{\beta}_t$ for our proposed model (solid black lines) and a standard TVP regression with random walk state equation (gray shaded area). Panel (b) shows the $16^{th}$/$84^{th}$ percentiles of the posterior distribution of $m_t$ (in solid black). The red lines denote the actual outcome of $\tilde{\beta}_t$.
\end{minipage}
\end{figure}

\begin{table}
\caption{Posterior probabilities for a given number of groups $G (= 30)$}\label{tab: G_postRW}
\begin{tabular}{rrrrrrrrrrrrr}
  \toprule
 $G_0=$& 12 & 13 & 14 & 15 & 16 & 17   & 18 &  19 &  20 &  21 &  22 &  23 \\ 
  \midrule
& 0.01 & 0.02 & 0.04 &0.07& 0.13& 0.18 &0.16 &0.15 &0.12 &0.07 &0.03& 0.02\\
 \bottomrule
\end{tabular}
\end{table}

The previous discussion has shown that our model works well if the DGP is characterized by relatively few breaks. In the next step, we test the model under a  less favourable DGP: we assume that the law of motion of $\tilde{\beta}_t$ is a random walk with a state innovation variance of one and $\tilde{\beta}_0 = 3$. The results are shown in \autoref{fig: simBetaRW}  and \autoref{tab: G_postRW}. Panel (a) shows that even when the DGP is characterized by many small breaks, our model is flexible enough to capture this behavior as well. This is because we essentially pool coefficients but also allow for idiosyncratic (i.e., time-specific) deviations from the common mean. If we consider panel (b) we observe that the mean process $m_t$ captures the bulk of the variation in $\tilde{\beta}_t$. \autoref{tab: G_postRW} suggests that even if we set $G=30$, the sparse finite mixture allocates substantial posterior mass to lower values of $G$ (with values of $G$ between $15$ and $21$), but still is able to retrieve over $80$ percent of the posterior mass. Hence, even if the true DGP is a random walk and $G$ is much smaller than $T$, our approach accurately recovers the full history of the latent states.


\section{An Application to US Inflation}\label{forecasting} 

\subsection{Data and selected in-sample features}\label{sec: insample}
Modeling and forecasting inflation is of great value for economic agents and policymakers. In most central banks, inflation is the main policy objective and the workhorse forecasting model is based on some version of the Phillips curve.  The practical forecasting of inflation is difficult \citep[see][]{sw_jmcb} and the persistence of low inflation in the presence of a closing output gap in recent years has led to a renewed debate about the usefulness of the curve as a policy instrument in the United States \citep[see, e.g.,][]{BallMazumder, coibion2015phillips}. 

There are three main issues when forecasting inflation. A first problem is that the theoretical literature relating to the Phillips curve and the determination of inflation includes a large battery of very different specifications, emphasizing domestic vs. international variables, forward vs. backward looking expectations or including factors such as labor market developments. The overall number of potential predictors can be quite large \citep[see][]{SW08}.
Second, within each econometric specification there is considerable uncertainty about which indicator should be used as a proxy for the economic cycle  \citep[see][]{Moretti2019}.
Third, there are structural breaks that make different variables and specifications more or less important at different times \citep[see][]{kk2012}. The Great Recession, for example, is universally considered as a structural break that requires appropriate econometric techniques.   

The mainstream literature has dealt with the curse of dimensionality which arises in TVP regressions with many predictors in several ways. Until recently, the two main approaches included principal components or strong Bayesian shrinkage. A comparison of the two approaches can be found in \cite{demol2008}.
Following \cite{RafteryDMA}, a second stream of research uses model combination to deal with the curse of dimensionality and the fact that models can change over time  \citep[e.g.,][]{kk2012}. 
Finally, a recent (but expanding) stream of literature forecasts inflation using machine learning techniques 
\citep{medeiros2019forecasting}. These methods, although useful, suffer from the ``black box problem''; while their accuracy compares well with other techniques, they are not able to show how the result is obtained and thus do not offer a simple interpretation.\footnote{A survey of these techniques is given in \cite{Hassani2015}.} 

For the reasons above, inflation forecasting is an ideal empirical application in which we can investigate the performance of our methods. An important criterion is the capacity of our approach to generalize  standard TVP models, which are less flexible because they are based on random walk or autoregressive specifications to determine the evolution of the states. A second challenge is the correct detection of well-known structural breaks. In addition, we assess the forecasting performance of our methods relative to alternative approaches.

Following  \citet{stock1999forecasting}, we define the target variable as follows
\begin{equation*}
y_{t+h} = \ln \left( \frac{P_{t+h}}{P_{t}} \right) - \ln \left( \frac{P_{t}}{P_{t-1}} \right),
\end{equation*}
with  $P_{t+h}$ denoting the price level (\texttt{CPIAUCSL}) in period $t+h$. Using this definition, we estimate a generalized Phillips curve involving $49$ covariates plus the lagged value of $y_t$ that cover different segments of the economy. Further information on the specific variables included and the way they are transformed is provided in Section B of the Online Appendix. The design matrix $\bm x_t$ includes $p=p_y=p_d=2$ lags and an intercept and thus features $K=101$ covariates. 

Before we use our model to perform forecasting, we provide some information on computation times, illustrate some in-sample features of our model  and briefly discuss selected posterior estimates of key parameters.


Table \ref{tab:realTime} shows empirical runtimes (in minutes) for estimating the different models for this large dataset.  
As highlighted in the beginning of Section \ref{artdata}, our approaches start improving upon FFBS-based algorithms in terms of computation time  if $K$ exceeds $50$, with the improvements increasing non-linearily in $K$. Hence, it is unsurprising that, for our present  application with $K=101$, our algorithm (without clustering) is almost {five times} faster than using FFBS and twice as fast as the efficient AWOL sampler. If clustering is added, our approach is still more than  three times faster than FFBS.  The additional computational complexity from using the clustering prior strongly depends on $G$. If $G$ is close to $T$ (which typically does not occur in practice and we thus do not consider this case), then the computation time increases and the advantage of using the SVD is diminished. This arises since  estimating the location parameters of the mixtures becomes the bottleneck in our MCMC algorithm.  {Finally, using a random walk state evolution equation (i.e., a lower triangular $\bm Z$) with a ridge-prior yields the strongest gains in terms of computational efficiency, being almost six times faster than FFBS and over twice as fast than the AWOL sampler.}



\begin{table}[t]
\centering
\caption{Runtime comparison of empirical exercise ($K = 101$; $ T = 212; G=30$) with $30,000$ draws from the posterior distribution}\label{tab:realTime}
\scriptsize
\begin{tabular}{rccccccccc}
\toprule
& \multicolumn{3}{c}{SVD} && FFBS & \texttt{shrinkTVP} && TIV \\  
\cmidrule{2-4} \cmidrule{6-7}
& WN (g-prior w. clustering) & WN (g-prior) & RW (ridge-prior) & & RW & RW && \\ 
  \midrule
Time (in minutes) & 103 & 76 & 64 & &377 & 150  && 16  \\ \bottomrule
\end{tabular}
\end{table}

To further illustrate  the properties of the estimated parameters in our SVD approach using the g-prior with clustering we now turn to a small-scale model. In this case, the number of coefficients is relatively small and features such as multipliers with an economic interpretation can be easily plotted. This model is inspired by the New Keynesian Phillips curve (NKPC). The dependent variable is inflation and the right hand side variables include two lags of unemployment and inflation. We set $G=30$, thus allowing for a relatively large number of clusters.
 
\begin{figure}[t!]
\centering
\caption{Posterior means of multipliers}\label{fig:longNKPC}
\begin{minipage}[t]{0.49\textwidth}
\centering 
(a) SVD with g-prior with clustering
\end{minipage}
\begin{minipage}[t]{0.49\textwidth}
\centering 
(b) TVP-RW-FFBS
\vspace{10pt}
\end{minipage}
\begin{minipage}[t]{\textwidth}
\centering 
$t + 1$
\end{minipage}
\begin{minipage}[t]{0.49\textwidth}
\centering
\includegraphics[scale=.5]{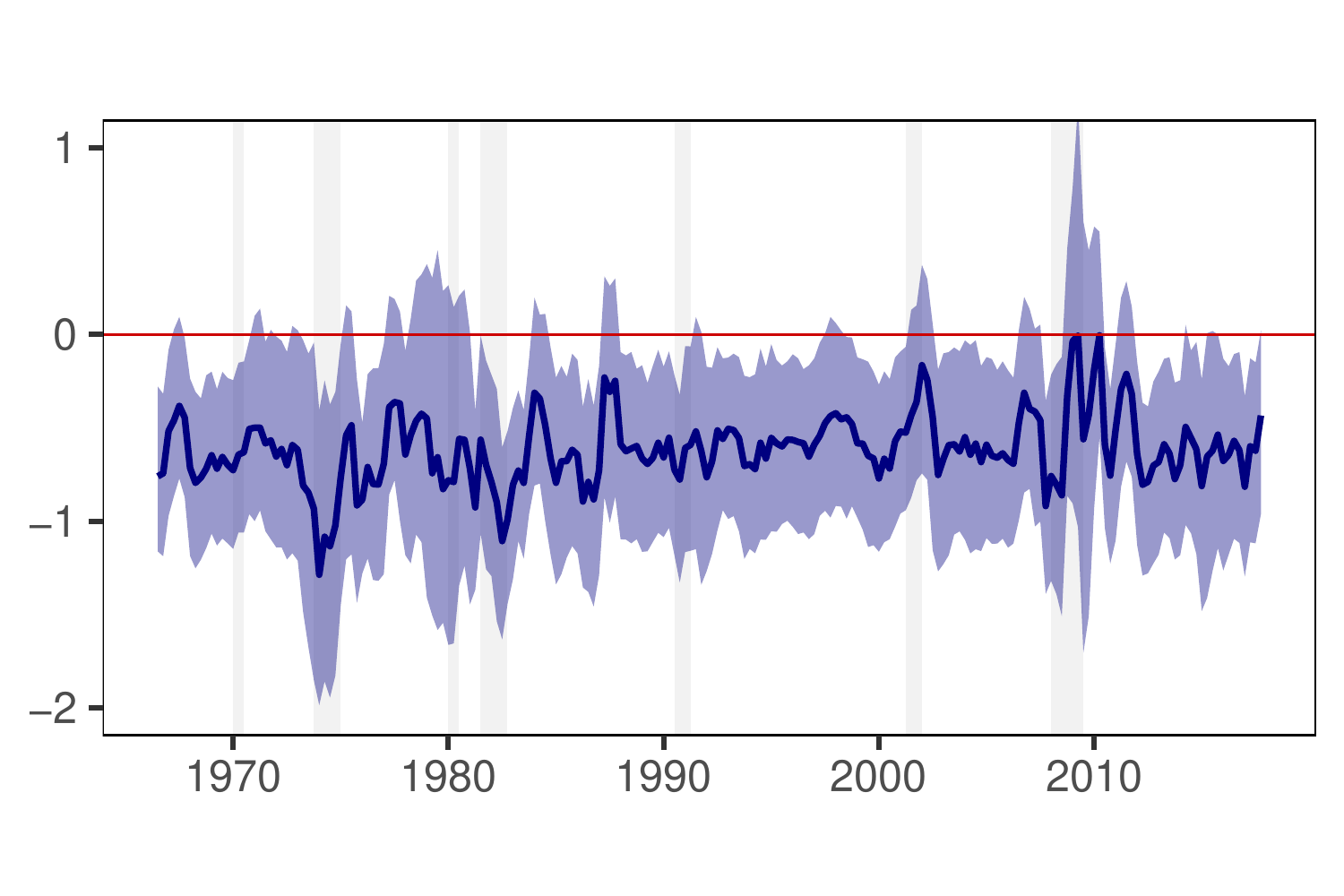}
\end{minipage}
\begin{minipage}[t]{0.49\textwidth}
\centering
\includegraphics[scale=.5]{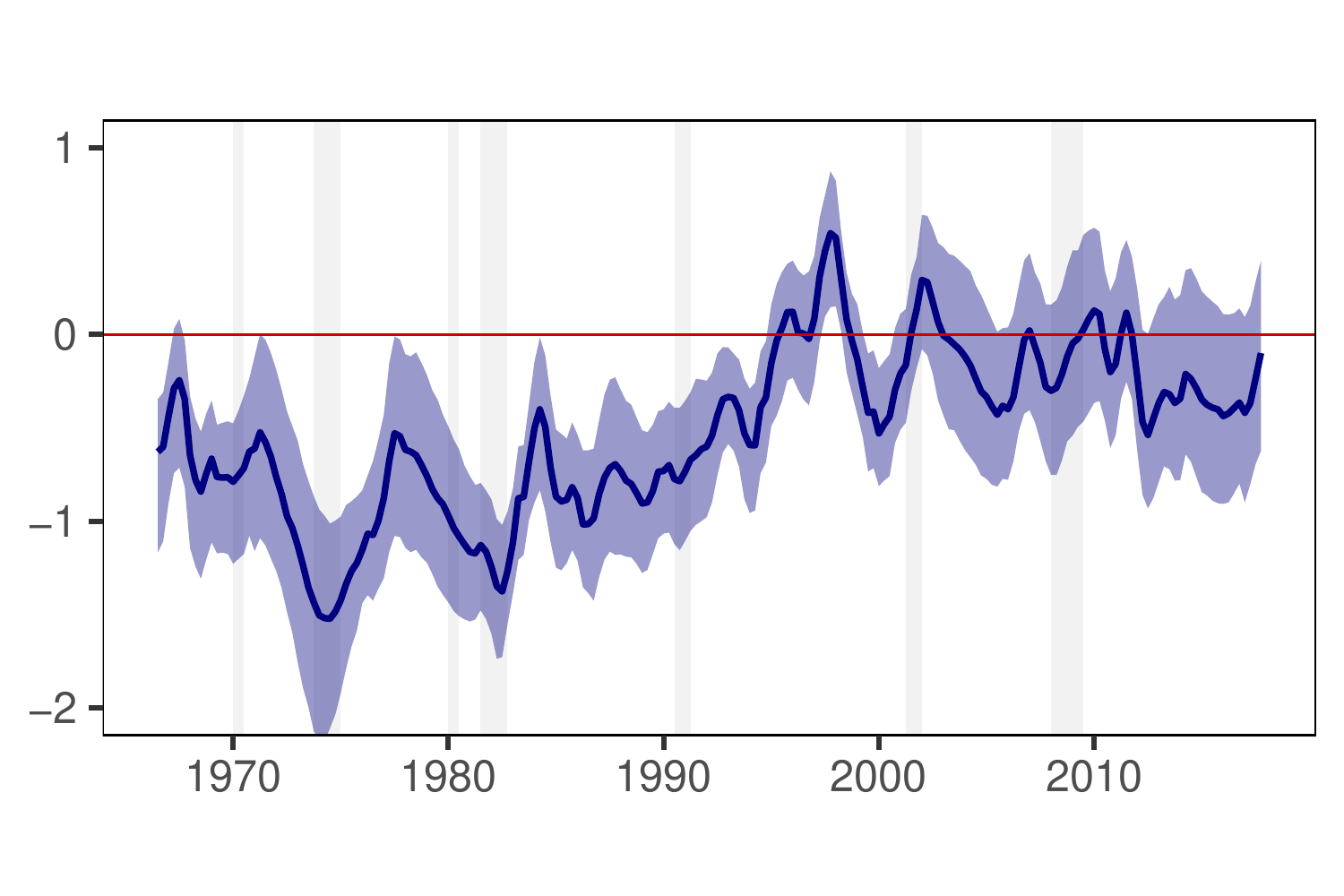}
\end{minipage}
\begin{minipage}[t]{\textwidth}
\centering 
$t + 4$
\end{minipage}
\begin{minipage}[t]{0.49\textwidth}
\centering
\includegraphics[scale=.5]{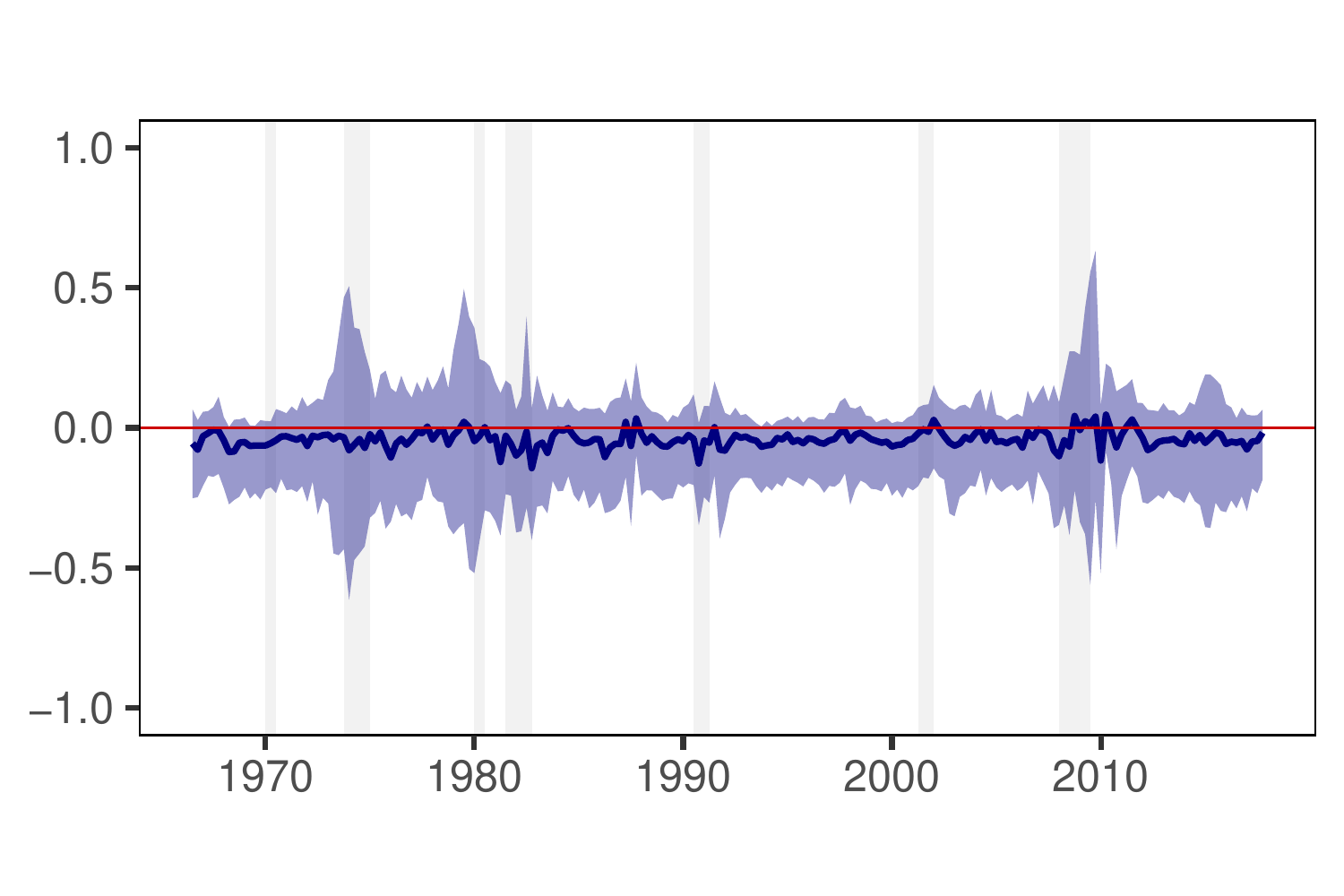}
\end{minipage}
\begin{minipage}[t]{0.49\textwidth}
\centering
\includegraphics[scale=.5]{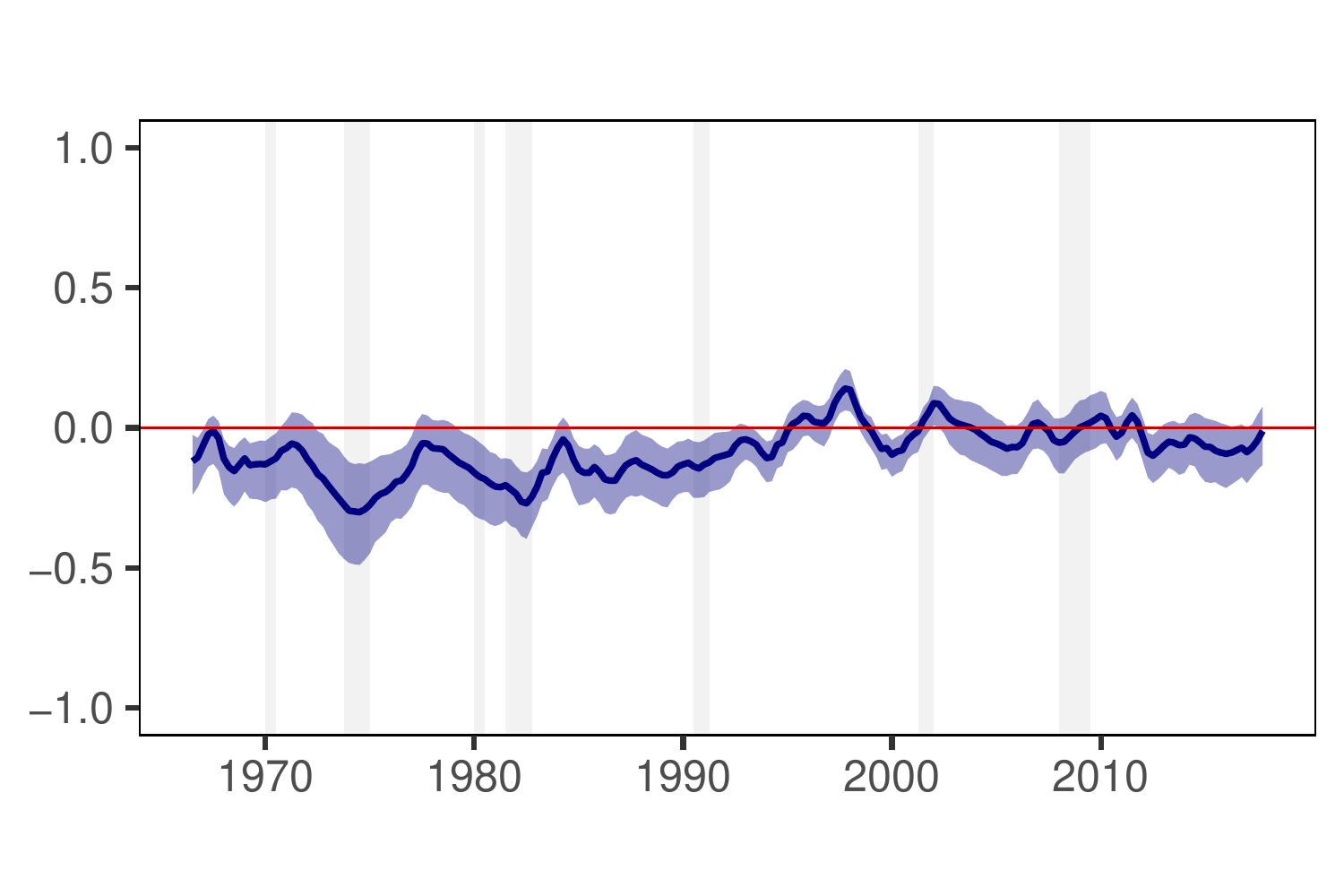}
\end{minipage}
\begin{minipage}[t]{\textwidth}
\centering 
Long run 
\end{minipage}
\begin{minipage}[t]{0.49\textwidth}
\centering
\includegraphics[scale=.5]{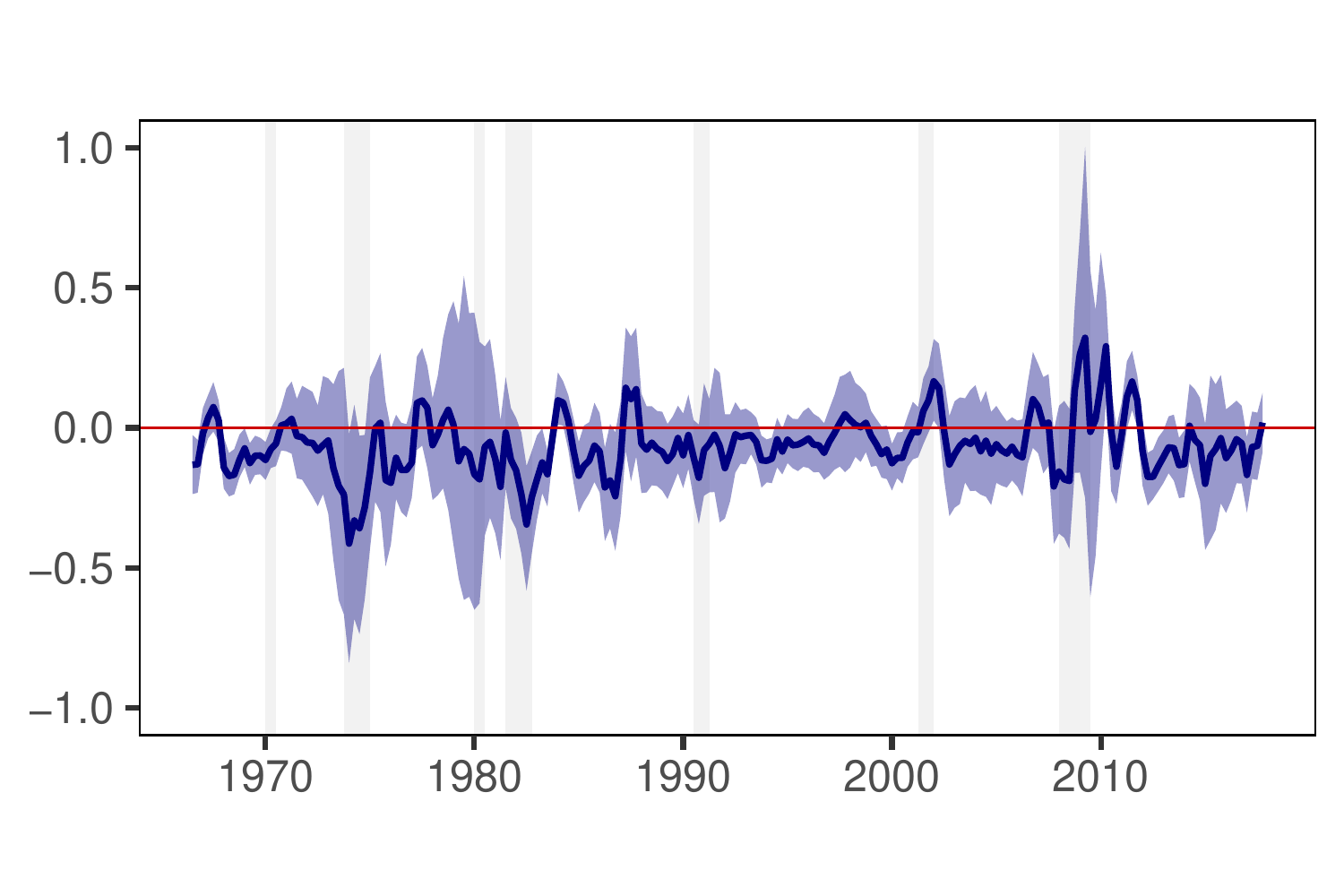}
\end{minipage}
\begin{minipage}[t]{0.49\textwidth}
\centering
\includegraphics[scale=.5]{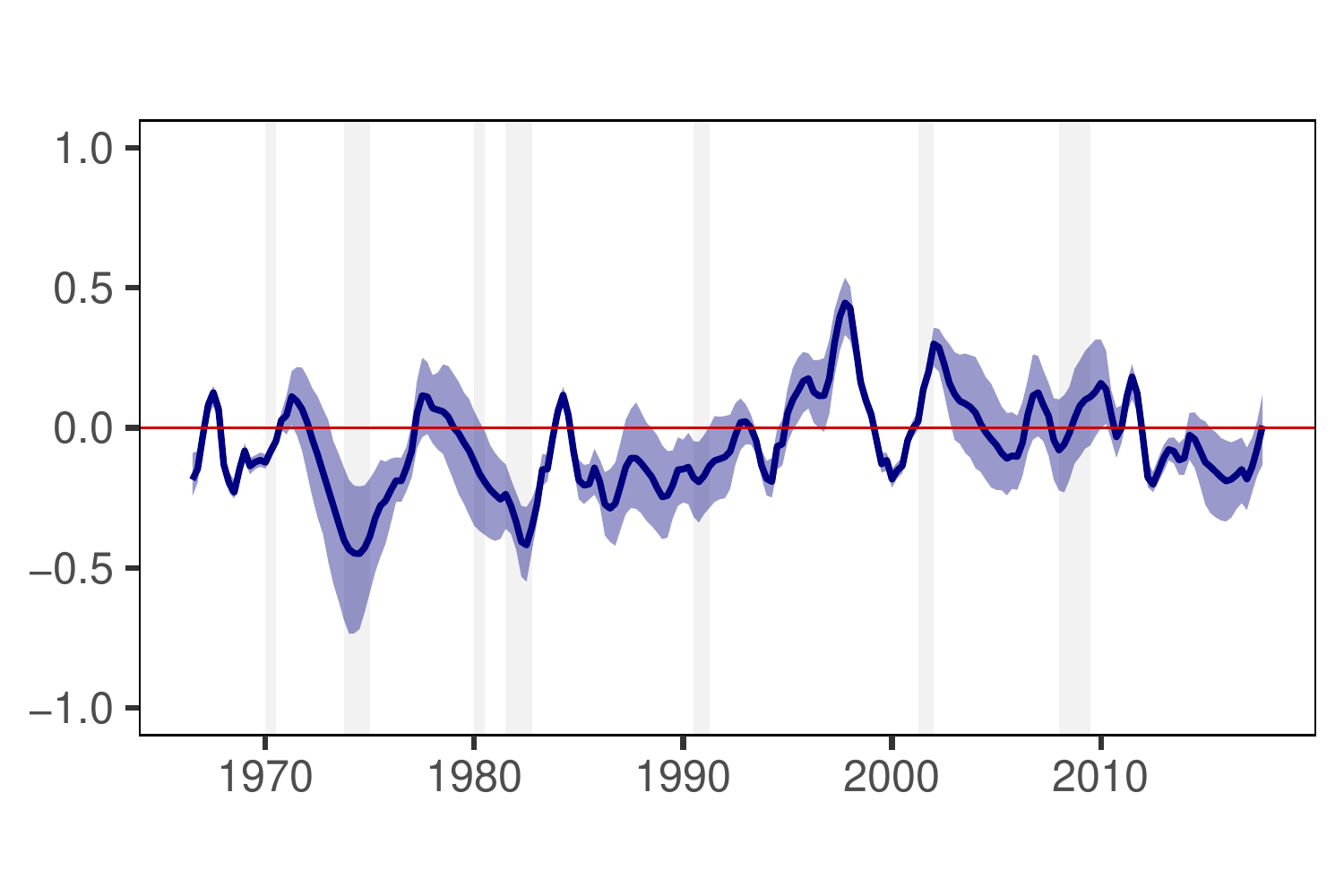}
\end{minipage}

\begin{minipage}{\linewidth}
\footnotesize \textbf{Notes}: Blue shaded areas are $68$\% credible intervals and gray shaded areas denote NBER recessions. 
\end{minipage}
\end{figure}

  \autoref{fig:longNKPC} plots multipliers (i.e., the cumulative effect on inflation of a change in unemployment at various horizons). A comparison of SVD to TVP-RW-FFBS shows many similarities. For instance, both models are saying an increase in unemployment has a negative effect on inflation in the very short term for much of the time. This is what the NKPC would lead us to expect. However, for SVD this negative effect remains for most of the time after the financial crisis whereas for TVP-RW-FFBS it vanishes and the NKPC relationship breaks down. Another difference between the two approaches can be seen in many recessions where the estimated effect changes much more abruptly using our approach than with TVP-RW-FFBS. This illustrates the great flexibility of our approach in terms of the types of parameter change allowed for. And this flexibility does not cost us much in terms of estimation precision in the sense that the credible intervals for the two approaches have similar width. 

\begin{figure} 
\centering 
\caption{Posterior distribution of number of non-empty clusters ($G_0$)}\label{fig:Xi}
\includegraphics[scale=.7]{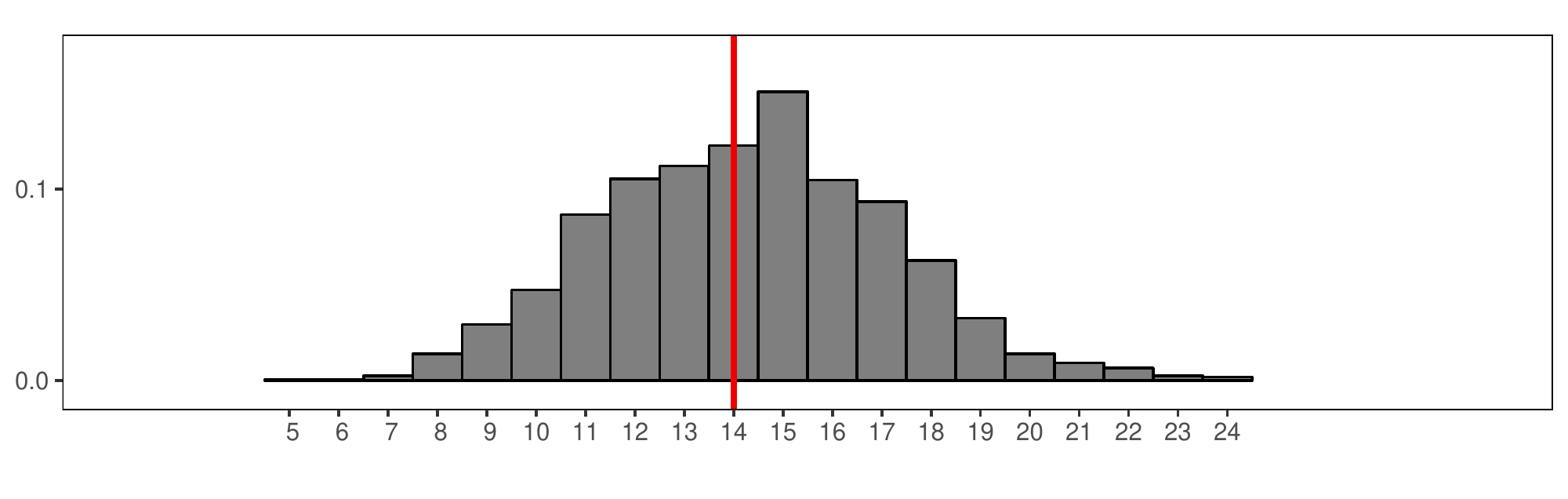}
\begin{minipage}{\linewidth}
\footnotesize \textbf{Notes}:  $G_0$ refers to the non-empty groups with $G = 30$. The red line denotes the median of $G_0$. 
\end{minipage}
\end{figure}

\autoref{fig:Xi} displays the posterior of $G_0$, the number of clusters selected by the algorithm. The posterior is spread over a range of values, although almost all of the posterior probability is associated with a number of clusters between ten and $20$. $G_0=1$ implies that $\tilde{\bm \beta}_t$ is centered around a non-zero value that is time-invariant  and there is  little posterior evidence in this figure indicating support for this. This is the lower bound on the number of clusters. The upper bound on the number of clusters is 30, but the posterior probability lies in a region far below $30$ indicating that the algorithm is successfully finding parsimonious representations for the time variation in parameters. It is worth stressing that these statements hold for the small NKPC model. For the large model with $K=101$, we find the number of clusters to be even smaller. In this case the posterior mode is eight clusters. This inverse relationship between $K$ and $G_0$ is to be expected. That is, as model size increases, more of the variation over time can be captured by the richer information set in $\bm x_t$, leaving less of a role for time variation in coefficients. Our clustering algorithm automatically adjusts to this effect. 

\subsection{Forecasting evidence}\label{sec: forecastingInf}
The forecasting design adopted is recursive. We consider an initial estimation period from $1965$Q$1$ to $1999$Q$4$. The remaining observations ($2000$Q$1$ to $2018$Q$4$) are used as a hold-out period to evaluate our forecasting methods. After obtaining $h \in \{1, 4\}$-step-ahead predictive distributions for a given period in the hold-out, we include this period in the estimation sample and repeat this procedure until we reach the end of the sample. In order to compute longer horizon forecasts, we adopt the direct forecasting approach \citep[see e.g.,][]{stock2002macroeconomic}. To assess forecasting accuracy, we use root mean square forecast errors (RMSEs) for point forecasts and log predictive likelihoods (LPLs, these are averaged over the hold-out period) for density forecasts. We evaluate the statistical significance of the forecasts relative to random walk (RW) forecasts using the \cite{diebold1995comparing} test. 

We compare {four variants of our SVD approach (i.e., the Minnesota prior, the g-prior with and without clustering and the SVD model with a random walk-type state evolution, labeled TVP-RW-SVD)} to alternatives which vary in their treatment of parameter change and in the number of explanatory variables. With regards to parameter change, we consider the time-invariant (TIV) model (which sets $\tilde{\bm \beta}_t=\bm 0$ for all $t$) and the TVP-RW-FFBS approach which has random walk parameter change. Moreover, as an alternative treatment of the TVPs we consider the model of \cite{chan2020reducing} which introduces a factor structure in the latent states (labeled TVP-fac-FFBS). 

With regards to the number of explanatory variables, we consider models with two lags of all $50$ of them (labeled FULL in the tables), none of them as well as some specifications which contain a subset of them. To be specific, we present results for all these models using the NKPC specification discussed in the preceding sub-section (labeled NKPC in the tables).  We also have versions of the model where the intercept is the only explanatory variable, thus leading to an unobserved components model (labeled UCM in the tables).\footnote{For the SVD versions of the UCM models, we only present results for the g-prior with clustering as the other priors imply white noise behavior for inflation which is not sensible.} 

In addition, we include some simple benchmarks that have been used elsewhere in the literature. These include a constant coefficient AR(2) model,  a TVP-AR(2)  and an AR(2) augmented with the two lags of the first three principal components of $\bm d_{t}$ (this is labeled PCA3). This model is closely related to the diffusion index model of \cite{stock2002macroeconomic}. Additionally, we also compress the data to three dimensions using targeted random compressions \citep[labeled TARP, see][]{mukhopadhyay2020targeted}. For each of these two dimension reduction techniques we also present forecasts for a TVP-RW-FFBS model. All models considered include stochastic volatility.

Table \ref{tab:LPS} contains our main set of forecasting results. Note first that, with some exceptions, the FULL models do best, indicating that there is information in our $K=50$ variables useful for inflation forecasting. If we focus on results for the FULL models, it can be seen that, for $h=1$ all of the approaches forecast approximately as well as each other. But for $h=4$ there are substantial improvements provided by our SVD approaches relative to the competitors. At this forecast horizon, it is interesting to note that the very parsimonious UCM version of the TVP-RW-FFBS provides point forecasts that are almost as good as those provided by the FULL SVD approaches. However, the density forecasts provided by the UCM are appreciably worse than those provided by the SVD approaches. The FULL SVD approaches are also beating  approaches based on dimension reduction (PCA, TARP), even if we allow for time-variation in the coefficients for these models. 

Comparing the results of our SVD-based models {with a block-diagonal $\bm Z$} to the ones which constrain the state evolution (i.e., TVP-RW-FFBS, TVP-fac-FFBS and TVP-RW-SVD) sheds light on how much the increased flexibility improves forecasting accuracy. In terms of one-step-ahead forecasts we find that our flexible approaches yield very similar forecasts to the ones of TVP regressions with random walk state equations. This is consistent with the statement that for short-term forecasting, our model yields forecasts which are competitive to established methods in the literature. When we consider multi-step-ahead forecasts we find pronounced improvements in terms of point and density forecasts for the FULL and NKPC models. Notice the better performance of TVP-fac-FFBS and TVP-RW-SVD relative to TVP-RW-FFBS. {In the latter case, this is driven by the ridge-type prior which strongly shrinks the TVPs towards zero whereas in the former case, the better performance can be attributed to the parsimonious factor structure on the TVPs.}

\begin{table}[t]
{\tiny
\begin{center}
\caption{\label{tab:LPS} Forecasting Performance of SVD Approaches Relative to Benchmarks}
\begin{tabular}{llllcll}
\toprule
\multicolumn{1}{l}{\bfseries }&\multicolumn{3}{c}{\bfseries Specification}&\multicolumn{1}{c}{\bfseries }&\multicolumn{2}{c}{\bfseries Forecast horizon}\tabularnewline
\cmidrule{2-4} \cmidrule{6-7}
\multicolumn{1}{l}{}&\multicolumn{1}{c}{TVP/TIV}&\multicolumn{1}{c}{Type}&\multicolumn{1}{c}{$\kappa$}&\multicolumn{1}{c}{}&\multicolumn{1}{c}{1-step}&\multicolumn{1}{c}{4-steps}\tabularnewline
\midrule
{\scshape AR(p)}&&&&&&\tabularnewline
~~&TIV&Benchmark&&&0.90&0.77***\tabularnewline
~~&&&&&(0.08)&(0.23***)\tabularnewline
~~&TVP-RW-FFBS&Benchmark&&&0.90&0.75***\tabularnewline
~~&&&&&(0.08)&(0.26***)\tabularnewline
\midrule
{\scshape FULL}&&&&&&\tabularnewline
~~&TIV&Benchmark&&&0.82*&0.61**\tabularnewline
~~&&&&&(0.15)&(0.37)\tabularnewline
~~&TVP-fac-FFBS&Benchmark&&&0.83**&0.63\tabularnewline
~~&&&&&(0.17**)&(0.25)\tabularnewline
~~&TVP-RW-FFBS&Benchmark&&&0.78*&0.92\tabularnewline
~~&&&&&(0.16)&(0.01)\tabularnewline
\cmidrule{2-7}
~~&TVP-RW-SVD&ridge-prior&0.001&&0.81**&0.62**\tabularnewline
~~&&&&&(0.14)&(0.43***)\tabularnewline
~~&TVP-WN-SVD&g-prior&0.1&&0.80***&0.59**\tabularnewline
~~&&&&&(0.15**)&(0.42*)\tabularnewline
~~&TVP-WN-SVD&g-prior (clustering)&0.05&&0.80***&0.57***\tabularnewline
~~&&&&&(0.17**)&(0.48***)\tabularnewline
~~&TVP-WN-SVD&Minnesota&0.1&&0.82**&0.61**\tabularnewline
~~&&&&&(0.16*)&(0.37*)\tabularnewline
\midrule
{\scshape NKPC}&&&&&&\tabularnewline
~~&TIV&Benchmark&&&0.91&0.82***\tabularnewline
~~&&&&&(0.06)&(0.12)\tabularnewline
~~&TVP-RW-FFBS&Benchmark&&&0.92&0.86\tabularnewline
~~&&&&&(0.07)&(-0.28*)\tabularnewline
\cmidrule{2-7}
~~&TVP-WN-SVD&g&0.001&&0.89&0.79***\tabularnewline
~~&&&&&(0.07)&(0.13)\tabularnewline
~~&TVP-WN-SVD&g-prior (clustering)&0.001&&0.90&0.80***\tabularnewline
~~&&&&&(0.07)&(0.12)\tabularnewline
~~&TVP-WN-SVD&Minnesota&0.001&&0.91&0.81***\tabularnewline
~~&&&&&(0.05)&(0.13)\tabularnewline
\midrule
{\scshape PCA3}&&&&&&\tabularnewline
~~&TIV&Benchmark&&&0.92&0.83***\tabularnewline
~~&&&&&(0.06)&(0.18***)\tabularnewline
~~&TVP-RW-FFBS&Benchmark&&&0.88&0.86\tabularnewline
~~&&&&&(0.09)&(0.05)\tabularnewline
\midrule
{\scshape TARP}&&&&&&\tabularnewline
~~&TIV&Benchmark&&&0.99&0.85***\tabularnewline
~~&&&&&(0.01)&(0.15***)\tabularnewline
~~&TVP-RW-FFBS&Benchmark&&&0.92***&0.82\tabularnewline
~~&&&&&(0.14***)&(0.17)\tabularnewline
\midrule
{\scshape UCM}&&&&&&\tabularnewline
~~&TVP-RW-FFBS&Benchmark&&&0.86***&0.59**\tabularnewline
~~&&&&&(0.16)&(0.16)\tabularnewline
~~&TVP-WN-SVD&g-prior (clustering)&1&&0.88*&0.71\tabularnewline
~~&&&&&(0.08)&(0.14)\tabularnewline
\bottomrule
\end{tabular}
\end{center}}
\begin{minipage}[c]{\linewidth}
\footnotesize{\textbf{Notes}: The table shows RMSEs with LPL's in parentheses below. Asterisks indicate statistical significance for each model relative to a random walk at the $1$ ($^{***}$), $5$ ($^{**}$) and $10$ ($^{*}$) percent significance levels. }
\end{minipage}
\end{table}

With two different forecast horizons and two different forecast metrics, we have four possible ways of evaluating any approach. For three of these, the FULL SVD approach using the g-prior with clustering performs best. The only exception to this is for RMSEs for $h=1$, although even here FULL SVD with g-prior is the second best performing approach. The improvements relative to our other SVD approaches which do not involve clustering are small, but are consistently present. This indicates the benefits of the clustering prior.

 In general, the TIV approaches do well (for $h=4$ even better than TVP-RW-FFBS) in terms of point forecasts, but the density forecasts produced by our SVD approaches are slightly better. This suggests there is only a small amount of time-variation in this data set, but that our SVD approach (particularly when we add the hierarchical clustering prior) is effectively {capturing structural breaks} in a manner that the random walk evolution of the TVP-RW-FFBS {and TVP-RW-SVD} cannot. 

\begin{figure}[t!]
\centering
\caption{Evolution of log predictive Bayes factor relative to RW}
\begin{minipage}[t]{\textwidth}
\centering 
(a) One-step-ahead
\vspace{10pt}
\end{minipage}
\begin{minipage}[t]{0.32\textwidth}
\centering 
\footnotesize \textit{TIV}
\end{minipage}
\begin{minipage}[t]{0.32\textwidth}
\centering 
\footnotesize \textit{TVP-RW-FFBS}
\end{minipage}
\begin{minipage}[t]{0.32\textwidth}
\centering 
\footnotesize \textit{TVP-SVD}
\end{minipage}
\begin{minipage}[t]{0.32\textwidth}
\includegraphics[width=\textwidth]{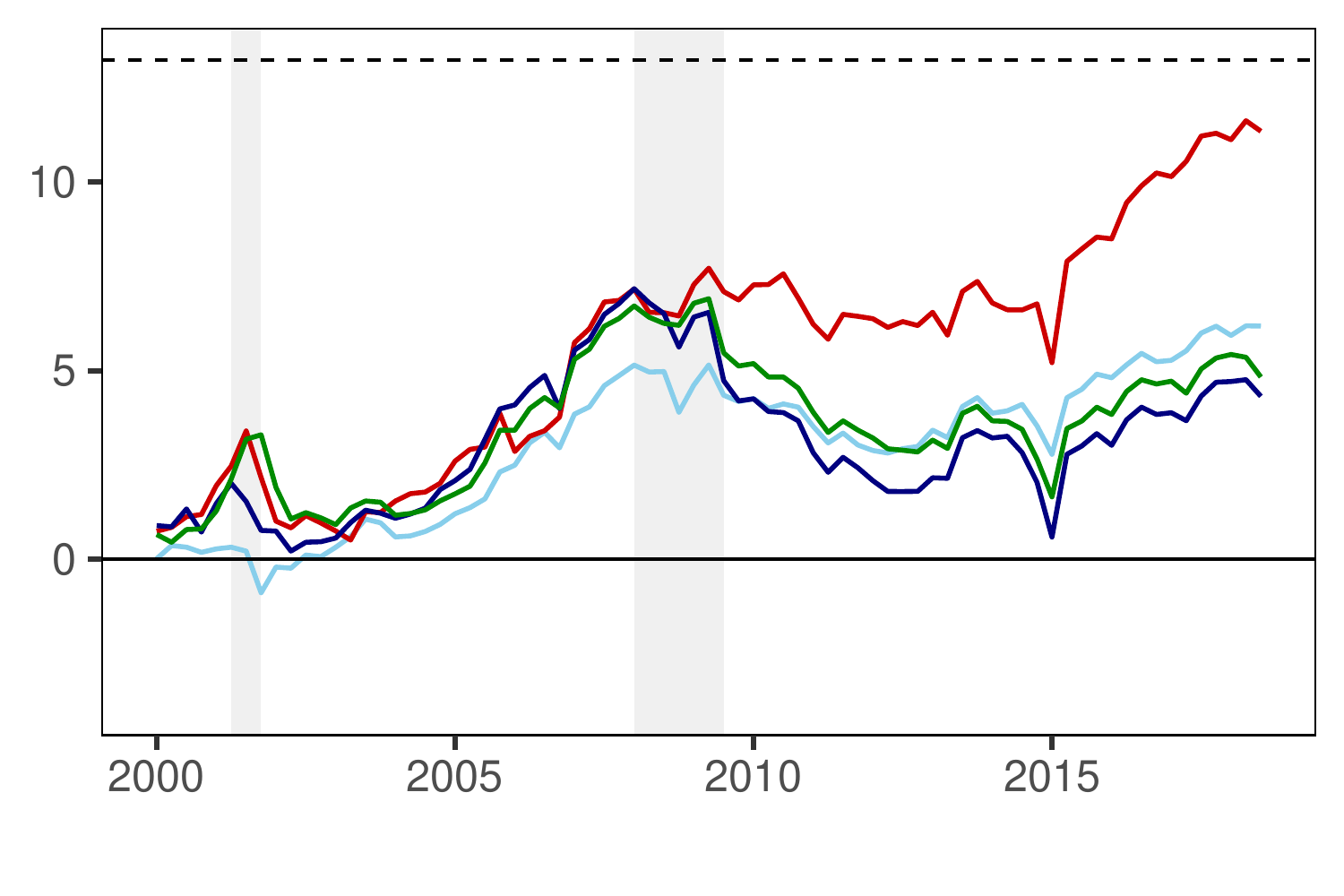}
\end{minipage}
\begin{minipage}[t]{0.32\textwidth}
\includegraphics[width=\textwidth]{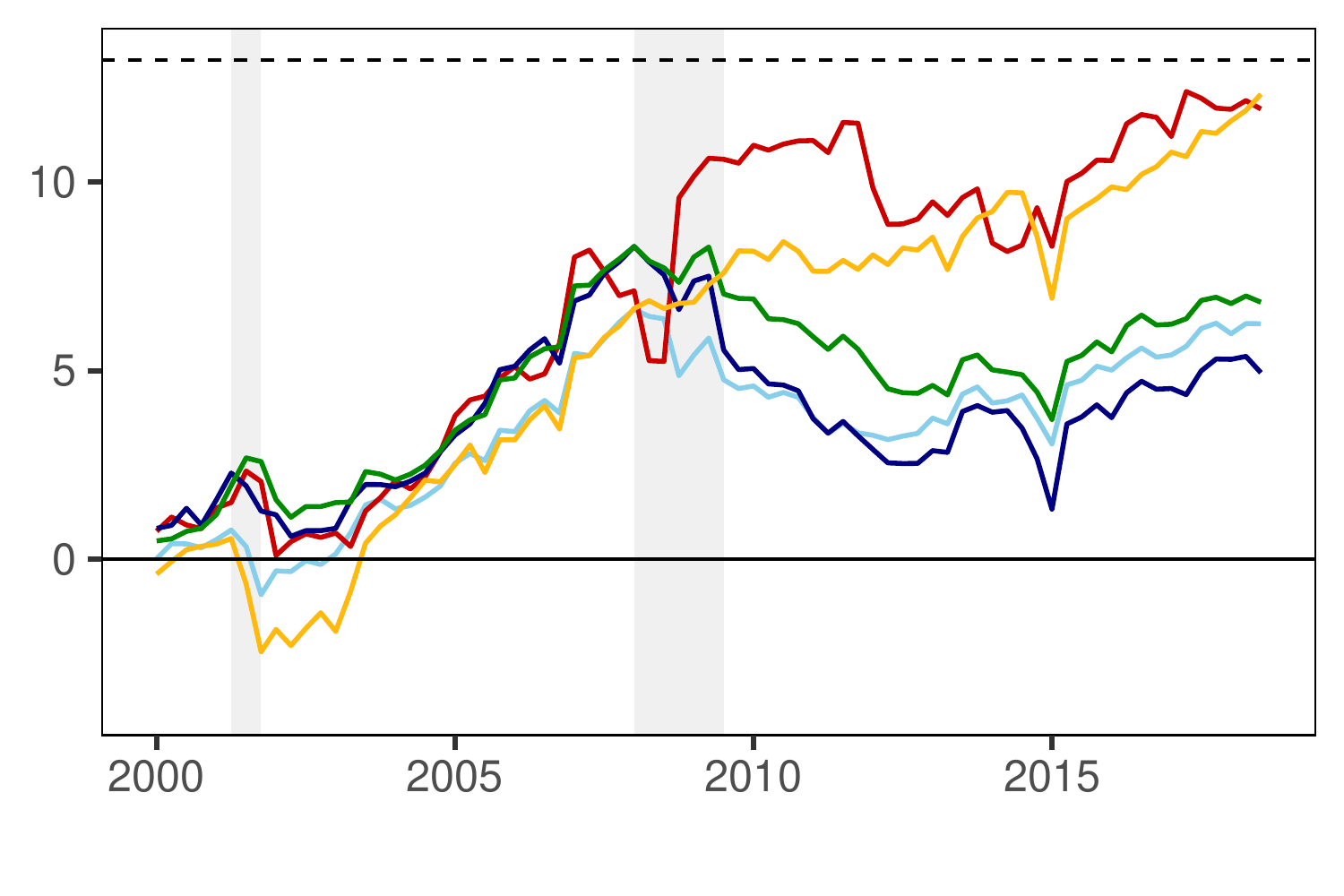}
\end{minipage}
\begin{minipage}[t]{0.32\textwidth}
\includegraphics[width=\textwidth]{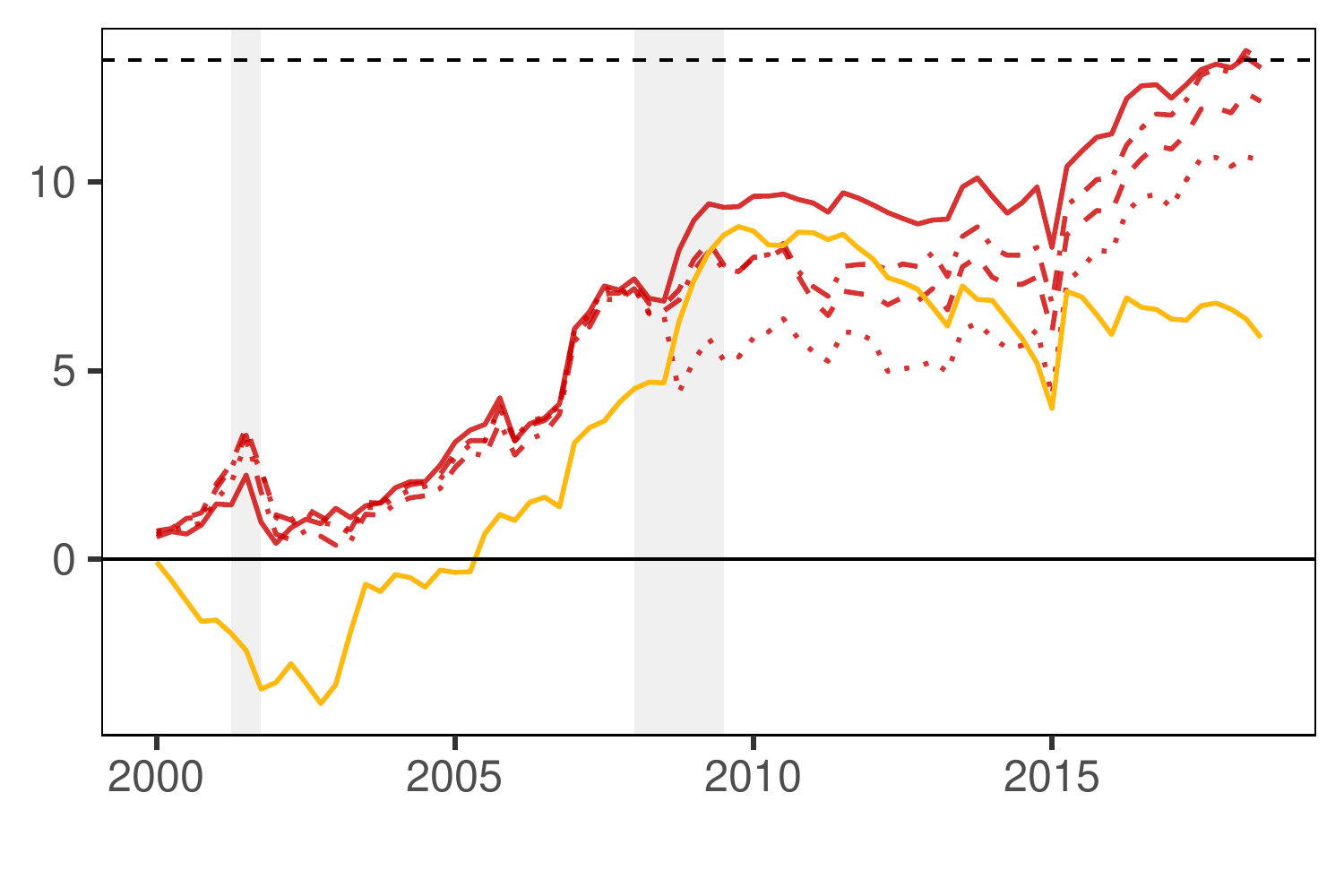}
\end{minipage}
\begin{minipage}[t]{\textwidth}
\centering 
(b) Four-step-ahead
\vspace{10pt}
\end{minipage}
\begin{minipage}[t]{0.32\textwidth}
\centering 
\footnotesize \textit{TIV}
\end{minipage}
\begin{minipage}[t]{0.32\textwidth}
\centering 
\footnotesize \textit{TVP-RW-FFBS}
\end{minipage}
\begin{minipage}[t]{0.32\textwidth}
\centering 
\footnotesize \textit{TVP-SVD}
\end{minipage}
\begin{minipage}[t]{0.32\textwidth}
\includegraphics[width=\textwidth]{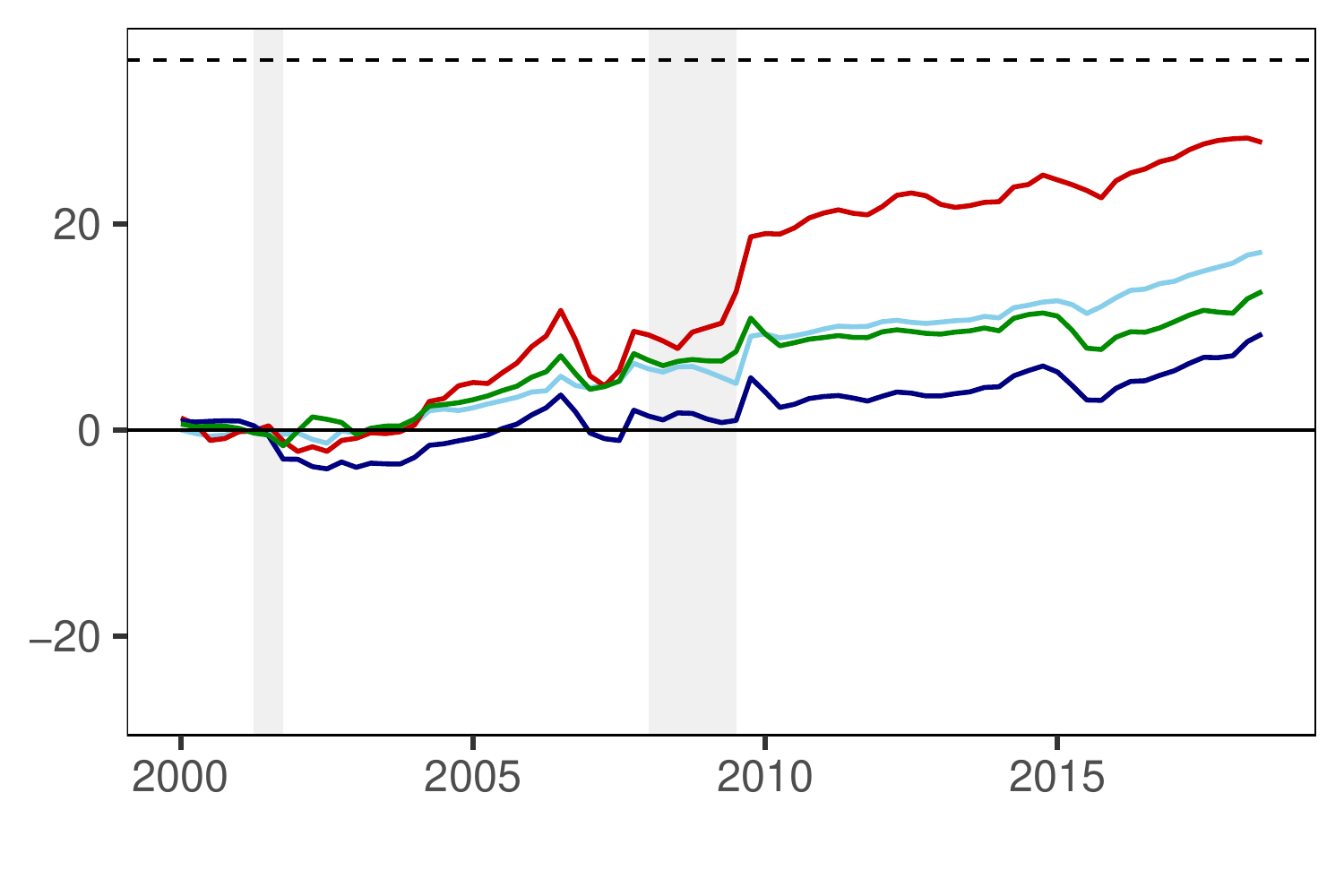}
\end{minipage}
\begin{minipage}[t]{0.32\textwidth}
\includegraphics[width=\textwidth]{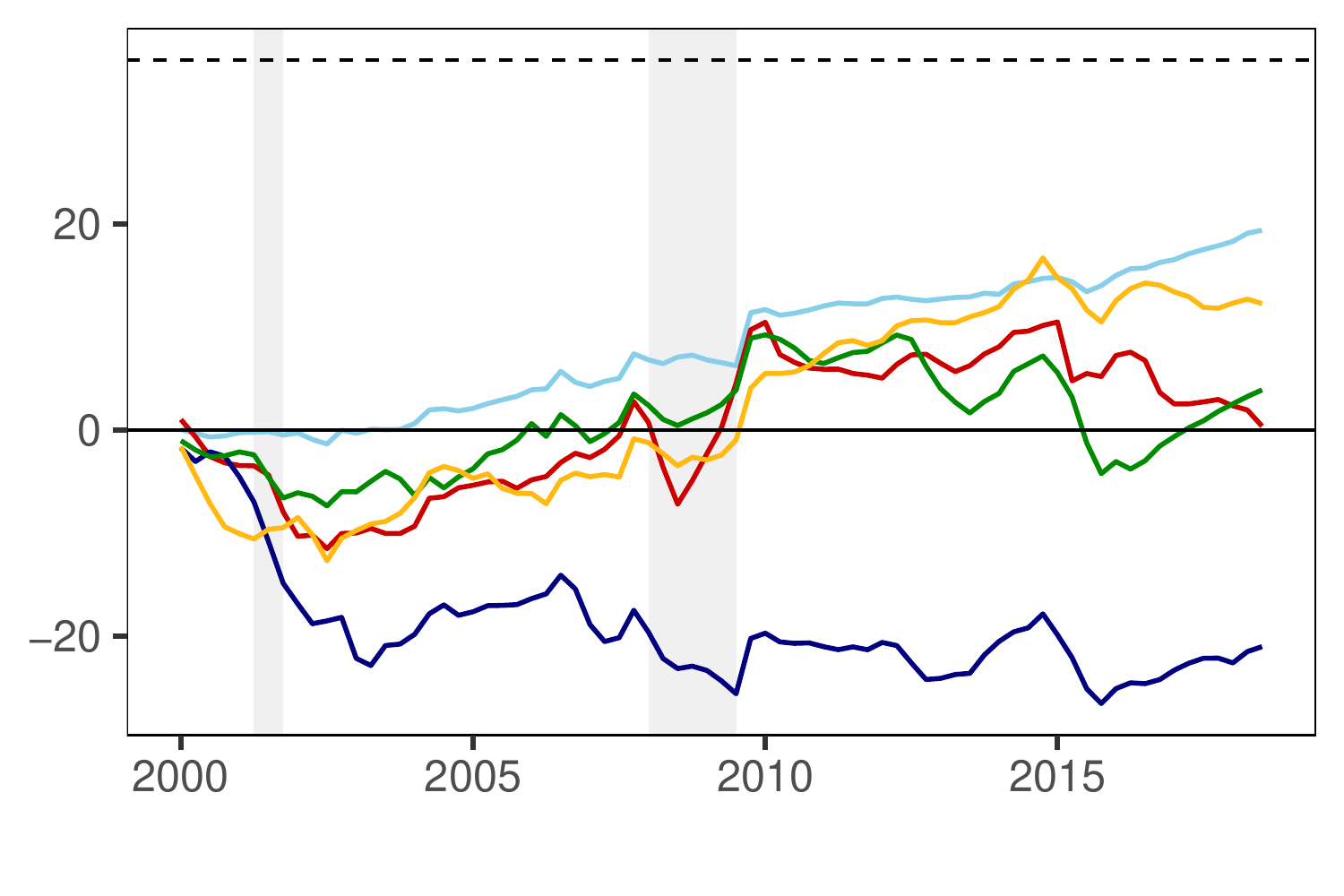}
\end{minipage}
\begin{minipage}[t]{0.32\textwidth}
\includegraphics[width=\textwidth]{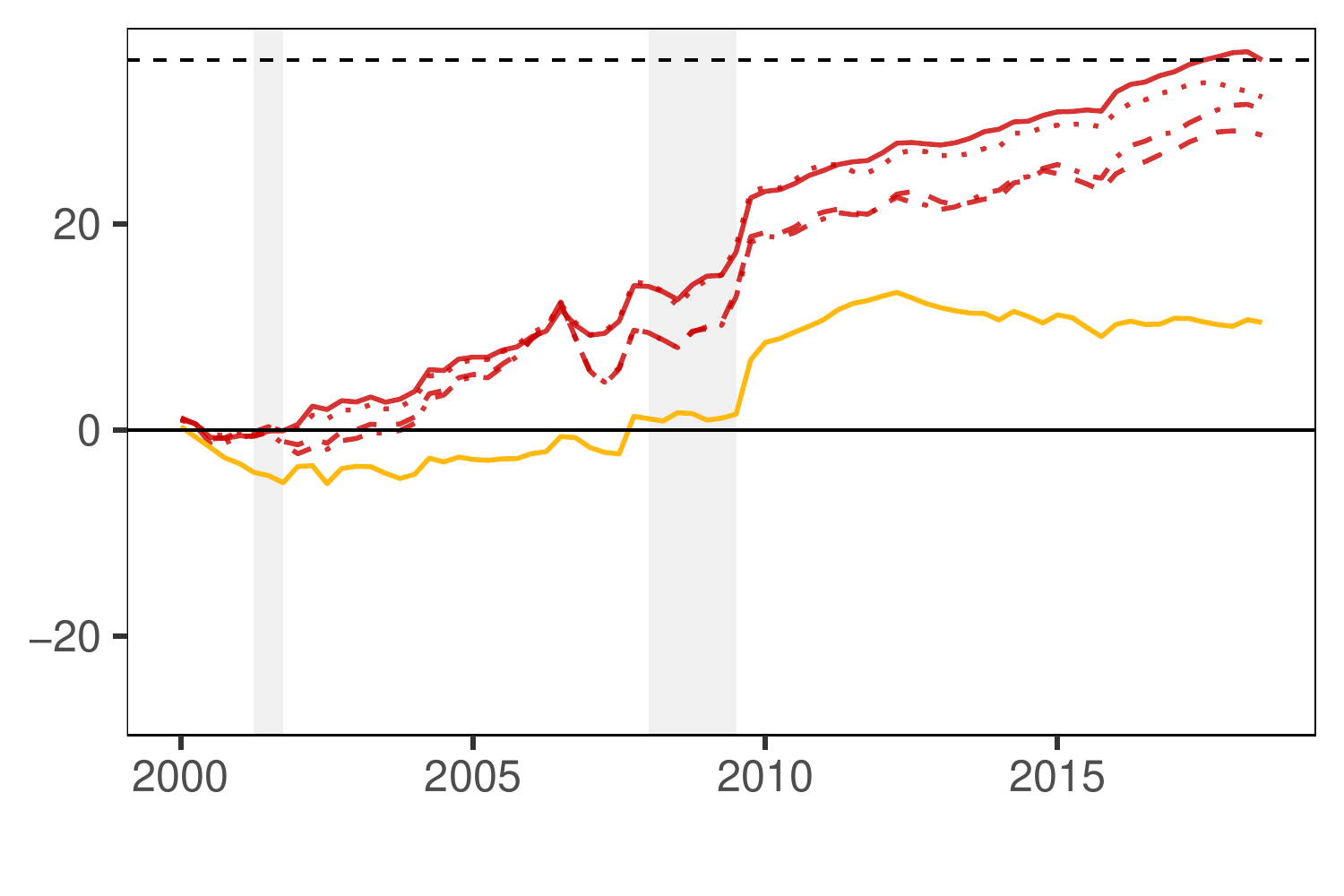}
\end{minipage}
\begin{minipage}[t]{1\textwidth}
\centering
\includegraphics[width=0.4\textwidth]{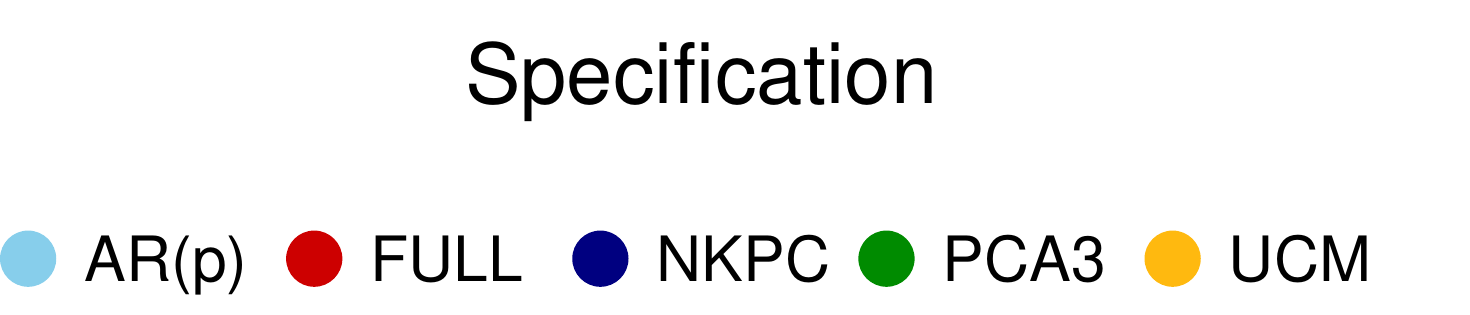}
\end{minipage}
\begin{minipage}{\linewidth}
\footnotesize \textbf{Notes}: The log predictive Bayes factors are cumulated over the hold-out. For the TVP-SVD models the solid line refers to the g-prior with clustering, the dashed line to the Minnesota prior and the dot-dashed line to the g-prior without clustering (each with block-diagonal $\bm Z$), while the dotted line refers to the ridge-prior (with lower triangular $\bm Z$). The dashed black lines refer to the maximum Bayes factor at the end of the hold-out sample. The gray shaded areas indicate the NBER recessions in the US. 
\end{minipage}
\label{lpl}
\end{figure}

Figure \ref{lpl} provides evidence of forecast performance over time for selected models used in this forecasting exercise. The lines in this figure are cumulated log predictive Bayes factors relative to a random walk.

One pattern worth noting is that the benefits of using the FULL model increase after the beginning of the financial crisis. This is true not only for our SVD models, but also for the TIV model. However, notice that during the crisis, the slope of the line associated with the FULL SVD approach becomes steeper, indicating that the model strongly outperforms the RW for that specific time period. This potentially arises from the fact that during recessions, we typically face abrupt structural breaks in the regression parameters and our approach is capable of detecting them. 

To examine how our model performs in turbulent times we focus on forecast accuracy in the Great Recession. It is worthwhile to keep in mind that inflation was fairly stable through $2008$Q$3$. $2008$Q$4$ and $2009$Q$1$ were the periods associated with a substantial fall in inflation. Subsequently, inflation became more stable again. Accordingly, it is particularly interesting to look at $2008$Q$4$ and $2009$Q$1$ as periods of possible parameter change.  We find that the FULL SVD approach performs comparable to a no-change benchmark model. The simple RW model can be expected to handle a one-off structural break well in the sense that it will forecast poorly for the one period where the break occurs and then immediately adjust to the new lower level of the series. Our FULL SVD approach handles the $2008$Q$4$ and $2009$Q$1$ period about as well as the RW. Subseqeuently, its forecasts improve relative to a RW. This improvement occurs in the middle of the Great Recession for $h=1$ and a bit later for $h=4$. In contrast, the TVP-RW-FFBS and TVP-RW-SVD models with the large data set experience a big drop in forecasting performance at the beginning of the Great Recession and tend not to outperform the random walk after $2010$. However, both do well in late $2009$. We conjecture that this pattern of performance reflects two things. First, similarly to our SVD based models which do not constrain the state evolution, both allow for structural breaks, but are slow to adjust to them. Second, they overfit the data and, thus, provide  wide predictive distributions. In the latter half of 2009, after the structural break had occured, when there was still uncertainty about the new pattern in inflation, having this wider predictive distribution benefitted forecast performance. 

This discussion provides evidence that our model works well under stressful conditions. The main mechanism driving the strong forecast performance is that the prior variance is allowed to adapt over time and if uncertainty increases (i.e., $\sigma_t^2$ becomes large), the prior variances increase as well and thus make larger jumps in the parameters more probable.

It can also be seen that our SVD approaches with block diagonal $\bm Z$ tend to perform similarly to one another and never forecast very poorly. This contrasts with the TVP-RW-FFBS and TVP-RW-SVD models which sometimes forecast well, but sometimes yield imprecise forecasts (see, e.g., results for $h=4$ using the NKPC data set). 

Overall, we find our SVD approaches, and in particular the version that uses the clustering prior, to exhibit the best forecast performance among a set of popular benchmarks. And it is worth stressing that they are computationally efficient and, thus, scaleable. The reason this application uses $K=101$ explanatory variables as opposed to a much larger number is due to our wish to include the slower TVP-RW-FFBS approach so as to offer a comparison with the most popular TVP regression model. If we were to have omitted this comparison, we could have chosen $K$ to be much larger. 

Finally, a brief word on prior sensitivity is in order. The two key (hyper)parameters of our model are $\kappa$ and $G$. In Section \ref{sec: empAppendix} of the Online Appendix we carried out an extensive prior robustness analysis. In this analysis we find that the precise choice of $\kappa$ plays a limited role for predictive performance unless it is set too large. This statement holds for large models but, to a somewhat lesser extent, also for smaller models. In the smaller models, we find that predictive performance is slightly more sensitive to the choice of $\kappa$ and the researcher thus has to select this hyperparameter with some care. When it comes to the choice of $G$, we find that as long as it is not set too small, forecasting accuracy does not change substantially. This finding indicates that our shrinkage prior on $\pi$ successfully empties out irrelevant clusters if $G$ is large. In an extreme case, that is, if $G$ is set too small a priori, we lose important information on how states evolve over time and this is deleterious for predictive accuracy.


\section{Conclusions}\label{concl}
In many empirical applications in macroeconomics, there is strong evidence of parameter change. But there is often uncertainty about the form the parameter change takes. Conventional approaches to TVP regression models have typically made specific assumptions on how the states evolve over time (e.g., random walk or structural break). In the specification used in this paper, no restriction is placed on the form that the parameter change can take. However, our very flexible specification poses challenges in terms of computation and surmounting over-parameterization concerns. We have addressed the computational challenge through using the SVD of the high-dimensional set of regressors. We show how this leads to large simplifications since key matrices become diagonal or have banded forms. The over-parameterization worries are overcome through the use of hierarchical priors and, in particular, through the use of a sparse finite mixture representation for the time-varying coefficients. 

In artificial data, we demonstrate the speed and scaleability of our methods relative to standard approaches. In an inflation forecasting exercise, we show how our methods can uncover different forms of time-variation in parameters than other approaches. Furthermore, they forecast well. Since our approach is capable of quickly adjusting to changing economic conditions and outliers, it might also be well suited when applied to macroeconomic forecasting  in extreme periods such as the Covid-19 pandemic.

\footnotesize{\setstretch{0.83}
\addcontentsline{toc}{section}{References}
\bibliographystyle{custom.bst}
\bibliography{tvpsvd.bib}
}
\clearpage
\normalsize{\setstretch{1.5}}
\begin{appendices}
\setcounter{page}{1}
\begin{center}
\LARGE\textbf{Online Appendix}\\[0.5em]
\Large\textbf{Fast and Flexible Bayesian Inference in Time-varying \\ \vspace{-0.5em} Parameter Regression Models}\\[1em]
\large\MakeUppercase{Niko Hauzenberger}$^1$, \MakeUppercase{Florian Huber}$^1$, \MakeUppercase{\\ \vspace{-0.5em} Gary Koop}$^2$, and \MakeUppercase{Luca Onorante}$^{3,4}$\\
$^1$ \textit{University of Salzburg}\\
\vspace{-0.5em} $^2$ \textit{University of Strathclyde} \\
\vspace{-0.5em} $^3$\textit{Joint Research Centre, European Commission} \\
\vspace{-0.5em} $^4$ \textit{European Central Bank}
\end{center}

\numberwithin{equation}{section}
\setcounter{section}{0}
\section{Full conditional Posterior Simulation}\label{app:post}
\subsection{Full Conditional Posterior Distributions}
In this section, we provide details on the full conditional posterior distributions of the model described in Section \ref{sec:model}. We start by outlining the relevant full conditionals for the time-invariant part of the model. The conditional posterior of the time-invariant coefficients $\bm \gamma$ follows a multivariate Gaussian distribution:
\begin{equation}
\bm \gamma|Data, \tilde{\bm \beta}, \bm \sigma^2, \bm \tau, \psi \sim \mathcal{N}\left(\bar{\bm \gamma}, \bm V_{\gamma} \right) \label{eq: post_gamma},
\end{equation}
with 
\begin{align*}
\bm V_{\gamma} =& (\tilde{\bm X}'\tilde{\bm X} + \bm D_{\gamma}^{-1})^{-1},\\
\bar{\bm \gamma} =& \bm V_{\gamma}(\tilde{\bm{X}}'\tilde{\bm y}).
\end{align*}
We let $\bm \sigma^2=(\sigma^2_1, \dots, \sigma^2_T)'$ denote a $T$-dimensional vector of volatilities, $\bm \tau=(\tau_1, \dots, \tau_K)'$ stores the $K$ local scaling parameters of the NG prior  while the $T \times K$-dimensional matrix $\tilde{\bm X}$ is obtained by stacking the rows of $\bm x_t$ and normalizing by dividing each row by $\sigma_t$.

The local scaling parameters follow a generalized inverse Gaussian (GIG) distribution \citep{griffin2010inference}:\footnotetext{The $\text{GIG}(a,b,c)$ is parameterized as $p(x)\propto x^{a -1} \exp\{-(bx + c/x)/2\}$.}
\begin{equation}
\tau_j|\gamma_j, \psi \sim \text{GIG}\left(\vartheta - \frac{1}{2}, \vartheta \psi, \gamma_j^2 \right),  \text{ for} \hspace{2pt} j = 1, \dots, K. \label{eq: post_tau}
\end{equation}

The full conditional posterior distribution of the global shrinkage parameter is defined as 
\begin{equation}
\psi|\tau_1, \dots, \tau_K \sim \mathcal{G} \left(a_0 + \vartheta K, b_0 + \frac{\vartheta}{2} \sum_{k = 1}^{K}\tau_j \right).\label{eq: post_lambda}
\end{equation}

To update $\bm \theta$, irrespective of the prior on $\tilde{\bm \beta}$ adopted, we use a RWMH step.  Due to  the hierarchical nature of the model, the likelihood  $p(\tilde{\bm \beta}|\bm \Sigma, \bm b_0,\bm D_0)$ does not depend on the data and is given by
\begin{equation}
p(\tilde{\bm \beta}|\bm \Sigma, \bm b_0,\bm D_0) =  f_{\mathcal{N}}(\tilde{\bm \beta} | \bm b_0, \bm \Sigma \bm V_{\tilde{\beta}}).\label{eq: post_GG}
\end{equation}
This conditional distribution is then combined with the appropriate Uniform prior and easy to evaluate since $\bm \Sigma \bm V_{\tilde{\beta}}$ is a diagonal matrix. As a proposal distribution for $\bm \theta$, we use a log-Normal distribution:
\begin{equation*}
\log \bm \theta^* = \log \bm \theta^{(a)} + \sigma_\theta \bm \zeta, \quad \bm \zeta \sim \mathcal{N}(\bm 0, \bm I).
\end{equation*}
Hereby, we let $\bm \theta^*$ and $\bm \theta^{(a)}$ denote the proposed and previously accepted value of $\bm \theta$, respectively. Moreover, $\sigma_\theta$ is a scaling factor that is specified such that the acceptance rate of the MH algorithm is between 20 and 40\%. This is achieved by adjusting $ \sigma_\theta$ over the first 25\% of the burn-in stage.


With the clustering prior, the algorithm becomes slightly more complicated and the following steps need to be added. 

The posterior distribution of the mixture probabilities follows a Dirichlet distribution:
\begin{equation}
\bm w|\bm \delta \sim Dir(\pi_1, \dots, \pi_G), \label{eq: post_weights}
\end{equation}
with $\pi_g = \pi + T_g$, where $T_{g}$ denotes the number of $\tilde{\bm \beta}_t$'s assigned to group $g$, and $\bm \delta = (\delta_1, \dots, \delta_T)'$.

It can be shown that the regime indicators $\delta_t~(t=1,\dots, T)$ follow a Multinomial distribution with 
\begin{equation}
\text{Pr}(\delta_t = g|w_g, \bm \mu_g, \sigma_t, \bm \Psi) \propto w_g f_{\mathcal{N}}(\tilde{\bm \beta}_{t}|\bm \mu_g, \sigma_t \bm \Psi), \quad \text{for} \hspace{5pt} g = 1, \dots, G.
\label{eq: post_delta}
\end{equation}
The full conditional posterior of $\bm \mu = \text{vec}(\bm \mu_1, \dots, \bm \mu_G)$ follows a multivariate Gaussian distribution with diagonal variance-covariance matrix:
\begin{equation}
\bm \mu|\bm \Pi, \bm \sigma, \bm \delta, \bm \mu_0  \sim \mathcal{N}(\overline{\bm \mu}, (\bm I_G \otimes \bm \Psi) \odot \overline{\bm V_\mu}), \label{eq: post_mu}
\end{equation}
where the posterior variance and mean are given by, respectively:
\begin{align*}
\overline{\bm V_\mu} &= \left(\bm I_K \otimes \bm Q' \bm Q + \bm I_G \otimes  \bm \Pi^{-1} \right)^{-1},\\
\overline{\bm \mu} &= \overline{\bm V_\mu} \left(\bm I_K \otimes \bm Q' \tilde{\bm \beta}^*+\bm \iota_G \otimes  \bm \Pi^{-1} \bm \mu_0\right).
\end{align*}
We let $\bm Q$ denote a $T \times G$ matrix with $t^{th}$ row given by $\bm Q_t = (I(\delta_t=1)/ \sigma_t, \dots, I\left(\delta_t=G)/ \sigma_t\right)$, $\tilde{\bm \beta}^*$ is simply $\tilde{\bm \beta}$ normalized by dividing through $\sigma_t$, and $\bm \iota_G$ is a $G$-dimensional vector of ones. Notice that it is straightforward to sample from  \autoref{eq: post_mu} because all involved quantities are easily vectorized.

 Similarly, the full conditional posterior distribution of the common mean $\bm \mu_0$ follows a  Gaussian:
\begin{equation}
 \bm \mu_0|\bm \Pi, \bm \mu \sim \mathcal{N}\left(\frac{\sum_{g = 1}^{G} \bm \mu_g}{G},  \frac{1}{G} \bm \Pi \right). \label{eq: post_mu0}
\end{equation} 
 The full conditional posterior distribution of the elements $\upsilon_1, \dots, \upsilon_K$ of the common variance-covariance is defined as a GIG distribution
\begin{equation}
\upsilon_j|\bm R, \bm \mu \sim \text{GIG}\left( c_0 - \frac{G}{2}, 2 c_1, \frac{\sum_{g = 1}^{G}(\mu_{gj} - \mu_{0j})^2}{R_j} \right).\label{eq: post_upsilon}
\end{equation} 
Here, $\mu_{gj}~(g = 1, \dots, G)$ denotes the $j^{th}$ element of group-specific means $\bm \mu_{g}$, $\mu_{0j}$ is the $j^{th}$ element of the common mean $\bm \mu_0$ and $R_j$ corresponding to the $j^{th}$ element of $\bm R$.  

{
Finally, a brief word on how to sample the log-volatilities is in order. Since we assume  dependence between $\tilde{\bm \beta}_t$ and $h_t$ we have to modify the standard algorithm of \cite{kastner2014ancillarity}. This is achieved by integrating out $\tilde{\bm \beta}_t$ analytically. This works irrespective of whether we use a block-diagonal $\bm Z$ and our mixture model or if we assume that the states evolve according to a random walk (i.e., a lower triangular $\bm Z$). In what follows, we will illustrate our approach under the assumption 
that $\tilde{\bm \beta_t}$ arises from a sparse finite mixture model. Plugging \autoref{eq: priorBETAT} into the $t^{th}$ equation of (\ref{static}) and rewriting yields:
\begin{equation*}
    \hat{y}_t - \bm x_t'\bm m_t  =  \sigma_t \underbrace{(\bm x_t' \bm \Psi^{1/2} \bm \nu_t + \eta_t)}_{\hat{\eta}_t}
\end{equation*}
where $\hat{\eta_t} \sim \mathcal{N}(0, \varsigma^2_t)$ with $\varsigma^2_t = \bm x_t' \bm \Psi \bm x_t + 1$. Dividing by $\varsigma_t$ yields:
\begin{equation*}
    \frac{\hat{y}_t - \bm x_t'\bm m_t }{\varsigma_t} = \sigma_t u_t, \quad u_t \sim \mathcal{N}(0, 1),
\end{equation*}
which is the observation equation of a  SV model and standard algorithms such as the one proposed in \cite{kastner2014ancillarity} can be used.}

\subsection{Posterior Computation}\label{sec: posterior}
In this sub-section, we briefly summarize the main steps of the algorithm. Our algorithm cycles between the following steps.  
\begin{enumerate}
\item Draw $\bm \gamma$ from a multivariate Gaussian distribution (see \autoref{eq: post_gamma}). 
\item Draw the local shrinkage parameters $\tau_j~(j=1,\dots, K)$ from a generalized inverse Gaussian (GIG) distribution (see \autoref{eq: post_tau}).
\item Draw the global shrinkage parameter $\psi$ from a Gamma distribution (see \autoref{eq: post_lambda}).
\item Draw $\tilde{\bm \beta}$ from a $TK$-dimensional Gaussian distribution (see \autoref{eq: betatilde}) and the discussion below.
\item Draw the volatilities $\sigma_1, \dots, \sigma_T$ as well as the parameters of the state equation of $h_t$ using a modified version of the algorithm proposed  in \cite{kastner2014ancillarity}.\footnote{This is implemented in the \texttt{R} package \texttt{stochvol}.} The modification is necessary because of the dependence of the prior on $\tilde{\bm \beta}_t$ on $\sigma_t$. 
\item Draw $\bm \theta$ using a random walk Metropolis-Hastings (RWMH) step. 
\item Draw the weights $\bm w$ from a Dirichlet distribution (see \autoref{eq: post_weights}).
\item Draw $\delta_t$ for each $\tilde{\bm \beta}_t$ from a Multinomial distribution (see \autoref{eq: post_delta}).
\item Draw $\bm \mu_0$ from a multivariate Gaussian distribution (see \autoref{eq: post_mu0}).
\item Draw $\upsilon_j~(j=1,\dots, K)$ from a GIG distribution (see \autoref{eq: post_upsilon}).
\end{enumerate}
For the non-clustered approaches, the final four steps are not required.  After specifying appropriate starting values, we repeat the following steps $30, 000$ times and discard the first $10,000$ draws as burn-in. 
\newpage
\section{Data Description}
\renewcommand{\thetable}{B\arabic{table}}
\setcounter{table}{0}
\label{app:data}
\begin{table}[!htbp]
{\scriptsize
\begin{center}
\caption{Data is obtained from the FRED data base of the Federal Reserve of St. Louis \label{tab:data-descr}}
\begin{tabular}{lllr}
\toprule
\multicolumn{1}{l}{\ }&\multicolumn{1}{c}{\ FRED.Mnemonic}&\multicolumn{1}{c}{\ Description}&\multicolumn{1}{c}{\ Trans I(0)}\tabularnewline
\midrule
&\textbf{CPIAUCSL}& \textbf{Consumer Price Index for All Urban Consumers:  All Items}&$\bm 5$\tabularnewline
&GDPCTPI&Gross Domestic Product: Chain-type Price Index&$5$\tabularnewline
&PCECTPI&Personal Consumption Expenditures: Chain-type Price Index &$5$\tabularnewline
&GDPC1&Real Gross Domestic Product&$5$\tabularnewline
&PCECC96&Real Personal Consumption Expenditures&$5$\tabularnewline
&FPIx&Real private fixed investment &$5$\tabularnewline
&INDPRO&IP:Total index Industrial Production Index (Index 2012=100)&$5$\tabularnewline
&CUMFNS&Capacity Utilization:  Manufacturing (SIC) (Percent of Capacity)&$1$\tabularnewline
&PAYEMS& Emp:Nonfarm All Employees: Total nonfarm (Thousands of Persons)&$5$\tabularnewline
&CE16OV&Civilian Employment (Thousands of Persons)&$5$\tabularnewline
&UNRATE&Civilian Unemployment Rate (Percent)&$1$\tabularnewline
&UNRATESTx&Unemployment Rate less than 27 weeks (Percent)&$1$\tabularnewline
&UNRATELTx&Unemployment Rate for more than 27 weeks (Percent)&$1$\tabularnewline
&LNS14000012&Unemployment Rate - 16 to 19 years (Percent)&$1$\tabularnewline
&LNS14000025&Unemployment Rate - 20 years and over, Men (Percent)&$1$\tabularnewline
&LNS14000026&Unemployment Rate - 20 years and over, Women (Percent)&$1$\tabularnewline
&UEMPLT5&Number of Civilians Unemployed - Less Than 5 Weeks (Thousands of Persons)&$5$\tabularnewline
&UEMP5TO14&Number of Civilians Unemployed for 5 to 14 Weeks (Thousands of Persons)&$5$\tabularnewline
&UEMP15T26&Number of Civilians Unemployed for 15 to 26 Weeks (Thousands of Persons)&$5$\tabularnewline
&UEMP27OV&Number of Civilians Unemployed for 27 Weeks and Over (Thousands of Persons)&$5$\tabularnewline
&AWHMAN&Average Weekly Hours of Production and Nonsupervisory Employees:  Manufacturing&$1$\tabularnewline
&CES0600000007&Average Weekly Hours of Production and Nonsupervisory Employees:  Goods-Producing&$1$\tabularnewline
&HOUST&Housing Starts: Total: New Privately Owned Housing Units Started&$5$\tabularnewline
&PERMIT&New Private Housing Units Authorized by Building Permits&$5$\tabularnewline
&IPDBS&Business Sector:  Implicit Price Deflator (Index 2012=100)&$5$\tabularnewline
&CPILFESL&Consumer Price Index for All Urban Consumers:  All Items Less Food \& Energy&$5$\tabularnewline
&WPSFD49207&Producer Price Index by Commodity for Finished Goods &$5$\tabularnewline
&PPIACO&Producer Price Index for All Commodities &$5$\tabularnewline
&WPSFD49502&Producer Price Index by Commodity for Finished Consumer Goods &$5$\tabularnewline
&WPSFD4111&Producer Price Index by Commodity for Finished Consumer Foods&$5$\tabularnewline
&PPIIDC&Producer Price Index by Commodity Industrial Commodities &$5$\tabularnewline
&WPSID61&Producer Price Index by Commodity Intermediate Materials:  Supplies \& Components&$5$\tabularnewline
&WPU0561&Producer Price Index by Commodity for Fuels and Related Products and Power&$5$\tabularnewline
&OILPRICEx&Real Crude Oil Prices:  West Texas Intermediate (WTI) - Cushing, Oklahoma&$5$\tabularnewline
&WPSID62&Producer Price Index:  Crude Materials for Further Processing &$5$\tabularnewline
&PPICMM&Producer Price Index:  Commodities:  Metals and metal products:  Primary nonferrous metals&$5$\tabularnewline
&CPIAPPSL&Consumer Price Index for All Urban Consumers:  Apparel&$5$\tabularnewline
&CPITRNSL&Consumer Price Index for All Urban Consumers:  Transportation&$5$\tabularnewline
&CPIMEDSL&Consumer Price Index for All Urban Consumers:  Medical Care&$5$\tabularnewline
&CES2000000008x&Real Average Hourly Earnings of Production and Nonsupervisory Employees: Construction&$5$\tabularnewline
&CES3000000008x&Real Average Hourly Earnings of Production and Nonsupervisory Employees: Manufacturing&$5$\tabularnewline
&COMPRNFB&Nonfarm Business Sector:  Real Compensation Per Hour (Index 2012=100)&$5$\tabularnewline
&CES0600000008&Average Hourly Earnings of Production and Nonsupervisory Employees:&$5$\tabularnewline
&FEDFUNDS&Effective Federal Funds Rate (Percent)&$1$\tabularnewline
&TB3MS&3-Month Treasury Bill: Secondary Market Rate (Percent)&$1$\tabularnewline
&GS10&10-Year Treasury Constant Maturity Rate (Percent)&$1$\tabularnewline
&GS10TB3Mx&10-Year Treasury Constant Maturity Minus 3-Month Treasury Bill, secondary market&$1$\tabularnewline
&M1REAL& Real M1 Money Stock&$5$\tabularnewline
&S.P.500&S\&P Common Stock Price Index:  Composite&$5$\tabularnewline
&S.P..indust&S\&P Common Stock Price Index:  Industrials&$5$\tabularnewline
\bottomrule
\end{tabular}
\end{center}}
\begin{minipage}[c]{\linewidth}
\footnotesize{\textbf{Notes}: 
 The column \texttt{Trans I(0)} denotes the transformation applied to each variable so as to make it stationary. The transformation codes are taken from \cite{mccracken2016fred} with $(1)$ implying no transformation applied and $(5)$ denoting growth rates, defined as log first differences $\log\left(\frac{x_t}{x_{t-1}
}\right)$. All variables are standardized by substracting the mean and dividing by the standard deviation.}
\end{minipage}
\end{table}

\newpage
\section{Empirical Appendix}\label{sec: empAppendix} 
\renewcommand{\thetable}{C\arabic{table}}
\renewcommand{\thefigure}{C\arabic{figure}}
\setcounter{table}{0}
\setcounter{figure}{0}

\subsection{MCMC Mixing and Convergence Properties}
In this sub-section we briefly analyze the mixing properties of our MCMC algorithm for the best performing TVP-SVD model in Table \ref{tab:LPS}, which features the full information set, a pooling prior and $\kappa = 0.05$. To this end, we report  inefficiency factors (IFs) and   \cite{raftery1992many} diagnostics for the TVPs and the error volatilities. We focus on these two quantities because these are the ones which we use to set up predictive distributions and, moreover, these are the ones we are interested in if we focus on functions of the parameters such as the multipliers reported in \autoref{fig:longNKPC}. 

\begin{table}[!htbp]
{\tiny
\begin{center}
\caption{Summary of MCMC diagnostics of posterior estimates \label{tab:mcmcdiag}}
\begin{tabular}{lcllllll}
\toprule
\multicolumn{1}{c}{}&\multicolumn{1}{c}{\bfseries}&\multicolumn{6}{c}{\bfseries Summary Statistics}\tabularnewline
\cmidrule{3-8}
\multicolumn{2}{l}{}&\multicolumn{1}{c}{Mean}&\multicolumn{1}{c}{Median}&\multicolumn{1}{c}{Min}&\multicolumn{1}{c}{Max}&\multicolumn{1}{c}{$5^{th}$ Perc.}&\multicolumn{1}{c}{$95^{th}$ Perc.}\tabularnewline
\midrule
Inefficiency factors (IF) &&&&&&& \tabularnewline
&&&&&&& \tabularnewline
$\{\bm \beta_{t}\}_{t=1}^{T}$ && 7.01  &  4.80 &  1.32 & 57.97 &  2.09 & 20.23  \tabularnewline
&&&&&&& \tabularnewline
$\{\sigma^2_{t}\}_{t=1}^{T}$ && 8.31 &  8.09 &  4.32 & 12.54 &  5.94 & 10.88   \tabularnewline
\midrule
\cite{raftery1992many}'s diagnostics &&&&&&& \tabularnewline
&&&&&&& \tabularnewline
$\{\bm \beta_{t}\}_{t=1}^{T}$ && 355 & 350 &  150 & 1380 & 164 & 726\tabularnewline
&&&&&&& \tabularnewline
$\{\sigma^2_{t}\}_{t=1}^{T}$ && 525  & 502 & 221  & 968  &338  &756  \tabularnewline
\bottomrule
\end{tabular}
\end{center}}
\begin{minipage}[c]{\linewidth}
\footnotesize{\textbf{Notes}: The table shows the inefficiency factors, specified as the inverse of the relative effective sample size, and the \cite{raftery1992many}'s diagnostics of the number of runs to obtain the $2.5^{th}$ percentile with $95\%$ probability and $2.5\%$ accuracy.}
\end{minipage}
\end{table}

The upper part of Table \ref{tab:mcmcdiag} shows the IFs. Since our state space is enormous, we report  summary statistics across coefficients and time. In principle, inefficiency factors below $20$ are considered acceptable \citep{primiceri2005}.  For both the TVPs and the error variances we find IFs which are on average considerably below $10$. For the volatilities, the maximum of IFs is $12.54$. For the TVPs, we find a maximum inefficiency factor  of around $58$. Notice, however, that the $95^{th}$ percentile is close to $20$. Thus, according to the IFs, we find that our MCMC algorithm mixes rapidly and yields draws which display a relatively modest amount of autocorrelation across successive MCMC draws.

The bottom part of Table \ref{tab:mcmcdiag} shows the \cite{raftery1992many} diagnostics. This metric measures the number of iterations of the MCMC algorithm necessary to achieve a certain level of precision. These numbers are well below the total number of iterations (with a maximum of 1380 for $\bm \beta_t$ and $968$ for $\sigma^2_t$). This suggests that our algorithm performs well empirically  according to both diagnostics adopted.


\subsection{Prior Robustness}
Our priors are hierarchical and involve few prior hyperparameters which must be selected by the researcher. For most of these, we can draw upon existing papers such as \cite{malsiner2016} to provide suggestions for sensible benchmark values. However, the choice of the prior hyperparameters $G$ and $\kappa$ defined in \autoref{kappa} do not fall into this category and, hence, it is worth offering some additional discussion of prior robustness relating to $G$ and $\kappa$ in this appendix. Strictly speaking, our mixture model implies that $G$ should be set to a very large value, translating into an overfitting mixture model. But the key question is how forecasting performance changes when $G$ is set too small or whether a too large value could negatively impact forecast accuracy due to potential overfitting. The second hyperparameter, $\kappa$, determines the upper bound of the grid for the parameters which enter the prior covariance matrix.

We start our analysis by considering how $\kappa$ affects forecasting accuracy by repeating  our forecasting exercise using the SVD approach combined with our set of  priors and three different model sizes, but allow a range of values for $\kappa$. Results are given in \autoref{tab:LPS1}. Note first that for $\kappa=1$, which is the largest value we considered, no results are given for the FULL or NKPC models. For these cases, $\kappa=1$ led to severe over-fitting problems and resulting poor forecast performance (e.g., LPLs of minus infinity). Clearly this value does not induce adequate shrinkage in larger models and care must be taken to avoid such regions of the hyperparameter space. {It is worth noting that for the TVP-RW-SVD specification values of $\kappa$ exceeding $0.05$ are already too large. This is because $\kappa$ effectively determines the upper bound of the size of a single shock within a given point in time and hence, larger values imply excessive amounts of time-variation and elevated risks of overfitting.}

However, provided $\kappa$ is kept small, we find a high degree of prior robustness. If we consider point forecast performance, we find that as long as the upper bound is specified between $0.01$ and $0.1$, point forecasts are only slightly affected by the choice of $\kappa$. No clear patterns emerge. For one-step-ahead forecasts using the FULL model estimated using the g-prior, relative RMSEs seem to be inversely related to $\kappa$. But this does not carry over to the four-step-ahead horizon. For longer-run forecasts, we observe that accuracy is largest if $\kappa$ is set equal to $0.05$. For the NKPC model, a similar U-shaped pattern arises, indicating that the optimal value of $\kappa$ should be between $0.005$ and $0.05$. The only model that strongly profits from using a larger scaling paramater is the UCM. Here, we find that the best forecasting performance is obtained if $\kappa =1$, a choice that seriously distorts predictive accuracy for larger models. In the case of the Minnesota prior, the specific choice of $\kappa$ plays a limited role, with point forecasts of the full model being indistinguishable from each other for the one-quarter-ahead horizon and quite similar for the one-year-ahead horizon. 

When the full predictive distribution is considered, we also find only limited differences in predictive accuracy for varying values of $\kappa$ for models which feature a block-diagonal $\bm Z$. In this case, and for the large-scale model, differences are typically quite small. This suggests that the precise value of $\kappa$, as long as it is not specified too large, plays only a minor role in impacting forecasting accuracy. Notice, however, that this does not carry over to smaller models and models which feature a lower triangular $\bm Z$. In these cases, we find that using a smaller $\kappa$ improves LPLs for both priors and forecast horizons. 

When we consider the performance of the UCM models based on the full predictive density, we find that forecast accuracy appreciably increases with $\kappa$. This contrasts with the findings for the other models. This is due to the fact that with small values of $\kappa$, it becomes increasingly difficult to control for unobserved heterogeneity and the resulting model approaches a white noise specification. 

\begin{figure}[!ht]
\centering
\caption{Evolution of one-step-ahead log predictive Bayes factors relative to RW
\label{fig::svd_forc1}}
\begin{minipage}[t]{\textwidth}
\centering 
(a) with clustering 
\vspace{5pt}
\end{minipage}
\begin{minipage}[t]{0.49\textwidth}
\centering
\includegraphics[width=1\textwidth]{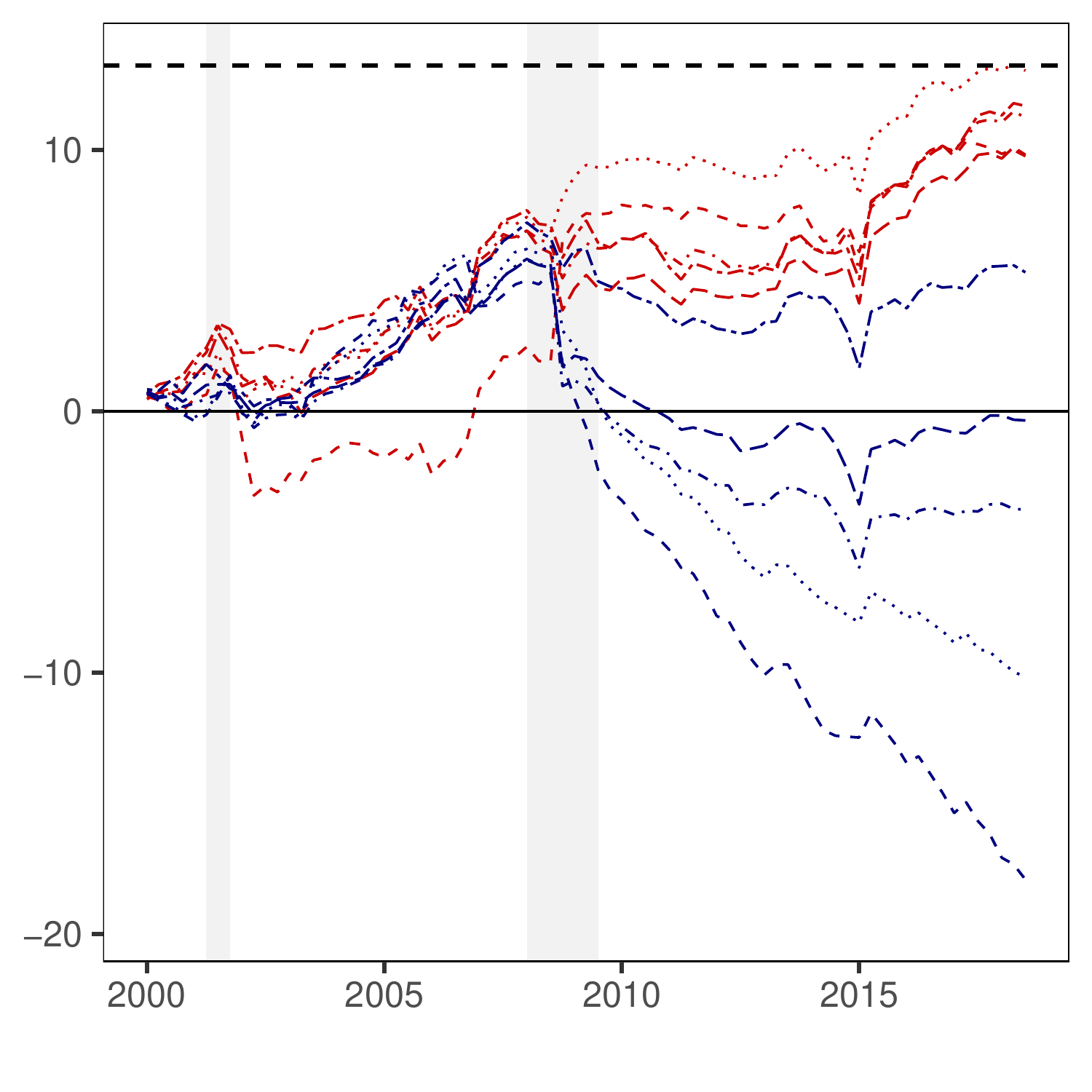}
\end{minipage}
\begin{minipage}[t]{0.49\textwidth}
\centering
\includegraphics[width=1\textwidth]{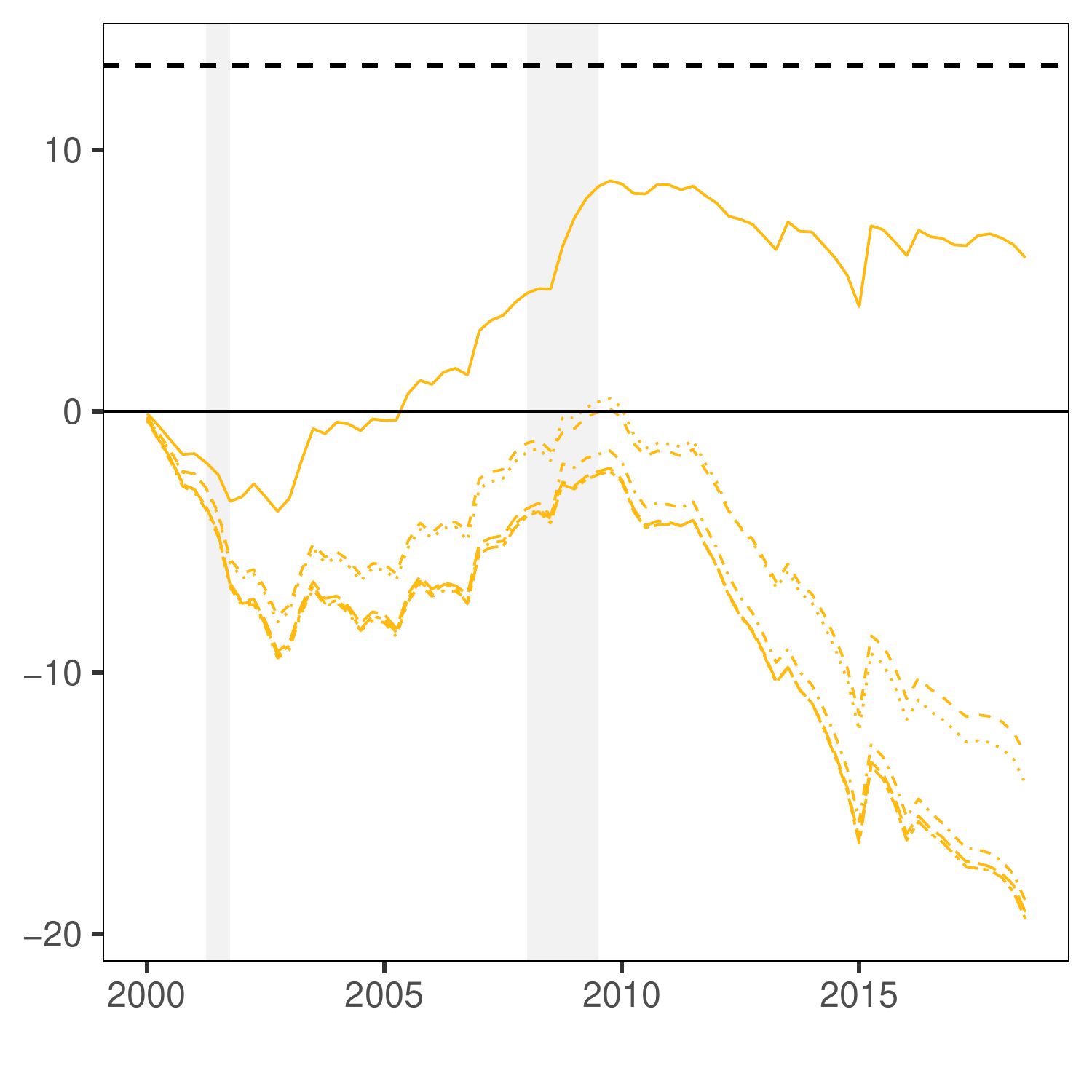}
\end{minipage}
\begin{minipage}[t]{\textwidth}
\vspace{5pt}
\centering 
(b) without clustering 
\vspace{5pt}
\end{minipage}
\begin{minipage}[t]{0.49\textwidth}
\centering 
\textit{g-prior} 
\end{minipage}
\begin{minipage}[t]{0.49\textwidth}
\centering 
\textit{Minnesota} 
\end{minipage}
\begin{minipage}[t]{0.49\textwidth}
\centering
\includegraphics[width=1\textwidth]{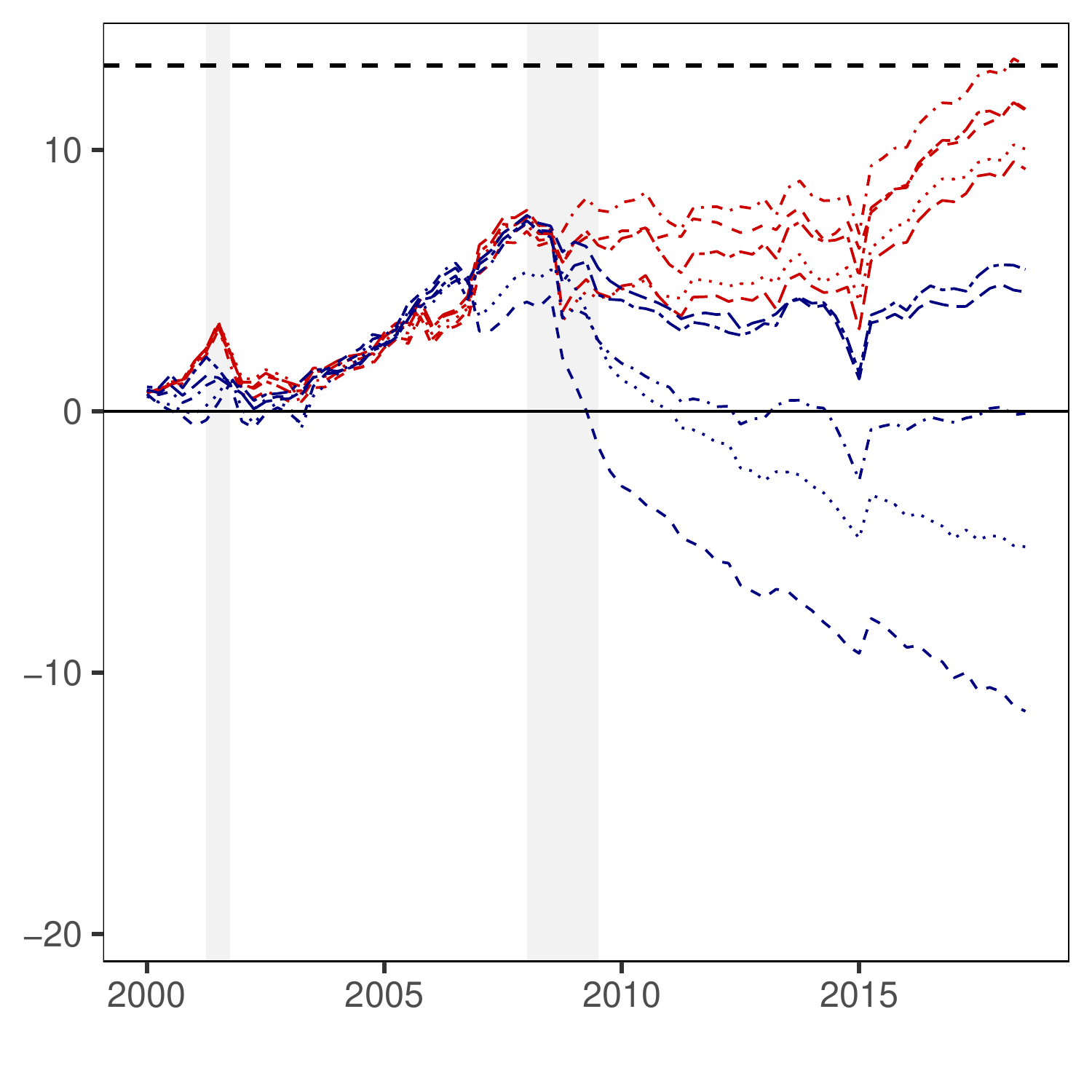}
\end{minipage}
\begin{minipage}[t]{0.49\textwidth}
\centering
\includegraphics[width=1\textwidth]{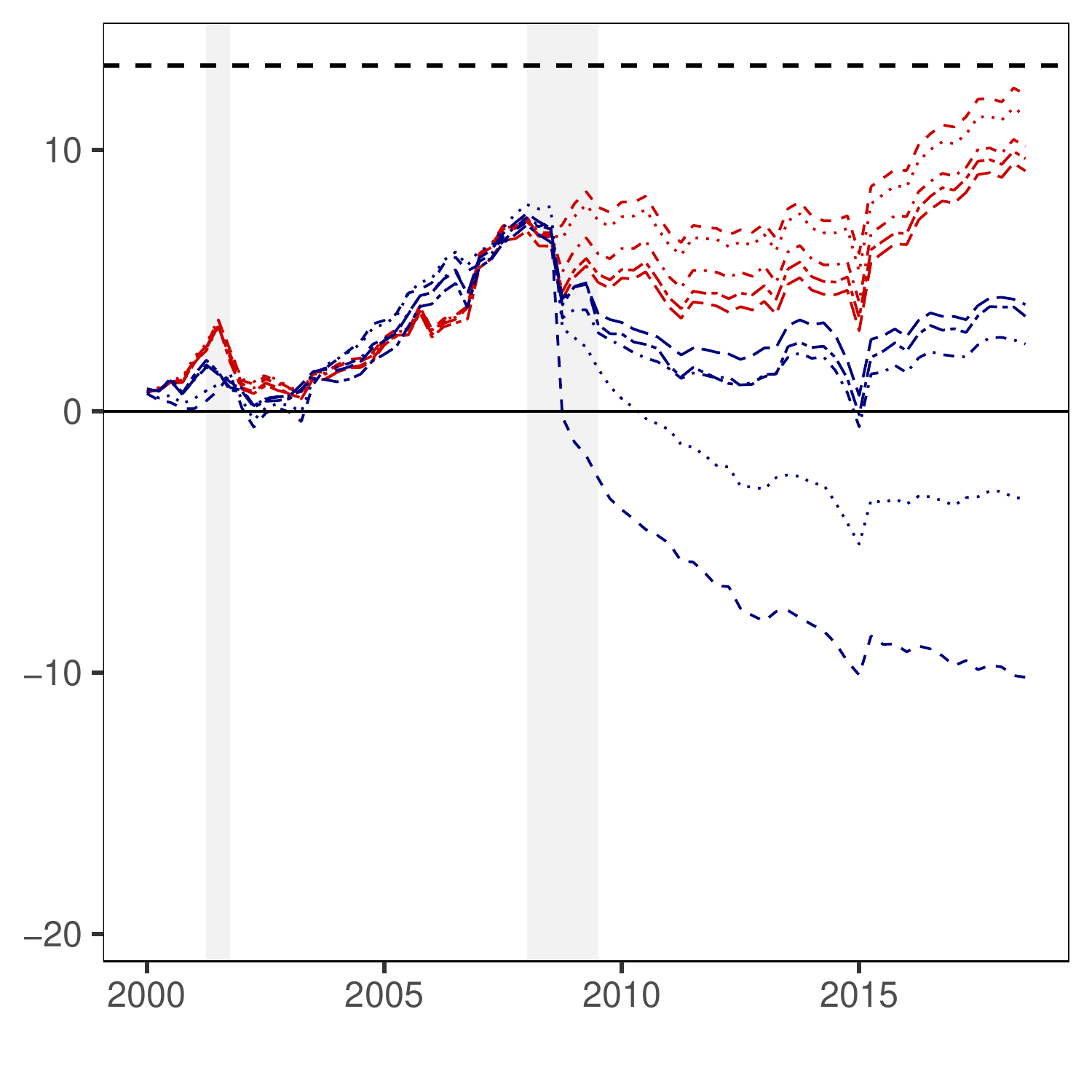}
\end{minipage}
\begin{minipage}[t]{1\textwidth}
\centering
\includegraphics[width=0.5\textwidth]{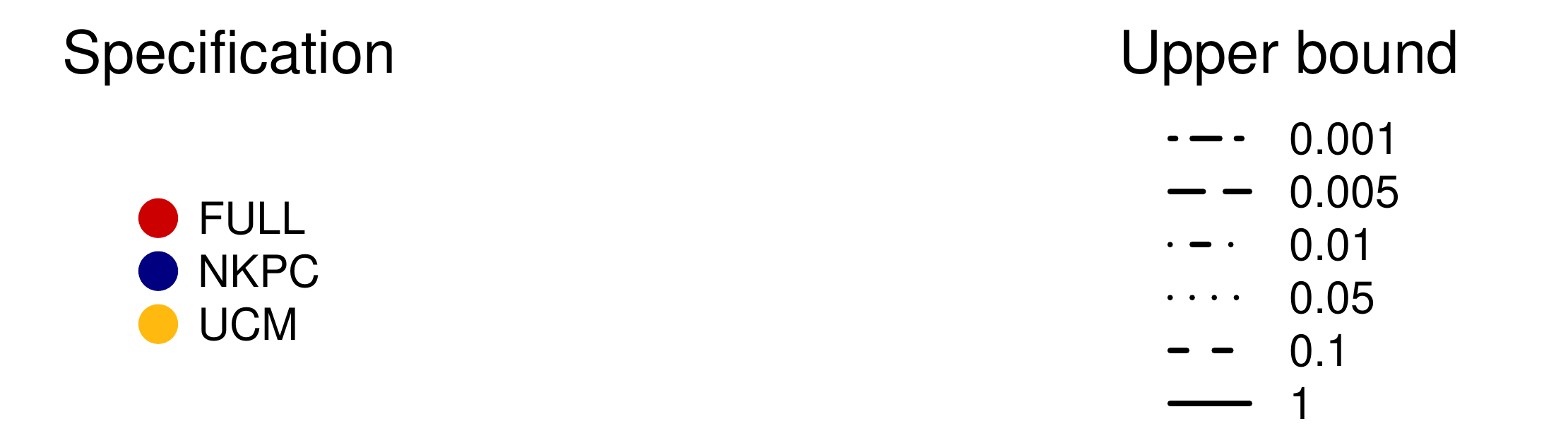}
\end{minipage}
\begin{minipage}{\linewidth}
\vspace{5pt}
\footnotesize \textbf{Notes}: The log predictive Bayes factors are cumulated over the hold-out sample. The blue lines refer to the maximum Bayes factor at the end of the hold-out sample. The gray shaded areas indicate the NBER recessions in the US.
\end{minipage}
\end{figure}

\begin{figure}[!ht]
\centering
\caption{Evolution of four-step-ahead log predictive Bayes factors relative to RW for different values of $\kappa$
\label{fig::svd_forc4}}
\begin{minipage}[t]{\textwidth}
\centering 
(a) with clustering 
\vspace{5pt}
\end{minipage}
\begin{minipage}[t]{0.49\textwidth}
\centering
\includegraphics[width=1\textwidth]{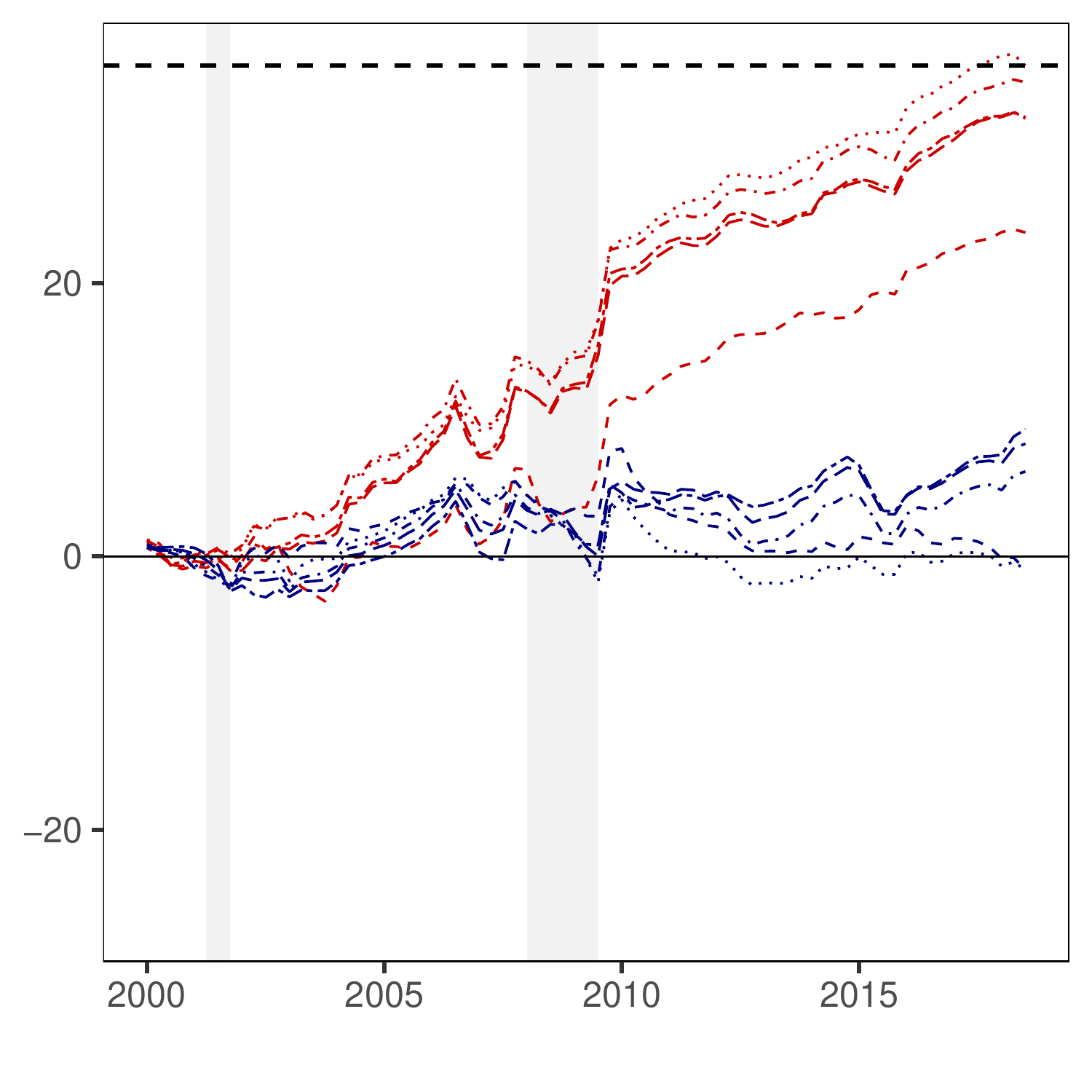}
\end{minipage}
\begin{minipage}[t]{0.49\textwidth}
\centering
\includegraphics[width=1\textwidth]{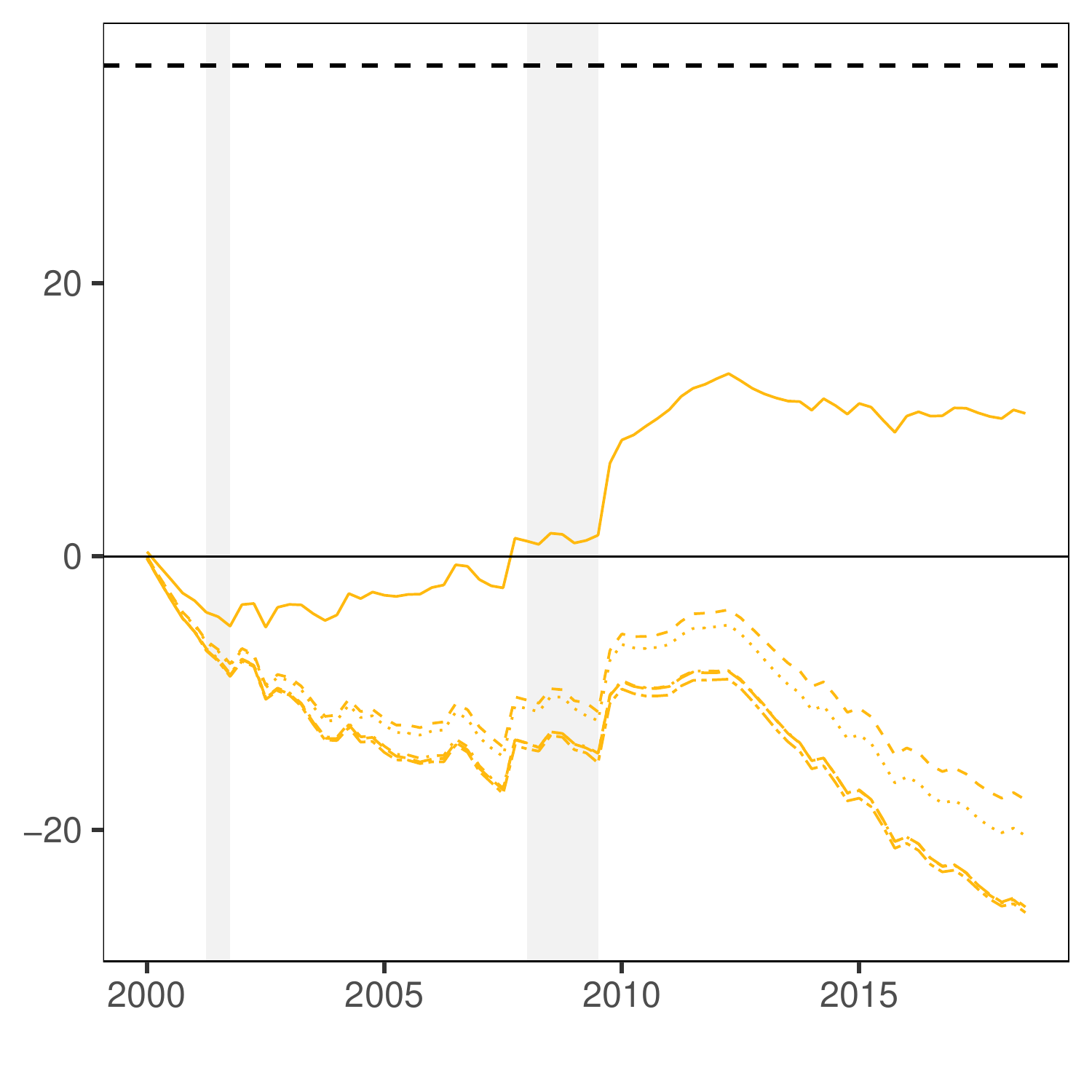}
\end{minipage}
\begin{minipage}[t]{\textwidth}
\vspace{5pt}
\centering 
(b) without clustering
\vspace{5pt}
\end{minipage}
\begin{minipage}[t]{0.49\textwidth}
\centering 
\textit{g-prior} 
\end{minipage}
\begin{minipage}[t]{0.49\textwidth}
\centering 
\textit{Minnesota} 
\end{minipage}
\begin{minipage}[t]{0.49\textwidth}
\centering
\includegraphics[width=1\textwidth]{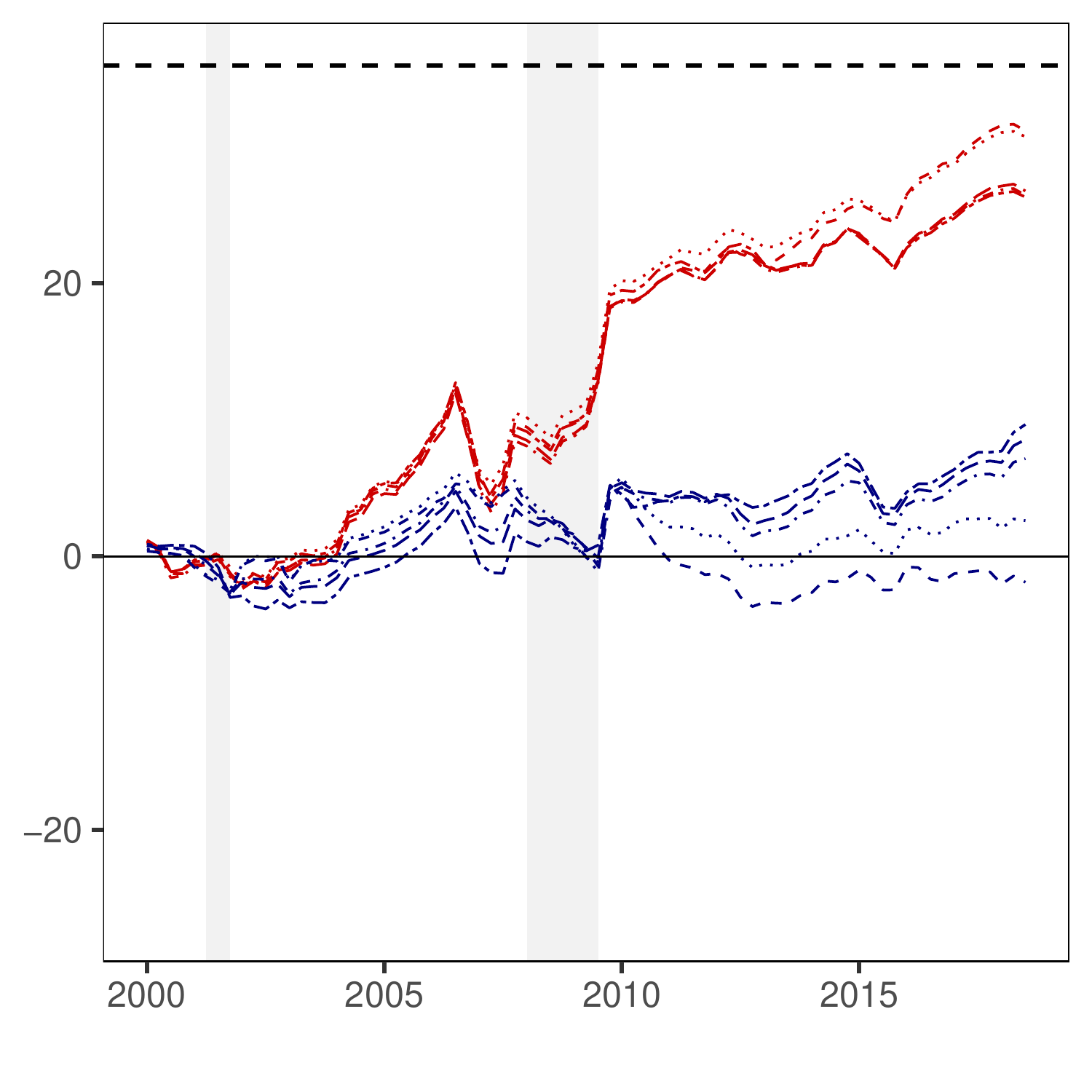}
\end{minipage}
\begin{minipage}[t]{0.49\textwidth}
\centering
\includegraphics[width=1\textwidth]{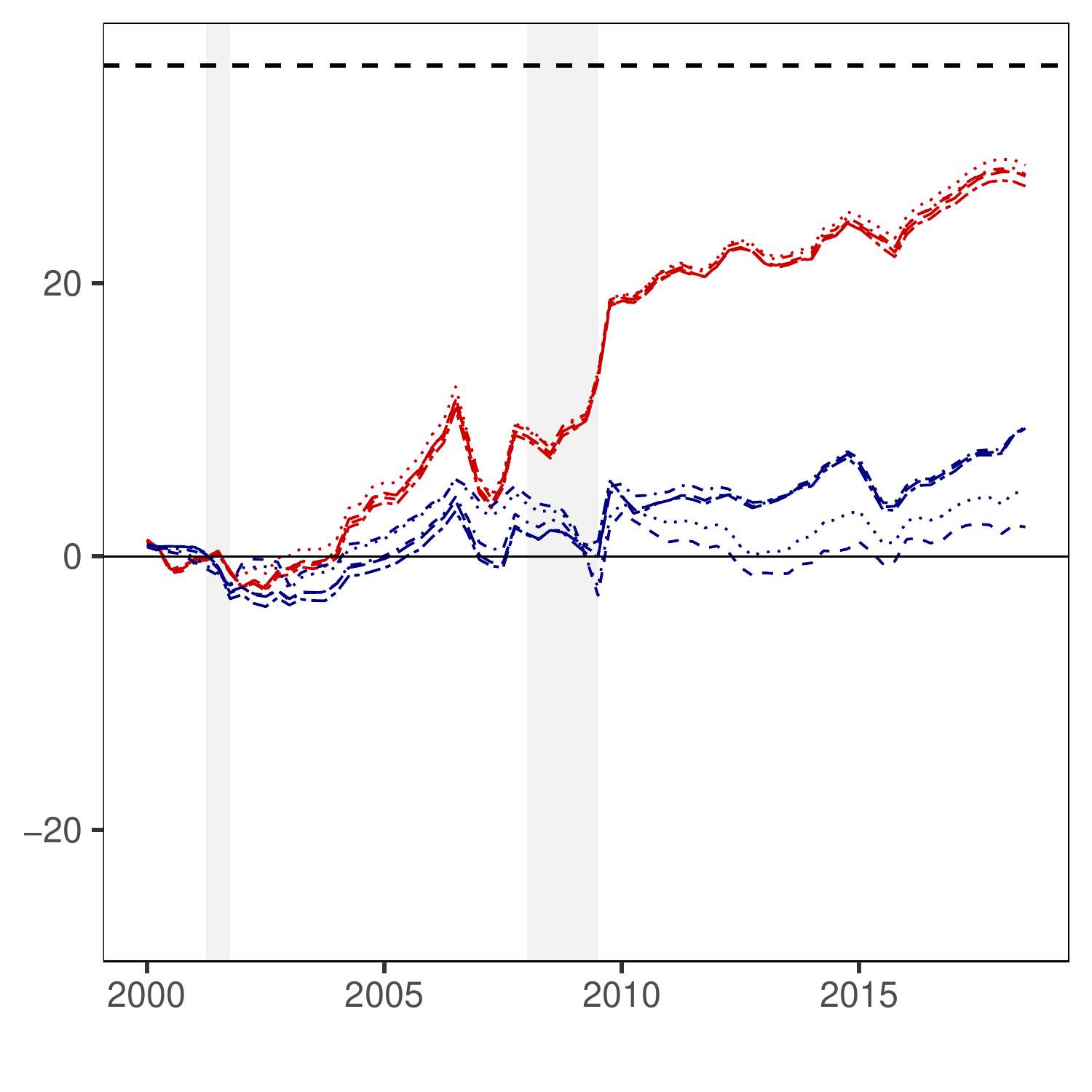}
\end{minipage}
\begin{minipage}[t]{1\textwidth}
\centering
\includegraphics[width=0.5\textwidth]{Figures/Forecasting_US/legend_svd.pdf}
\end{minipage}
\begin{minipage}{\linewidth}
\vspace{5pt}
\footnotesize \textbf{Notes}: The log predictive Bayes factors are cumulated over the hold-out sample. The dashed black lines refer to the maximum Bayes factor at the end of the hold-out sample. The gray shaded areas indicate the NBER recessions in the US.
\end{minipage}
\end{figure}

Figures \ref{fig::svd_forc1} and \ref{fig::svd_forc4} plot the cumulative Bayes factors comparing a variety of approaches and values for $\kappa$ (which is labeled "upper bound" in the figures) against the random walk for $h=1$ and $h=4$, respectively. These reinforce the preceding discussion. For the FULL model, containing all the regressors, lines for different values of $\kappa$ are plotted in red on the figures. Note that all the red lines are similar to one another. Note too, the increasingly good performance of the FULL models after the financial crisis for $h=4$. For the smaller data sets and, in particular, the UCM model there is much more sensitivity to the choice of $\kappa$. For the UCM model, it is interesting to note that the deterioration in forecast performance associated with poor choices of $\kappa$ occurs largely after the financial crisis. For the NKPC specifications, when $h=1$ some choices of $\kappa$ lead to poor forecast performance. But for $h=4$, with the NKPC specification, we are finding much more robustness. 

To sum up this discussion, we find that the precise value of $\kappa$ plays only a limited role  for forecasting accuracy when the large model is adopted, provided we avoid large values of $\kappa$ which clearly lead to over-fitting problems. In contrast, in smaller models, the precise value of $\kappa$ has a bigger impact on  forecasting accuracy and the researcher needs to carefully select this hyperparameter.

\begin{table}[!tbp]
{\tiny
\begin{center}
\caption{Forecasting Performance for Different Values of $\kappa$ \label{tab:LPS1}}
\begin{tabular}{lllcllllll}
\toprule
\multicolumn{1}{l}{\bfseries }&\multicolumn{2}{c}{\bfseries Specification}&\multicolumn{1}{c}{\bfseries }&\multicolumn{6}{c}{\bfseries $\kappa$}\tabularnewline
\cmidrule{2-3} \cmidrule{5-10}
\multicolumn{1}{l}{}&\multicolumn{1}{c}{Horizon}&\multicolumn{1}{c}{Information set}&\multicolumn{1}{c}{}&\multicolumn{1}{c}{0.001}&\multicolumn{1}{c}{0.005}&\multicolumn{1}{c}{0.01}&\multicolumn{1}{c}{0.05}&\multicolumn{1}{c}{0.1}&\multicolumn{1}{c}{1}\tabularnewline
\midrule
\multicolumn{10}{c}{\bfseries TVP-WN-SVD}\tabularnewline
{\scshape g-prior (no clustering)}&&&&&&&&&\tabularnewline
~~&One-step-ahead&FULL&&0.81**&0.82**&0.80**&0.79***&0.80***&\tabularnewline
~~&&&&(0.15*)&(0.12)&(0.18**)&(0.13)&(0.15**)&\tabularnewline
~~&&NKPC&&0.89&0.88&0.89&0.94&1.02&\tabularnewline
~~&&&&(0.07)&(0.06)&(0.00)&(-0.07)&(-0.15**)&\tabularnewline
~~&Four-step-ahead&FULL&&0.60**&0.59**&0.61**&0.58**&0.59**&\tabularnewline
~~&&&&(0.35)&(0.36)&(0.35)&(0.41*)&(0.42*)&\tabularnewline
~~&&NKPC&&0.79***&0.78*&0.76*&0.85&0.95&\tabularnewline
~~&&&&(0.13)&(0.11)&(0.10)&(0.03)&(-0.03)&\tabularnewline
\midrule
{\scshape g-prior (clustering)}&&&&&&&&&\tabularnewline
~~&One-step-ahead&FULL&&0.82**&0.81**&0.80***&0.80***&0.76**&\tabularnewline
~~&&&&(0.16*)&(0.13)&(0.15*)&(0.17**)&(0.13)&\tabularnewline
~~&&NKPC&&0.90&0.88*&0.88*&0.96&1.31&\tabularnewline
~~&&&&(0.07)&(0.00)&(-0.05)&(-0.14*)&(-0.24***)&\tabularnewline
~~&&UCM &&1.10**&1.10**&1.11**&1.05&1.01&0.88*\tabularnewline
~~&&&&(-0.26***)&(-0.26***)&(-0.25**)&(-0.19**)&(-0.18*)&(0.08)\tabularnewline
~~&Four-step-ahead&FULL&&0.60***&0.60***&0.59***&0.57***&0.69**&\tabularnewline
~~&&&&(0.43***)&(0.43***)&(0.46***)&(0.48***)&(0.32***)&\tabularnewline
~~&&NKPC&&0.80***&0.77*&0.76&0.85&1.00&\tabularnewline
~~&&&&(0.12)&(0.11)&(0.08)&(-0.01)&(-0.01)&\tabularnewline
~~&&UCM &&1.07*&1.06*&1.06*&1.00&0.96&0.71\tabularnewline
~~&&&&(-0.35**)&(-0.34**)&(-0.34**)&(-0.27*)&(-0.24)&(0.14)\tabularnewline
\midrule
{\scshape Minnesota}&&&&&&&&&\tabularnewline
~~&One-step-ahead&FULL&&0.82**&0.82**&0.82**&0.82*&0.82**&\tabularnewline
~~&&&&(0.13)&(0.12)&(0.14)&(0.15)&(0.16*)&\tabularnewline
~~&&NKPC&&0.91&0.88&0.88*&0.95&1.08&\tabularnewline
~~&&&&(0.05)&(0.05)&(0.03)&(-0.04)&(-0.14)&\tabularnewline
~~&Four-step-ahead&FULL&&0.61**&0.61**&0.62**&0.61**&0.61**&\tabularnewline
~~&&&&(0.36)&(0.37)&(0.37*)&(0.38*)&(0.37*)&\tabularnewline
~~&&NKPC&&0.81***&0.78***&0.78**&0.75&0.77&\tabularnewline
~~&&&&(0.13)&(0.12)&(0.12)&(0.06)&(0.03)&\tabularnewline
\midrule
\multicolumn{10}{c}{\bfseries TVP-RW-SVD}\tabularnewline
{\scshape ridge-prior}&&&&&&&&&\tabularnewline
~~&One-step-ahead&FULL&& 0.81** &  0.82* &  0.82* &  0.87 & \tabularnewline
~~&&&&(0.14) & (0.10) & (0.04) & (-0.04)&&\tabularnewline
~~&Four-step-ahead&FULL&&  0.62** &  0.61* &  0.62* &  0.64 &\tabularnewline
~~&&&&(0.43***) & (0.43***) & (0.42***) & (0.35**)&\tabularnewline
\bottomrule
\end{tabular}
\end{center}}
\begin{minipage}[c]{\linewidth}
\footnotesize{\textbf{Notes}: The table shows RMSEs with LPL's in parentheses below. Asterisks indicate statistical significance for each model relative to a random walk at the $1$ ($^{***}$), $5$ ($^{**}$) and $10$ ($^{*}$) percent significance levels. }
\end{minipage}
\end{table}

Next, we consider how $G$ impacts predictive accuracy. We focus on the FULL models and set $\kappa=0.05$. Table \ref{tab:LPS-G} shows the forecast results for $G \in \{5, 10, 15, 30, 50\}$. When we focus on point forecasts we see little differences across results. In all cases, RMSEs are very close to each other. This indicates that even if we set $G$ to a too small value, the model produces (almost) identical point forecasts to the case that $G=30$. When we focus on LPLs we find interesting differences. In the case that $G=5$, our model performs poorly and is outperformed by the random walk benchmark. Increasing $G$ from $5$ to $10$ yields much better density forecasts. It is worth noting that further increases in $G$ improve predictive performance slightly but these improvements seem to vanish once $G$ reaches values over $15$.

\begin{figure}[!ht]
\centering
\caption{Evolution of log predictive Bayes factors relative to RW for different values of G
\label{fig::G_forc}}
\begin{minipage}[t]{0.49\textwidth}
\centering 
(a) One-step-ahead
\vspace{5pt}
\end{minipage}
\begin{minipage}[t]{0.49\textwidth}
\centering 
(b) Four-step-ahead
\vspace{5pt}
\end{minipage}
\begin{minipage}[t]{0.49\textwidth}
\centering
\includegraphics[width=1\textwidth]{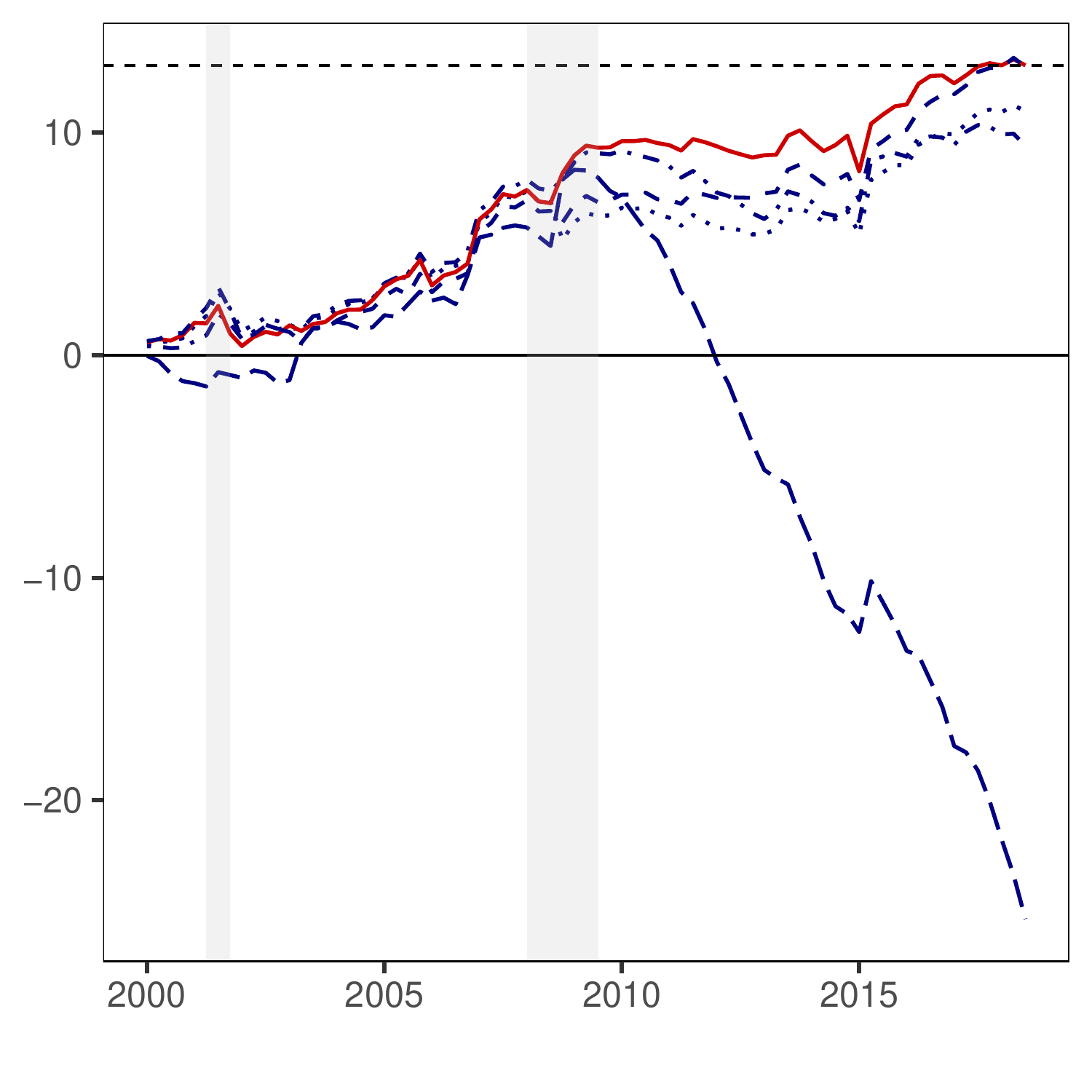}
\end{minipage}
\begin{minipage}[t]{0.49\textwidth}
\centering
\includegraphics[width=1\textwidth]{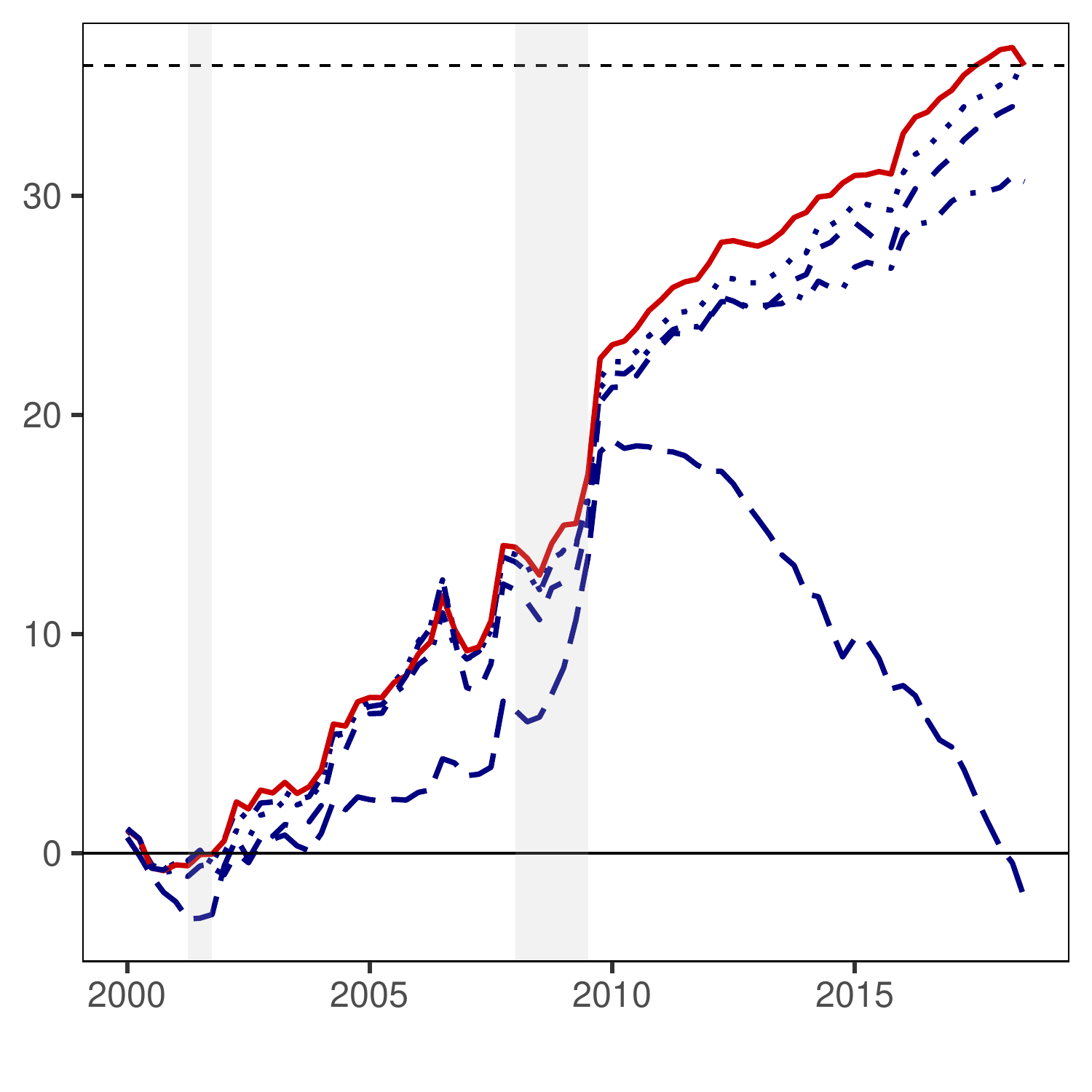}
\end{minipage}
\begin{minipage}[t]{1\textwidth}
\centering
\includegraphics[width=0.35\textwidth]{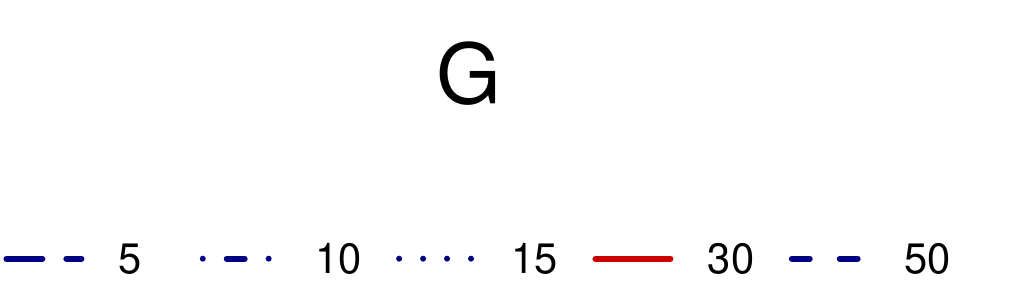}
\end{minipage}
\begin{minipage}{\linewidth}
\vspace{5pt}
\footnotesize \textbf{Notes}: The log predictive Bayes factors are cumulated over the hold-out sample. We focus on the best performing TVP-WN-SVD model with a pooling prior (i.e., full information set (FULL) and $\kappa = 0.05$). The dashed black lines refer to the maximum Bayes factor at the end of the hold-out sample. The gray shaded areas indicate the NBER recessions in the US.
\end{minipage}
\end{figure}

This discussion suggests that while the specific choice of $G$ has no strong impact on point forecasts, larger values of $G$ seem to yield more appropriate predictive intervals which might help during recessionary periods, as evidenced in \autoref{fig::G_forc}. In this figure we observe that for $G=5$, predictive performance was very close to the ones obtained from using larger values of $G$ up to the global financial crisis. After the recession, forecast performance deteriorates in the $G=5$ case. These decreases are mainly driven by predictive intervals which are too narrow.

\begin{table}[!tbp]
{\tiny
\begin{center}
\caption{Forecasting Performance for Different Values of $G$ \label{tab:LPS-G}}
\begin{tabular}{llclllll}
\toprule
\multicolumn{1}{l}{\bfseries }&\multicolumn{1}{c}{\bfseries Horizon}&\multicolumn{1}{c}{\bfseries }&\multicolumn{5}{c}{\bfseries Maximum number of groups $G$}\tabularnewline
\cmidrule{2-2} \cmidrule{4-8}
\multicolumn{1}{l}{}&\multicolumn{1}{c}{}&\multicolumn{1}{c}{}&\multicolumn{1}{c}{5}&\multicolumn{1}{c}{10}&\multicolumn{1}{c}{15}&\multicolumn{1}{c}{30}&\multicolumn{1}{c}{50}\tabularnewline
\midrule
{\scshape }&&&&&&&\tabularnewline
~~&One-step-ahead&&0.79***&0.78***&0.78***&0.80***&0.80***\tabularnewline
~~&&&(-0.34***)&(0.13**)&(0.15**)&(0.17**)&(0.17**)\tabularnewline
~~&Four-step-ahead&&0.58***&0.58***&0.58***&0.57***&0.59***\tabularnewline
~~&&&(-0.03)&(0.41***)&(0.47***)&(0.48***)&(0.45***)\tabularnewline
\bottomrule
\end{tabular}
\end{center}}
\begin{minipage}[c]{\linewidth}
\footnotesize{\textbf{Notes}: The table shows RMSEs with LPL's in parentheses below for the best performing TVP-WN-SVD model with a pooling prior (i.e., full information set (FULL) and $\kappa = 0.05$). Asterisks indicate statistical significance for each model relative to a random walk at the $1$ ($^{***}$), $5$ ($^{**}$) and $10$ ($^{*}$) percent significance levels. }
\end{minipage}

\end{table}

\end{appendices}

\end{document}